\documentclass[12pt]{article}
\usepackage[figuresright]{rotating}
\usepackage{mathrsfs}
\usepackage{natbib,graphicx,setspace,lscape,longtable,xr}
\usepackage{mathrsfs,amsmath,amsthm,amssymb,color}
\usepackage{natbib,epsfig,graphicx,pdfpages}
\usepackage{rotating}
\usepackage{float}
\usepackage{bm}
\usepackage{ulem}
\usepackage{CJK}
\usepackage{ragged2e}
\usepackage{setspace}
\usepackage{threeparttable}
\usepackage{multirow}
\usepackage{enumitem}
\usepackage{hyperref}
\usepackage{etoolbox}
\usepackage{subfig}
\usepackage{booktabs}
\usepackage{bigstrut}
\bibpunct{(}{)}{;}{a}{,}{,}

\newtheorem{theorem}{Theorem}
\newtheorem{lemma}{Lemma}
\newtheorem{proposition}{Proposition}
\newcommand{\mylabel}[2]{#2\def\@currentlabel{#2}\label{#1}}
\newcommand{\csection}[1]
{\begin{center}
\stepcounter{section}
{\bf\large\arabic{section}. #1}
\end{center}
}

\newcommand{\scsection}[1]
{\begin{center}
{\bf\large #1}
\end{center}
}

\newcommand{\csubsection}[1]{
\begin{center}
\stepcounter{subsection}
{\it\arabic{section}.\arabic{subsection}. #1}
\end{center}
}

\newcommand{\scsubsection}[1]{
\begin{center}
\stepcounter{subsection}
{\it #1}
\end{center}
}

\def\tr{\mbox{tr}}

\def\beq{\begin{equation}}
\def\eeq{\end{equation}}
\def\beqr{\begin{eqnarray}}
\def\eeqr{\end{eqnarray}}
\def\beqrs{\begin{eqnarray*}}
\def\eeqrs{\end{eqnarray*}}
\def\bet{\begin{theorem}}
\def\eet{\end{theorem}}
\def\bel{\begin{lemma}}
\def\eel{\end{lemma}}
\def\bep{\begin{proposition}}
\def\eep{\end{proposition}}
\def\bg{\begin{figure}[tbph]\begin{center}}
\def\eg{\end{center}\end{figure}}

\def\bc{\begin{center}}
\def\ec{\end{center}}

\def\wh{\widehat}

\def\SE{\widehat{\mbox{SE}}}

\def\mR{\mathbb{R}}
\def\mS{\mathbb S}

\def\mE{\mathcal E}

\def\mS{\mathcal S}

\def\mY{\mathbb{Y}}

\def\var{\mbox{var}}

\def\argmax{\mbox{argmax}}
\def\rank{\mbox{rank}}
\def\diag{\mbox{diag}}

\renewcommand{\arraystretch}{1.3}

\usepackage[a4paper,bindingoffset=0.2in,%
left=1in,right=1in,top=1in,bottom=1in,%
footskip=.25in]{geometry}

\numberwithin{equation}{section}

\begin{document}
\begin{CJK}{GBK}{song}
\begin{center}
{\bf\Large Subnetwork Estimation for Spatial Autoregressive Models in Large-scale Networks}\\
\bigskip

Xuetong Li$^{1}$, Feifei Wang$^{2,3}$\footnote{The corresponding author. Email: feifei.wang@ruc.edu.cn}, Wei Lan$^{4}$, and Hansheng Wang$^1$

{\it\small
$^1$ Guanghua School of Management, Peking University, Beijing, China;\\
$^2$ Center for Applied Statistics, Renmin University of China, Beijing, China;\\
$^3$ School of Statistics, Renmin University of China, Beijing, China;\\
$^4$ School of Statistics, Southwestern University of Finance and Economics, Chengdu, China.

}

\end{center}

\begin{singlespace}
\begin{abstract}
Large-scale networks are commonly encountered in practice (e.g., Facebook and Twitter) by researchers. In order to study the network interaction between different nodes of large-scale networks, the spatial autoregressive (SAR) model has been popularly employed. Despite its popularity, the estimation of a SAR model on large-scale networks remains very challenging. On the one hand, due to policy limitations or high collection costs, it is often impossible for independent researchers to observe or collect all network information. On the other hand, even if the entire network is accessible, estimating the SAR model using the quasi-maximum likelihood estimator (QMLE) could be computationally infeasible due to its high computational cost.
To address these challenges, we propose here a subnetwork estimation method based on QMLE for the SAR model. By using appropriate sampling methods, a subnetwork, consisting of a much-reduced number of nodes, can be constructed. Subsequently, the standard QMLE can be computed by treating the sampled subnetwork as if it were the entire network. This leads to a significant reduction in information collection and model computation costs, which increases the practical feasibility of the effort. Theoretically, we show that the subnetwork-based QMLE is consistent and asymptotically normal under appropriate regularity conditions. Extensive simulation studies, based on both simulated and real network structures, are presented.

\end{abstract}

\noindent {\bf KEY WORDS:} Large-scale Networks; Network Sampling; Quasi-maximum Likelihood Estimator; Spatial Autoregressive Model; Subnetwork

\end{singlespace}

\newpage

\csection{INTRODUCTION}

With the rapid development of the Internet, network data are commonly encountered in daily life. Conceptually, a network refers to a collection of nodes and the internode relationships (i.e., edges). For example, on social network platforms such as Facebook and Twitter, billions of users form a giant friendship network by their follower-followee relationships \citep{dunbar2015onoffline,zhu2017network}. Another typical example is third-party online platforms such as Amazon and Yelp, where the merchants and consumers on the platform form a transaction network based on purchasing or commenting behaviors \citep{huang2020two}. Other important network examples include collaboration networks \citep{2017A}, citation networks \citep{2016Coauthorship}and the protein-protein interaction networks \citep{Zhao2011Discovery}; see \cite{newman2006structure} and \cite{anselin2013spatial} for more network examples.

An important research problem for network data is the network dependence between different nodes. To address this issue, the spatial autoregressive (SAR) model has been popularly used \citep{Lee:2003,lee2004asymptotic,anselin2013spatial}. Basically, a SAR model assumes that the behavior of each node can be influenced by its connected neighbors, and a spatial autoregression parameter $\rho$ is used to describe the network effect \citep{Cliff:Ord:1981,Su:2012,Malikov:Sun:2017}.
Since its first proposal, the SAR model has been widely used not only in spatial statistics \citep{newman2006structure,anselin2013spatial}, but also for social network data analysis. For example, in marketing studies, the parameter $\rho$ is used to measure the social intercorrelation between different consumers. For example, \cite{yang2003modeling} used the SAR model to study the preference interdependence among individual consumers. \cite{aravindakshan2012spatiotemporal} applied the SAR model to investigate the optimal spatial allocation strategies for advertising budgets. In economic studies, the SAR model and its extensions, are also widely applied. For example, \cite{Lung2015Identification} investigated the international spillover of economic growth on a trade network constructed by OECD countries. \cite{QU2021180} studied US industry interactions via a production network.

To estimate a SAR model, a quasi-maximum likelihood estimator (QMLE) can be used. Specifically, by temporarily assuming a normal distribution for the residual variable, a normal log-likelihood function can be developed. By optimizing the resulting log-likelihood function, a QMLE can be obtained. As shown by a number of pioneering researchers, \citep{shi2017spatial,yang2017identification,zhu2020multivariate}, the resulting QMLE is consistent and asymptotically normal under appropriate regularity conditions, even if the normality assumption is violated. Except for the QMLE, various generalized
methods of moments (GMM) have been proposed \citep{2001Generalized, lee2007gmm,2010GMM}. Under appropriate regularity conditions, the resulting GMM estimator is also consistent and asymptotically normal. Compared with the QMLE, the GMM estimator is computationally more efficient. However, it could be statistically less efficient unless the optimal moment conditions are used \citep{lee2007gmm,2010GMM}.

Despite its popularity, the implementation of a SAR model could be difficult in practice. This is because modern networks are often very large in size. For example, as the leading social media platform, Facebook had 2.80 billion monthly active users in 2021 \citep{2021Mohsin}. As another example, Sina Weibo (the largest Twitter-type social media platform in China) had approximately 0.53 billion monthly active users by the first quarter of 2021 \citep{2021Weibo}.
For these large networks, it is often impossible for independent researchers to publicly collect information for all nodes in the network. This is because the entire network structure is not fully visible to the public, and the information collection cost could be prohibitive. \citep{maiya2011benefits,chen2013impact}. Even if the entire network is accessible (e.g., for network owners such as Facebook), estimating the SAR model on such a large network using QMLE could be computationally impossible. This is because a standard Newton-Raphson algorithm is often used to compute the QMLE. For each Newton-Raphson iteration, the inverse of a very large-scale matrix needs to be computed, which makes its computational complexity $O(N^3)$ for a network with $N$ nodes. This leads to extremely high computational costs for large-scale networks. Therefore, estimating the SAR model for large-scale networks with reasonable collection and computational costs has become a problem of great interest \citep{2017Estimating,huang2019least}.

A natural way to tackle this problem is network sampling. In other words, instead of working on an extremely entire large-scale network directly, we might consider studying a subnetwork, which consists of a much reduced number of nodes and internode relationships. Let $n$ be the size of the subnetwork, and we should have $n\ll N$. Then, collecting information on this subnetwork could be cost saving. After obtaining the subnetwork by some carefully designed sampling methods, we can treat the sampled subnetwork as if it were the entire network. Subsequently, the QMLE can be directly computed on the subnetwork, and the computational cost is also greatly reduced. However, it should be noted that sampling a subnetwork would inevitably break the network relationships between the nodes inside and outside of the subnetwork. As a result, to what extent the observed subnetwork structure can serve as a good approximation of the entire network structure is very questionable. In fact, \cite{chen2013impact} empirically demonstrated that the subnetwork estimator for $\rho$ could be greatly biased unless carefully designed sampling methods are applied. However, the statistical theory underlying this interesting phenomenon remains unknown.

In literature, there already exist previously works studying the estimation method for spatial models. An earlier notable work is \cite{Kelejian2010Spatial}, who propose instrumental variable estimators for spatial models with incomplete network information. However, given the requirement for instrumental variables, these methods cannot be directly applied to SAR models without covariates. \cite{2017Estimating} target on the SAR model with sampled network data. They first approximate the likelihood function by its first-order Taylor's expansion and then develop an approximate MLE and a more computationally efficient paired MLE for estimation. \cite{huang2019least} construct a least squares type objective function for the SAR model. A new LSE-based network sampling technique is also developed for analyzing large-scale social networks. However, to our best knowledge, no theoretical analysis for subnetwork estimation with QMLE has been studied. Given the good theoretical properties of QMLE, a deep understanding of its theoretical performance on subnetwork is extremely important. This inspires us to develop this work.

Specifically, in this work, we investigate the theoretical properties of the subnetwork estimator for the SAR model. Our study reveals that a subnetwork QMLE could be consistent and asymptotically normal under critical conditions. That is, the number of relationships occurring between the nodes inside and outside of the subnetwork should be small enough, compared with the number of relationships inside the subnetwork. We conducted extensive simulation studies to numerically verify the finite sample performance of the subnetwork QMLE. We find that by using appropriate sampling methods, the performance of the subnetwork QMLE is fairly satisfactory. This encouraging phenomenon was observed not only on simulated network structures but also on two large-scale real network structures. One was a social network collected from Sina Weibo, and has 557,818 nodes. The other was a public cell-cell similarity network, which contains 1,018,524 nodes. To the best of our knowledge, no previous literature has studied network sampling on such large networks. The outstanding performance on these two large-scale network structures powerfully demonstrates the wide applicability of our proposed subnetwork estimation method in practice.

The remainder of this paper is organized as follows. Section 2 introduces the subnetwork estimation method for the SAR model and discusses its theoretical properties. Section 3 presents extensive numerical experiments on two simulated network structures and two real large-scale network structures. All technical details are delegated to the supplementary materials. Section 4 concludes the paper with a brief discussion.

\csection{THE METHODOLOGY}

\csubsection{Subnetwork QMLE}

Consider a network with a total of $N$ nodes, which are indexed by $1\le i \le N$. To describe the network structure of $N$ nodes, we define an adjacency matrix $A=(a_{ij})\in\mR^{N\times N}$. Specifically, we define $a_{ij}=1$ if node $i$ follows node $j$, and $a_{ij}=0$ otherwise. We follow the convention to define $a_{ii}=0$ for $1\le i \le N$. This leads to a popularly used spatial weighting matrix $W=(w_{ij})\in \mR^{N\times N}$ with $w_{ij}=a_{ij}/d_{i}$ and $d_{i}=\sum_{j=1}^{N}a_{ij}>0$. For each node $i$ with $1\le i \le N$, we can collect a continuous response of interest denoted by $Y_i$. Define $\mY = (Y_1,...,Y_N)^\top \in \mR^N$ as the response vector. To investigate the network relationship among $\mY$, the spatial autoregression model (SAR) is often assumed to be constructed as follows \citep{lee2004asymptotic, 2010GMM}:
\beq \label{eq: SAR model}
\mY= \rho W\mY+\mE,
\eeq
where $\rho \in \mR$ is the spatial autocorrelation parameter, $\mE = \big(\varepsilon_1,...,\varepsilon_N\big)^{\top} \in \mR^N$ is the residual vector, and $\{\varepsilon_i\}$s are assumed to be independent and identically distributed with mean 0, variance $\sigma^2$ and finite fourth-order moment. By \eqref{eq: SAR model}, we could obtain the reduced form of the SAR model, i.e., $\mY=(I_{N}-\rho W)^{-1}\mE$, where $I_{N} \in \mR^{N\times N}$ is an identity matrix. Following the classical literature on SAR models, we assume $|\rho|<1$ to ensure that the matrix $(I_{N} - \rho W)$ is invertible \citep{anselin1998spatial,sun1999posterior,2015Estimating}. This leads to the following log-likelihood function:
\beqrs
\mathcal{L}_0(\theta) = -\frac{1}{2 \sigma^{2}} \Big\{ \big(I_{N}-\rho W\big)\mY \Big\}^\top \Big\{ \big(I_{N}-\rho W \big)\mY \Big\} -\frac{N}{2} \ln \Big( 2\pi\sigma^{2} \Big)+\ln \Big| I_{N}-\rho W \Big|.
\eeqrs
Given $\rho$, the log-likelihood function $\mathcal{L}_0(\theta)$ can be optimized with respect to $\sigma^2$. Define $\widetilde{\sigma}^2(\rho) = \argmax _{\sigma^2} \mathcal{L}_0(\theta)$. By simple calculations, we can derive the closed form of $\widetilde{\sigma}^2(\rho)$ as $\widetilde{\sigma}^2(\rho) = N^{-1} \mY^\top (I_{N}-\rho W)^\top  (I_{N} - \rho W) \mY$. Then, a profiled log-likelihood function can be derived as:
\beq
\label{eq:whole-likelihood}
\mathcal{L}(\rho) = -\frac{N}{2}\Big\{\ln \Big( 2\pi \Big)+1\Big\}-\frac{N}{2} \ln \Big\{ \widetilde{\sigma}^2(\rho) \Big\}+\ln \Big|I_{N}-\rho W \Big|.
\eeq
This leads to the profiled quasi-maximum likelihood estimator (QMLE) for $\rho$ as $\wh\rho_{\rm mle}=\argmax_{\rho} \mathcal{L}(\rho)$. By substituting $\wh\rho_{\rm mle}$ in $\widetilde{\sigma}^2(\rho)$, we obtain the profiled quasi-maximum likelihood estimator for $\sigma^2$ as:
$\wh{\sigma}_{\rm mle}^2 = N^{-1} \mY^\top \big(I_{N}-\wh\rho_{\rm mle} W\big)^\top  \big(I_{N} - \wh\rho_{\rm mle} W\big) \mY$.
It is notable that, to motivate the log-likelihood function for the SAR model, we follow the classical literature to first assume normality for the error terms \citep{lee2004asymptotic,su2010profile,yang2017identification}. However, although the normality assumption is needed to motivate the log-likelihood function, the statistical consistency and asymptotic normality of the resulting QMLE are free from this normality assumption.

We next consider how to compute $\wh\rho_{\rm mle}$. Note that once $\wh\rho_{\rm mle}$ is given, $\wh\sigma^2_{\rm mle}$ can be computed accordingly. To compute $\wh\rho_{\rm mle}$, we need to derive the first- and second-order derivatives of $\mathcal{L} (\rho)$ with respect to $\rho$, which are denoted by $\dot{\mathcal{L}} (\rho)$ and $\ddot{\mathcal{L}} (\rho)$, respectively. With the help of matrix derivatives \citep{minka2000old,selby1973standard}, we obtain:
\beqr
\label{eq: first_d}
\dot{\mathcal{L}} (\rho) &=&
\Big\{\mY^{\top} W^{\top} \Big(I_{N}-\rho W\Big) \mY \Big\} \big/ \widetilde{\sigma}^{2}(\rho) -\tr \Big\{ W \Big(I_{N}-\rho W \Big)^{-1} \Big\},\nonumber\\
\ddot{\mathcal{L}} (\rho)&=& 2 \Big\{\mY^{\top} W^{\top} \Big(I_{N}-\rho W \Big) \mY\Big\}^2 \big/ \Big\{ N\widetilde{\sigma}^{4} (\rho) \Big\}
- \mY^{\top} W^{\top} W \mY \big/ \widetilde{\sigma}^{2} (\rho) \nonumber
\\&& - \tr \Big\{ W \Big(I_{N}-\rho W \Big)^{-1} \Big\}^{2}.\nonumber\label{eq: second_d}
\eeqr
The detailed derivations are given in Appendix A.2. With the help of $\dot{\mathcal{L}} (\rho)$ and $\ddot{\mathcal{L}} (\rho)$, a standard Newton-Raphson algorithm can be used to compute $\wh\rho_{\rm mle}$. In general, the Newton-Raphson algorithm often takes a few iterations to converge. However, the computational cost associated with each iteration could be very high when $N$ is large. The high computational cost is mainly because of the calculation of $(I_{N}-\rho W)^{-1}$, which requires $O\big(N^3\big)$ computational complexity. This could be practically infeasible when the network size is extremely large. Thus, estimating a SAR model with a very large network size becomes a problem of great importance.

One way to relieve this computational burden is to compute QMLE on a carefully selected subnetwork. Let $\mathbb{S} = \big\{ 1,..., N \big\}$ be the entire network. Without loss of generality, assume that the first $n$ nodes are sampled to form a subnetwork recorded by $\mathcal{S}_1=\{1,...,n\}$. Then, the set of nonsampled nodes is defined as $\mathcal{S}_2=\{n+1,...,N\}$. Define the response vectors on $\mathcal{S}_1$ and $\mathcal{S}_2$ as $\mY_1=\{Y_i| i \in \mathcal{S}_1\} \in \mR^{n}$ and $\mY_2=\{Y_i| i \in \mathcal{S}_2\} \in \mR^{N-n}$, respectively. Then we have $\mY=(\mY_1^{\top},\mY_2^{\top})^{\top} \in \mR^N$. The residual vector $\mE \in \mR^{N}$ can be similarly split into $\mE_1 \in \mR^n$ and $\mE_2 \in \mR^{N-n}$. In addition, the spatial weighting matrix $W \in \mR^{N \times N}$ can be partitioned as $W = \big( W_{11}, W_{12}; W_{21}, W_{22} \big)$, where $W_{11}\in \mR^{n\times n}$ represents the network relationships in $\mathcal{S}_1$, $W_{12}\in \mR^{n\times (N-n)}$ represents the relationships from $\mathcal{S}_1$ to $\mathcal{S}_2$, $W_{21}\in \mR^{(N-n)\times n}$ represents the relationships from $\mathcal{S}_2$ to $\mathcal{S}_1$, and $W_{22}\in \mR^{(N-n)\times (N-n)}$ represents the network relationships occurring inside $\mathcal{S}_2$. The identity matrix $I_{N}$ can be similarly partitioned as $I_{N} = \big( I_{11}, O_{12} ; O_{21}, I_{22} \big)$, where $I_{11} \in \mR^{n \times n}$ and $I_{22}\in \mR^{(N-n) \times (N-n)}$ are both identity matrices, and $O_{12} \in \mR^{n\times (N-n)}$ and $O_{21}\in \mR^{(N-n) \times n}$ are both zero matrices.

It is notable that, although the subnetwork weighting matrix $W_{11}$ is a part of $W$, it can be directly computed when the subnetwork adjacency matrix $A_{11}$ is observed and the total number of followees (i.e., $d_i$ for $i \in \mathcal{S}_1$) is given. For most social networks (such as Facebook, Sina Weibo), the number of followees of each user is part of information on the website, which is easily accessible. By obtaining the weighting matrix $W_{11}$, we treat it as if it were the entire network weighting matrix for $\mathcal{S}_1$. Then, a working SAR model on the subnetwork is:
\beq
\label{eq: sub model}
\mY_1 = \rho W_{11} \mY_1 + \mE_1.
\eeq
Obviously, the model \eqref{eq: sub model} is not the true model for the response vector $\mY_1$ in the subnetwork. However, by treating the working model \eqref{eq: sub model} as if it were the true model, we can compute the subnetwork QMLE for $\rho$ and $\sigma^2$ using similar techniques for the entire network. Specifically, a working profiled log-likelihood function for the subnetwork $\mS_1$ can be spelled as:
\beq
\label{eq: sub loss}
\mathcal{L}_{\mS}(\rho) = - \frac{n}{2}\Big\{\ln \Big( 2\pi \Big) + 1 \Big\}-\frac{n}{2} \ln \Big\{ \widetilde{\sigma}^2_{\mS}(\rho) \Big\} + \ln \Big|I_{11}-\rho W_{11} \Big|,
\eeq
where $\widetilde{\sigma}^2_{\mS}(\rho) = n^{-1} \mY_1^{\top} (I_{11}-\rho W_{11})^\top  (I_{11} - \rho W_{11}) \mY_1$. Next, by optimizing $\mathcal{L}_{\mS}(\rho)$ with respect to $\rho$, a subnetwork QMLE for $\rho$ can be obtained by $\wh\rho_{\mS} =\argmax_{\rho} \mathcal{L}_{\mS}(\rho)$. By further substituting $\wh\rho_{\mS}$ in $\widetilde{\sigma}^2_{\mS}(\rho)$, we can obtain the subnetwork QMLE for $\sigma^2$ as $\wh\sigma^2_{\mS} = n^{-1}\mY_1^{\top} \big(I_{11}-\wh\rho_{\mS} W_{11} \big)^\top \big(I_{11} - \wh\rho_{\mS} W_{11} \big) \mY_1$.
Compared with the entire network estimator $\wh\rho_{\rm mle}$, the subnetwork estimator $\wh\rho_{\mS}$ is computationally more feasible because we usually have $n \ll N$.
However, the working log-likelihood function \eqref{eq: sub loss} is not the true log-likelihood function for the response vector $\mY_1$. Consequently, whether the resulting subnetwork estimator is consistent is very questionable. A deep theoretical understanding in this regard is needed.

\csubsection{The Asymptotic Theory}

As we mentioned before, the subnetwork estimator $\wh\rho_{\mS}$ is unlikely to be consistent under a general condition \citep{chen2013impact}. Therefore, under what conditions the subnetwork estimator could be consistent becomes an interesting problem to study. To this end, we develop in this work an asymptotic theory for the subnetwork estimator $\wh\rho_{\mS}$. We find that the following technical conditions are needed to guarantee the statistical performance of the subnetwork estimator.

\begin{itemize}

\item[(C1)] {\sc (Network Connectivity)} Consider a Markov chain with the state space $\mathbb{S}$ and the transition probability matrix $W$. (C1.1) Assume this Markov chain is irreducible and aperiodic. Then, there should exist a stationary distribution vector $\pi = (\pi_i)\in\mR^N$, which satisfies
$W^{\top} \pi=\pi$. (C1.2) Further assume $\pi^{\top}\pi = O\big( N^{-1 + \tau} \big)$, where $0 \le \tau < 1/2$.

\item[(C2)] {\sc (Network Structure)}
For the adjacent matrix $A=(a_{i j})$, assume there exists a fixed positive constant $C_{\max}$, such that $\|A\|_{\max} = \max_{1 \le j \le N} \sum_{i=1}^{N}\left|a_{i j}\right| \leq C_{\max}$ as $N \rightarrow \infty$.

\item[(C3)] {\sc (Subnetwork Structure)} For an arbitrary matrix $A \in \mR^{N\times N}$, define $\lambda_{\min}(A)$ and $\lambda_{\max}(A)$ to be the smallest and largest eigenvalues of $A$. Then we have the following conditions.
(C3.1) Assume $\lambda_{\max}\big(W_{12}^\top W_{12}\big) \lambda_{\max}\big(W_{21}^\top W_{21}\big) < \rho^{-2}c_{\min}^2$, where $c_{\min}$ is the minimal value of $\lambda_{\min} \big\{ \big(I_{ii}-\rho W_{ii}^\top\big) \big(I_{ii}-\rho W_{ii}\big)  \big\}$ for $i=1$ and 2.
(C3.2) Assume $\tr\big(W_{12}^{\top}W_{12} \big)+\tr\big(W_{21}^{\top}W_{21} \big)$ is of the order $o(\sqrt{n})$.

\end{itemize}

Conditions (C1) and (C2) for the entire network are regular assumptions in the SAR model \citep{huang2019least,zhu2020multivariate}.
Specifically, condition (C1) contains two subconditions. Condition (C1.1) requires that the entire network is fully connected within itself. Otherwise, the network can be decomposed into several disconnected and independent subnetworks, which could be studied separately. Condition (C1.2) requires that the network structure be reasonably balanced. Otherwise, by treating $W$ as a Markov transition probability matrix, the resulting stationary distribution could be heavily skewed. Then it would violate the condition $\pi^\top \pi=O\big(N^{-1+\tau}\big)$. As one can see, a fully balanced case with $\pi_i = 1/N$ for $1 \le i \le N$ naturally satisfies this condition.
Condition (C2) imposes the constraint on the network structure.
This means that there exists no superstar in the network.
Further, it implies that $W$ is uniformly bounded in
both row and column sums.
Similar condition has been used in the past literature; see for example the Assumption 5 in \cite{lee2004asymptotic}.
Condition (C3) imposes constraints on the matrices $W_{12}$ and $W_{21}$, which are related to the missing edges between the selected subnetwork $\mS_1$ and the unselected subnetwork $\mS_2$.
It implies that, to guarantee the subnetwork estimator $\wh\rho_{\mS}$ to be statistically consistent, the missing edges contained in $W_{12}$ and $W_{21}$ should be sufficiently small. All these conditions are numerically verified for the studied network examples in Section 3; see Appendix B for details.

With the help of conditions (C1) and (C2), we can prove that the entire network QMLE $\wh\rho_{\rm mle}$ is consistent and asymptotically normal, i.e., $\sqrt{N} \big( \wh{\rho}_{\rm mle} - \rho \big) \rightarrow_{d} N\big(0, \sigma_{2}^{-4} \sigma_{1}^2 \big)$, where $\sigma_{1}^2 = \lim_{N \rightarrow \infty} \var\big\{\dot{\mathcal{L}}(\rho)\big\}$ and $\sigma_{2}^2 = \lim_{N \rightarrow \infty} -E\big\{\ddot{\mathcal{L}}(\rho)\big\}$.
These results are in accordance with the existing literature \citep{huang2019least,ma2020approximate,zhu2020multivariate}.
With the additional help of condition (C3), we can further establish a similar asymptotic theory for the subnetwork estimator $\wh\rho_{\mS}$. The detailed results are given in the following theorem:
\bet
\label{Theorem 1}
Assume conditions (C1)-(C3) hold and $\mu_4=E(\varepsilon_i^4)$ exists. Define $M_{\mS}=W_{11}\big(I_{11}-\rho W_{11}\big)^{-1}$. Let $\sigma_{1\mS}^2 = \lim_{n \rightarrow \infty} \big(1-\mu_4/\sigma^2\big)\tr^2(M_{\mS})/n^2+ \tr\big(M_{\mS}^\top M_{\mS}\big)/n + \tr\big(M_{\mS}^2\big)/n + \big(\mu_4/\sigma^2 -3\big)\tr\big\{\diag^2\big(M_{\mS}\big)\big\}$ and $\sigma_{2\mS}^2 = \lim_{n \rightarrow \infty} 2\tr^2(M_{\mS})/n^2 - \tr\big(M_{\mS}^\top M_{\mS}\big)/n -\tr\big(M_{\mS}^2\big)/n$. Then we have
$\sqrt{n} \big( \wh{\rho}_{\mS} - \rho \big) \rightarrow_{d} N\big(0, \sigma_{2\mS}^{-4} \sigma_{1\mS}^2 \big)$ as $n \to \infty$.
\eet
\noindent
The detailed proof of Theorem \ref{Theorem 1} is given in Appendix A.3. Theorem \ref{Theorem 1} assumes $n \to \infty$, which implies
$N \to \infty$ since $n \leq N$. By Theorem \ref{Theorem 1}, we know that the subnetwork estimator $\wh{\rho}_{\mS}$ can be $\sqrt{n}$-consistent and asymptotically normal, as long as the appropriate technical conditions can be satisfied.

\csubsection{Network Sampling Methods}

As we mentioned before, the subnetwork estimator is unlikely to be consistent unless appropriate network sampling method is used.
We mainly study here three network sampling methods. They are, respectively: (1) the simple random sampling without replacement (SRS) method, (2) the snowball sampling (SNOW) method, and (3) the clustering sampling (CS) method. Other classic network sampling methods are also explored, and the detailed discussions are present in Appendix C.

The SRS method is arguably the most commonly used network sampling method in practice for various purposes \citep{Frank2005,leskovec2006sampling}. The implementation details of SRS are given below. Recall that $\mathbb{S}=\{1,...,N\}$ collects all the nodes from the entire network. Then SRS should directly sample $n$ nodes from $\mathbb{S}$ by the method of simple random sampling without replacement. Collect the selected $n$ nodes by $\mS^*=\{i_{1},i_{2},...,i_{n}\}$. Then the nodes in $\mS^*$ and their edges constitute the sampled subnetwork. As one can see, the SRS method is easy to implement. However, by sampling nodes independently, the relationships among nodes are not taken into consideration. This often leads to subnetworks with extremely sparse network structure. As a consequence, the resulting subnetwork estimator becomes seriously biased. This suggests that SRS should not be used for our subnetwork estimation and thus is not included for subsequent numerical studies.

We next consider the SNOW method, which is another popularly used network sampling method \citep{2017Estimating,huang2019least}. The implementation details are given below.
We start with $n_0 \ll n$ randomly selected seed nodes, which are collected by the seed node set $\mS^{\dag}$. Define $\mS^{*}$ to be the set containing all selected nodes. Then we have $\mS^{*}=\mS^{\dag}$. Next, all nodes connected with the seed nodes are selected. The set of connected nodes is defined as $\mS^{*}_{+}=\{j:a_{ij}=1, i\in \mS^{\dag}, j\notin \mS^{*}\}$. We then merge all the selected nodes together, which leads to the updated $\mS^{*}=\mS^{*}\bigcup \mS^{*}_{+}$. If the number of sampled nodes in $\mS^{*}$ is smaller than the target sample size $n$, then $\mS^{\dag}$ should be further updated by treating $\mS^{*}_{+}$ as the seed nodes.
This process is repeated until the number of sampled nodes in $\mS^{*}$ is equal to or larger than $n$. If $|\mS^{*}|$ is larger than $n$, some nodes in the last updating step will be randomly dropped from $\mS^{*}$ to make the final sample size exactly equal to $n$. Then the nodes contained in $\mS^{*}$ and their network relationships constitute the sampled subnetwork.

Lastly, we introduce the CS method, which has been suggested by an insightful anonymous referee. This method is inspired by the empirical observation that many real-world networks are cluster structured \citep{Cherifi2019}. Then the whole network can be decomposed into many small clusters. The nodes belonging to the same cluster are more likely to be connected with each other, as compared with the nodes from different clusters. Therefore, if the whole network is cluster structured, we are motivated to sample clusters directly. Specifically, assume $\mathbb{S}=\bigcup_{k=1}^K\mS_k$, where $K$ is the total number of clusters and $\mS_k$ represents the $k$th cluster with $1\leq k \leq K$. We then start with $\mS^*=\mS_k$ for an arbitrarily selected cluster. If the size of $\mS^*$ is smaller than the pre-defined subnetwork size $n$, then next cluster should be randomly sampled. Denote the newly sampled cluster by $\mS^{*}_{+}$. We then update $\mS^*=\mS^*\bigcup\mS^*_{+}$. This process should be replicated till the size of $\mS^*$ is no smaller than $n$. Some nodes from the last updating step should be randomly dropped so that the size of $\mS^*$ should be exactly equal to $n$. Then the sampled nodes in $\mS^*$ and their network relationships constitute the sampled subnetwork. Practically, the CS method can be easily implemented if the cluster structure is known in advance. Otherwise, various community detection (or node clustering) methods can be considered \citep{Newman2004,2011Stochastic,De2014Mixing,Qi2022}.

\csection{NUMERICAL STUDIES}

\csubsection{Simulation Setup and Performance Measures}

We investigate the finite sample performance of the subnetwork estimator in this section. Specifically, we consider four network structures, including two synthetic network structures and two real network structures. The two synthetic network structures are generated from the stochastic block model \citep{wang1987stochastic,nowicki2001estimation} and the latent space model \citep{hoff2002latent}.
One of the two real large-scale networks is collected from Sina Weibo, and the other is a public network dataset, which is available in the Stanford Large Network Dataset Collection (SNAP).
The detailed description of each network structure is subsequently given.

Once the network structure (i.e., $A$) is given, the row-normalized weighting matrix $W$ can be computed. Various specifications of $\rho$ (i.e., 0, 0.2, 0.4, 0.6) are considered. Subsequently, the response vector $\mY$ can be generated as $\mY=(I_{N}-\rho W)^{-1}\mE$ with $\mE=(\varepsilon_1,...,\varepsilon_N)^{\top} \in \mR^{N}$. Here $\varepsilon_i$s are independent and identically generated from a standard norm ({\sc NORM}) distribution or a centralized standard exponential ({\sc EXP}) distribution. To numerically generate $\mY$, the matrix $(I_{N} - \rho W)$ needs to be inverted. This could be computationally infeasible if the network size $N$ is very large (e.g., $N=10^5$). To solve the problem, we adopt the polynomial approximation method to generate $\mY$, which has been widely applied in the past literature \citep{Golgher2016,2017Estimating,huang2019least,ma2020approximate,zhu2020multivariate}. Specifically, we have $||W||_{\max}= 1$ since $W$ is row-normalized. Then under the condition $|\rho|<1$, we have $||\rho W||_{\max} < 1$. Next, by Lemma 2.3.3 in \cite{Golub1996Matrix}, we can obtain $(I_{N} - \rho W)^{-1} = \sum_{k=0}^{\infty} \rho^k W^k$. Consequently, we have $\mY = (I_{N} - \rho W)^{-1}\mE = \sum_{k=0}^{\infty} \rho^k W^k\mE$. This suggest that $\mY$ can be numerically approximated by $\mY \approx \sum_{k=0}^{m} \rho^k W^k\mE$ for some reasonably large $m$. Here $m$ can be selected by monitoring the approximation performance. Note that $W$ is a sparse matrix. Thus the computation cost of $W^k$ is extremely low. It makes this approximation method for generating $\mY$ computationally feasible. Our extensive numerical experiments suggest that this method works very well.

We next apply the aforementioned network sampling methods to generate the subnetwork. As one can expect that, network sampling method plays a significant role in determining the subnetwork estimator. In this regard, three network sampling methods have been introduced in Subsection 2.3. As we mentioned before, the SRS method often leads to extremely sparse subnetwork structure, which can hardly be used for subnetwork estimation. Therefore, we only study in this section the other two network sampling methods (i.e., SNOW and CS). The SNOW method is adopted for all network structures. However, the CS method is only applied in the case that the whole network is cluster structured. Therefore, we only apply the CS method on the two synthetic network examples (i.e., SBM and LSM), which are assumed bo be cluster structured. In addition, their cluster structures are assumed to be known in advance.

Once a subnetwork is generated, a subnetwork estimator $\wh\rho_\mS$ can be computed. For a reliable evaluation, each experiment is randomly replicated for a total of $M = 500$ times.
Let $\wh\rho_{\mS}^{(m)}$ denote the estimate obtained in the $m$-th replication. We then evaluate the bias as $\flat=\bar\rho-\rho$, where $\bar\rho=M^{-1}\sum_m\wh\rho_{\mS}^{(m)}$. Let $\SE^{(m)}$ be the estimated standard error computed in the $m$-th replication according to Theorem 1. Its average is then computed as $\SE=M^{-1}\sum_m \SE^{(m)}$. The true standard deviation of $\wh\rho_{\mS}$ is estimated by $\mbox{SE}=\{M^{-1}\sum_m (\wh\rho_{\mS}^{(m)}-\bar\rho)^2\}^{1/2}$.
We next construct a $95\%$ confidence interval for $\rho$ as: $\text{CI}^{(m)}=(\wh\rho_{\mS}^{(m)}-z_{0.975}\SE^{(m)},\wh\rho_{\mS}^{(m)}+z_{0.975}\SE^{(m)})$, where $z_{\alpha}$ is the $\alpha$-th lower quantile of a standard normal distribution. Then, the empirical coverage probability is computed as $\text{ECP}=M^{-1}\sum_{m}I\big(\rho \in \text{CI} ^{(m)}\big)$, where
$I(\cdot)$ is the indicator function. Lastly, the average CPU time is also computed and reported. All the details are summarized in Tables 1--3.

\csubsection{The Stochastic Block Model}

\begin{figure}[h]
	\centering
	\subfloat[\textit{Small Communities} ($N/K = 5$)]{
		\includegraphics[width=0.47\textwidth]{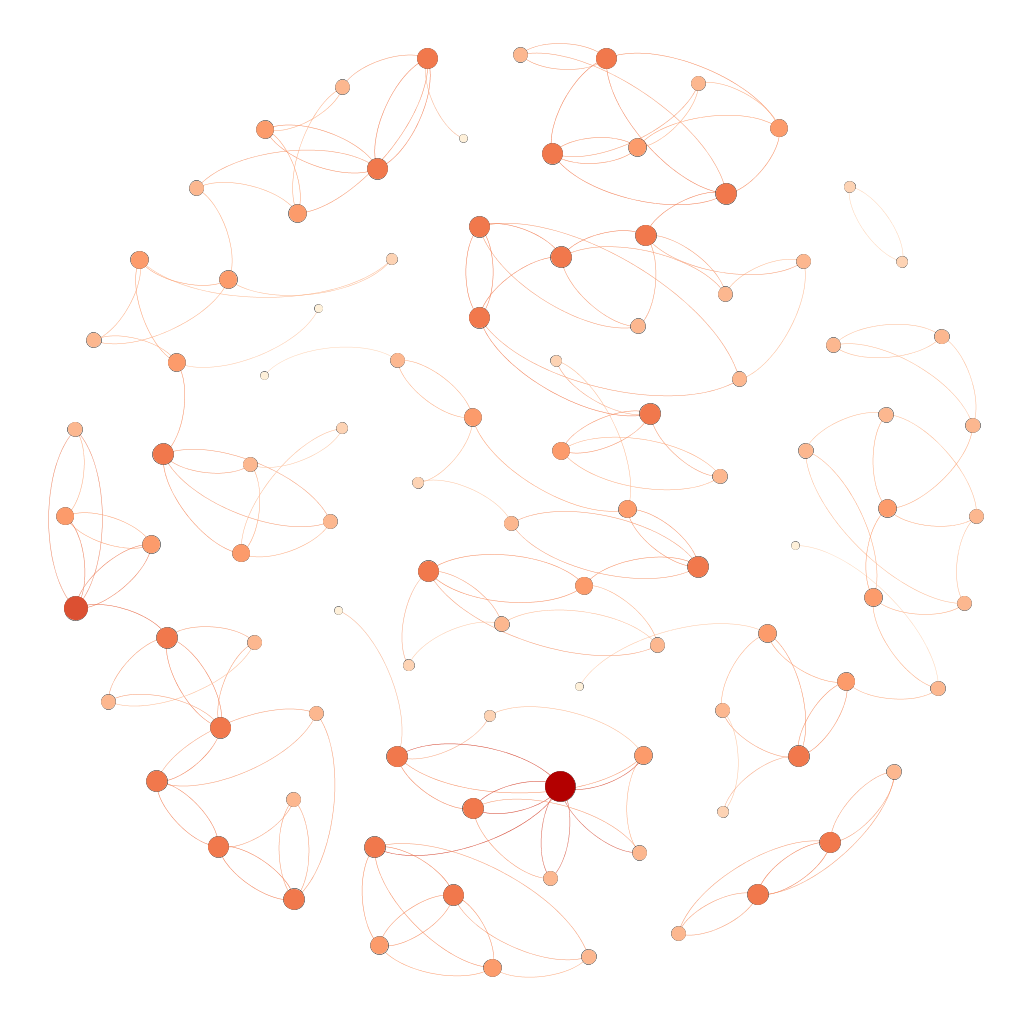}}\hfill
	\subfloat[\textit{Large Communities} ($N/K=25$)]{
		\includegraphics[width=0.47\textwidth]{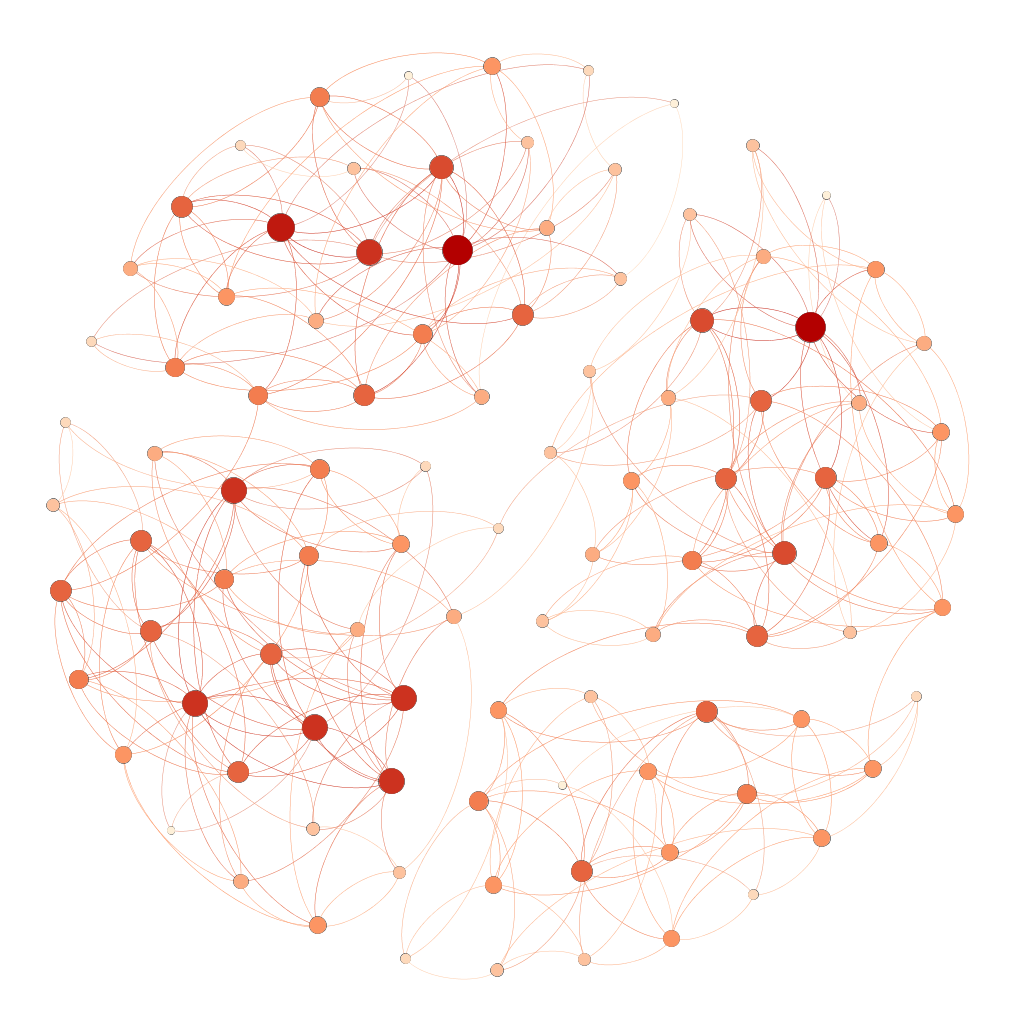}}\hfill
	\caption{ Visualization of two toy network examples generated from the stochastic block model with $N=100$. The left panel is an example with small communities (i.e., $N/K = 5$), while the right panel is an example with large communities (i.e., $N/K=25$).  In both networks, a dot denotes a node and a line represents an edge. A deeper color and a larger dot indicate a larger in-degree.
}
	\label{f:sbm}
\end{figure}

We start with our first type of synthetic network structure. That is the stochastic block model (SBM). It is a network structure that has been popularly used for community detection \citep{wang1987stochastic,nowicki2001estimation,zhao2012consistency,2018Community}.
Specifically, let $K$ be the total number of blocks. For the $i$-th node with $1\le i\le N$, let $c_i \in \{1,2,\cdots,K\}$ be its block membership, which is randomly assigned with equal probability $1 / K$.
Assume the edge between node $i$ and node $j$ is independently generated with $P(a_{ij}=1)=0.2N^{-1}$ if $c_i=c_j$, $P(a_{ij}=1)=0.2N^{-1.5}$ if $|c_i-c_j|=1$ and $P(a_{ij}=1)=0.2N^{-2}$ otherwise.
In this way, nodes within the same block are more likely to be connected with each other.
For visualization purpose, we generate two toy networks ($N=100$) from SBM with $N/K=5$ and $N/K=25$, respectively. The corresponding network structures are shown in Figure \ref{f:sbm}.

Next, we investigate the finite sample performance of the subnetwork estimator. For a comprehensive evaluation, three different network sizes are considered, i.e., $N=(10000,30000,50000)$.
For a fixed $N$, three different community sizes are also studied. They are, respectively, $N/K=10, 20$ and 50. Once the network adjacency matrix $A$ is generated, the row-normalized weighting matrix $W$ can be constructed. Therefore, the response vector $\mY$ can be generated by the approximation method described in the previous subsection. After data generation, the subnetwork estimation method is applied with the sample size $n$. Here we fix $n/N = 0.01$ for example. The detailed simulation results with the error terms following the {\sc EXP} error distribution are summarized in Table \ref{t:SBM}. The simulation results with the {\sc NORM} errors are quantitatively similar and thus not reported to save space.

\begin{landscape}
\begin{table}[t]
\caption{The detailed simulation results under the SBM network structure, the {\sc EXP} error distribution, and different network sizes, community sizes, and network coefficients. The bias $\flat$, estimated standard error $\wh{\rm SE}$, true standard error SE, and the empirical coverage probability ECP are reported for both the SNOW and CS methods. The average CPU computational time is also reported in seconds.}
\label{t:SBM}
\centering
\scriptsize
\begin{tabular}{cc|ccccc|ccccc|ccccc|ccccc}
\hline
\hline
\multirow{2}{*}{$N$} &\multirow{2}{*}{Method} &\multicolumn{5}{c|}{$\rho=0$} &\multicolumn{5}{c|}{$\rho=0.2$} &\multicolumn{5}{c|}{$\rho=0.4$} & \multicolumn{5}{c}{$\rho=0.6$} \\
\cline{3-22}
 & &$\flat$ &$\wh{\rm SE}$ &SE &ECP &CPU &$\flat$ &$\wh{\rm SE}$ &SE &ECP &CPU &$\flat$ &$\wh{\rm SE}$ &SE &ECP &CPU &$\flat$ &$\wh{\rm SE}$ &SE &ECP &CPU\\
\hline
\multicolumn{22}{c}{$N/K=10$}	\\
\multirow{2}{*}{10000} 	&SNOW	&0.000	&0.131	&0.129	&94.8\%	&0.01	&-0.005	&0.121	&0.120	&94.0\%	&0.01	&-0.008	&0.105	&0.105	&93.6\%	&0.01	&-0.006	&0.080	&0.093	&94.2\%	&0.02	\\
	&CS	&-0.004	&0.128	&0.118	&96.8\%	&0.02	&-0.008	&0.118	&0.109	&96.8\%	&0.02	&-0.012	&0.102	&0.095	&95.8\%	&0.01	&-0.012	&0.077	&0.078	&96.4\%	&0.03	\\
	&	&	&	&	&	&	&	&	&	&	&	&	&	&	&	&	&	&	&	&	&	\\
\multirow{2}{*}{30000} 	&SNOW	&0.000	&0.075	&0.078	&94.2\%	&0.04	&-0.002	&0.070	&0.073	&94.4\%	&0.04	&-0.004	&0.060	&0.063	&94.0\%	&0.04	&-0.004	&0.045	&0.048	&94.6\%	&0.05	\\
	&CS	&-0.006	&0.074	&0.072	&95.0\%	&0.04	&-0.007	&0.068	&0.066	&95.0\%	&0.04	&-0.008	&0.059	&0.057	&94.8\%	&0.05	&-0.007	&0.044	&0.044	&94.6\%	&0.05	\\
	&	&	&	&	&	&	&	&	&	&	&	&	&	&	&	&	&	&	&	&	&	\\
\multirow{2}{*}{50000} 	&SNOW	&-0.006	&0.058	&0.059	&96.0\%	&0.09	&-0.006	&0.054	&0.056	&95.2\%	&0.10	&-0.006	&0.046	&0.049	&94.6\%	&0.11	&-0.005	&0.035	&0.037	&93.4\%	&0.12	\\
	&CS	&-0.002	&0.057	&0.057	&96.0\%	&0.08	&-0.003	&0.053	&0.053	&95.8\%	&0.07	&-0.004	&0.045	&0.045	&95.0\%	&0.09	&-0.004	&0.034	&0.034	&94.6\%	&0.10	\\
\hline
\multicolumn{22}{c}{$N/K=20$}	\\
\multirow{2}{*}{10000} 	&SNOW	&-0.005	&0.137	&0.134	&96.2\%	&0.01	&-0.010	&0.131	&0.127	&95.8\%	&0.02	&-0.014	&0.118	&0.116	&95.8\%	&0.01	&-0.016	&0.096	&0.096	&95.4\%	&0.02	\\
	&CS	&-0.004	&0.136	&0.132	&95.4\%	&0.01	&-0.009	&0.130	&0.125	&94.6\%	&0.01	&-0.014	&0.116	&0.114	&94.2\%	&0.01	&-0.016	&0.094	&0.101	&94.8\%	&0.02	\\
	&	&	&	&	&	&	&	&	&	&	&	&	&	&	&	&	&	&	&	&	&	\\
\multirow{2}{*}{30000} 	&SNOW	&0.001	&0.079	&0.082	&93.0\%	&0.03	&-0.001	&0.075	&0.076	&95.4\%	&0.04	&-0.004	&0.067	&0.068	&95.4\%	&0.04	&-0.005	&0.054	&0.054	&94.6\%	&0.05	\\
	&CS	&0.000	&0.078	&0.075	&96.0\%	&0.04	&-0.003	&0.074	&0.071	&95.6\%	&0.03	&-0.006	&0.066	&0.064	&95.6\%	&0.05	&-0.007	&0.053	&0.052	&96.0\%	&0.05	\\
	&	&	&	&	&	&	&	&	&	&	&	&	&	&	&	&	&	&	&	&	&	\\
\multirow{2}{*}{50000} 	&SNOW	&-0.005	&0.061	&0.063	&93.6\%	&0.08	&-0.006	&0.058	&0.059	&94.2\%	&0.08	&-0.007	&0.052	&0.053	&94.4\%	&0.10	&-0.007	&0.042	&0.043	&94.4\%	&0.10	\\
	&CS	&0.000	&0.060	&0.058	&95.8\%	&0.08	&-0.001	&0.057	&0.055	&96.0\%	&0.09	&-0.003	&0.051	&0.050	&95.4\%	&0.09	&-0.004	&0.041	&0.040	&95.0\%	&0.09	\\
\hline
\multicolumn{22}{c}{$N/K=50$}	\\
\multirow{2}{*}{10000} 	&SNOW	&0.002	&0.168	&0.164	&95.2\%	&0.01	&-0.001	&0.164	&0.160	&95.2\%	&0.01	&-0.002	&0.156	&0.152	&95.6\%	&0.02	&0.008	&0.142	&0.139	&93.6\%	&0.03	\\
	&CS	&0.008	&0.142	&0.138	&95.2\%	&0.01	&0.002	&0.138	&0.134	&94.2\%	&0.01	&-0.005	&0.129	&0.126	&94.2\%	&0.01	&-0.012	&0.111	&0.116	&94.2\%	&0.02	\\
	&	&	&	&	&	&	&	&	&	&	&	&	&	&	&	&	&	&	&	&	&	\\
\multirow{2}{*}{30000} 	&SNOW	&0.004	&0.083	&0.081	&95.8\%	&0.04	&0.001	&0.081	&0.080	&95.6\%	&0.04	&-0.002	&0.075	&0.076	&94.6\%	&0.05	&-0.004	&0.065	&0.066	&93.8\%	&0.05	\\
	&CS	&0.006	&0.081	&0.081	&94.8\%	&0.05	&0.005	&0.079	&0.078	&95.2\%	&0.05	&0.003	&0.073	&0.071	&96.8\%	&0.04	&0.001	&0.063	&0.061	&96.4\%	&0.04	\\
	&	&	&	&	&	&	&	&	&	&	&	&	&	&	&	&	&	&	&	&	&	\\
\multirow{2}{*}{50000} 	&SNOW	&0.000	&0.063	&0.066	&93.8\%	&0.07	&-0.001	&0.061	&0.064	&93.4\%	&0.08	&-0.003	&0.057	&0.060	&92.4\%	&0.09	&-0.004	&0.049	&0.052	&92.2\%	&0.09	\\
	&CS	&0.003	&0.062	&0.059	&96.8\%	&0.08	&0.003	&0.061	&0.057	&96.8\%	&0.07	&0.002	&0.056	&0.052	&96.6\%	&0.07	&0.001	&0.048	&0.045	&96.2\%	&0.09	\\
\hline
\hline
\end{tabular}
\end{table}
\end{landscape}

By Table \ref{t:SBM}, we can draw the following conclusions. First, we find the bias of the subnetwork estimator $\wh\rho_{\mS}$ shrinks towards 0 as the network size $N$ diverges to infinity. In the meanwhile, both SE and $\SE$ decrease towards 0. This implies the subnetwork estimator should be statistically consistent. Second, the empirical coverage probabilities of all estimates are around the nominal level of 95\%. This result is not surprising since the difference between SE and $\wh{\rm SE}$ is quite small. This suggests that the standard error estimator (i.e., $\SE$) according to Theorem 1 can approximate the true standard error (i.e., SE) fairly well. Last, we find the performance of the SNOW and CS methods are very comparable. Tiny differences can be detected if $\rho$ is relatively large (e.g., $\rho=0.6$) and the community size is relatively large (e.g., $N/K=50$). However, this comparable performance is obtained under the assumption that the cluster structure is known in advance. Otherwise the performance of CS could be much worse.

\csubsection{A Latent Space Network}

\begin{figure}[h]
	\centering
	\subfloat[\textit{Small Clusters} ($N/K=5$)]{
		\includegraphics[width=0.47\textwidth]{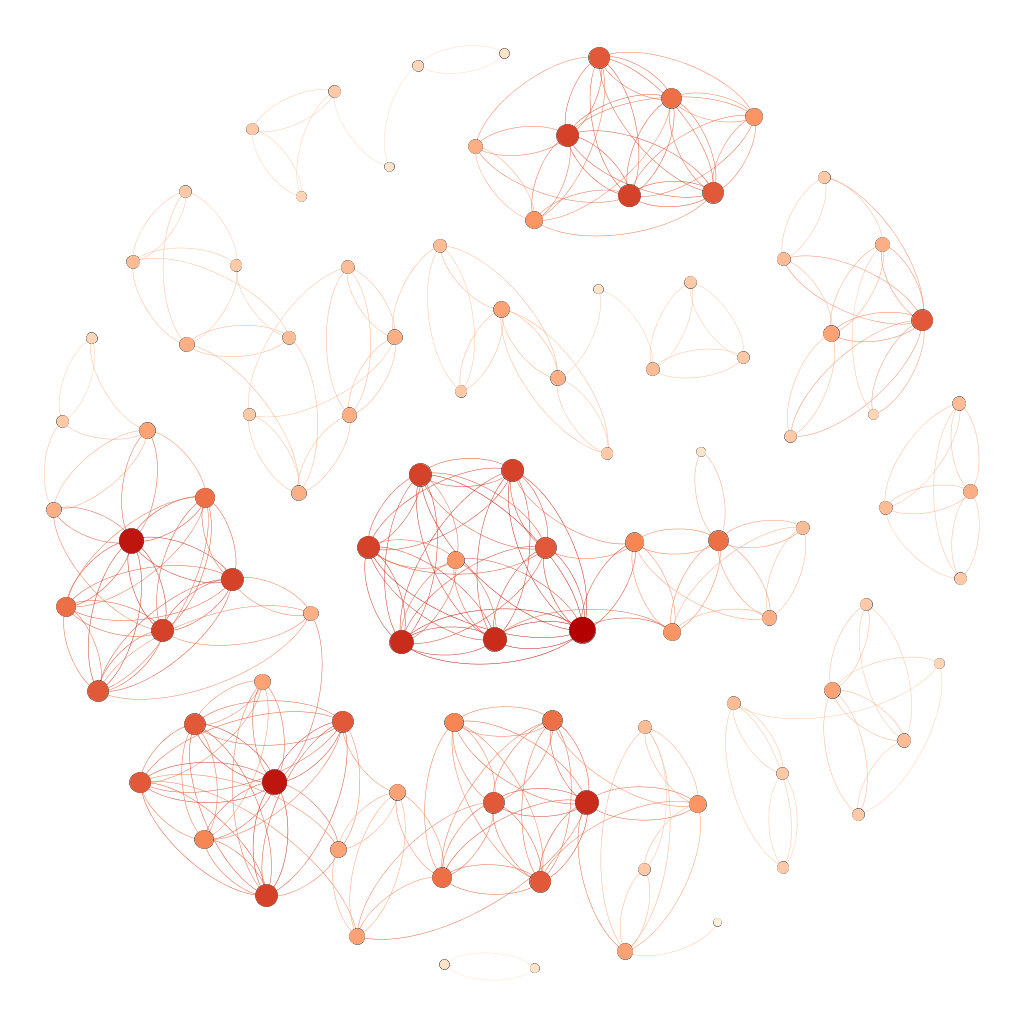}}\hfill
	\subfloat[\textit{Large Clusters} ($N/K=25$)]{
		\includegraphics[width=0.47\textwidth]{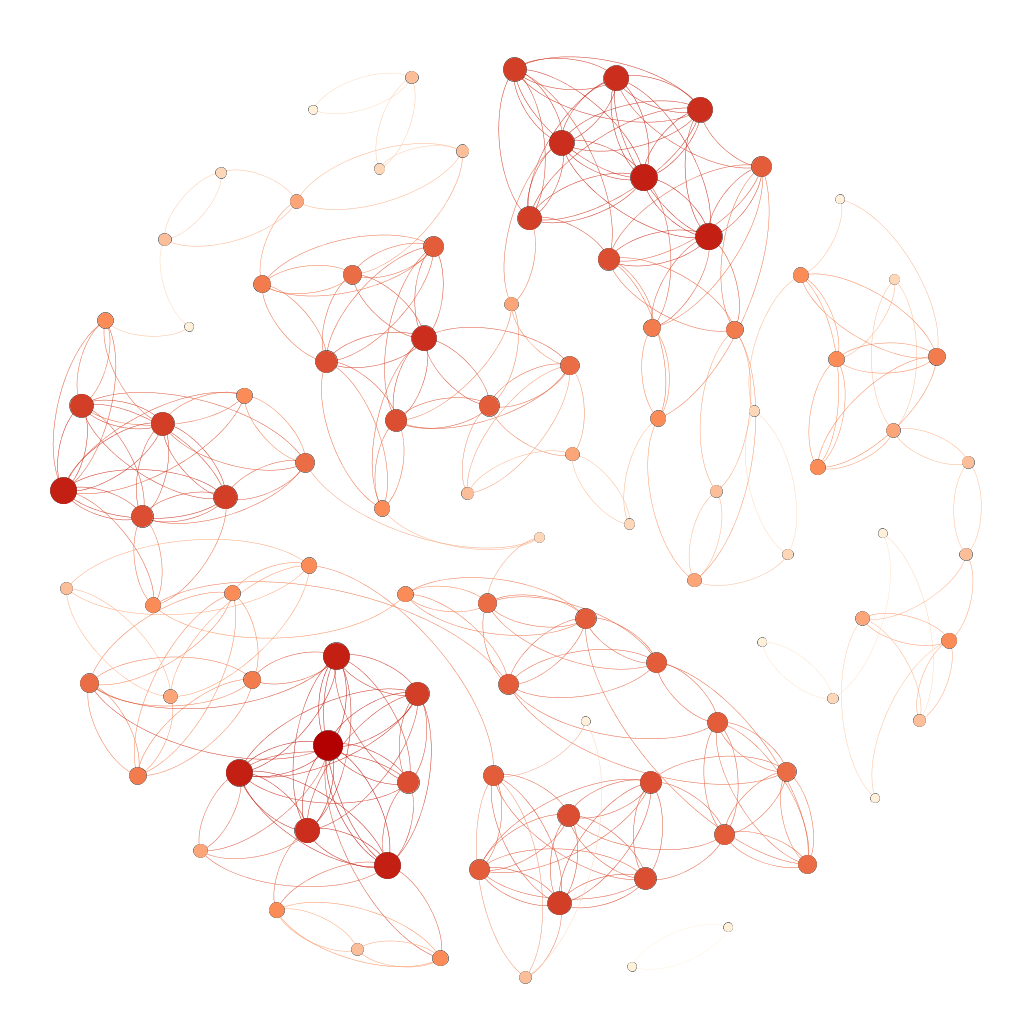}}\hfill
	\caption{Visualization of two toy network examples generated from the  latent space model with $N=100$. The left panel is an example with small cluster (i.e., $N/K=5$), while the right panel is an example with large clusters (i.e., $N/K=25$).
	In both networks, a dot denotes a node and a line represents an edge. A deeper color and a larger dot indicate a larger in-degree.}
	\label{f:lsm}
\end{figure}

We study another synthetic network structure, which is generated from the latent space model (LSM) of \cite{hoff2002latent}. This model assumes that the probability of a relationship between two different nodes depends on their positions in an unobserved ``latent space", along with some observed covariates.
Assume the network still has a cluster structure. Let $c_i \in \{1,2,\cdots,K\}$ be the cluster membership for the $i$-th node with $1\le i\le N$, which is randomly assigned with equal probability $1 / K$. For the $i$-th node, assume it has a latent position $Z_i \in \mR$, which is generated from a normal distribution with mean $\mu_{c_i}$ and variance $\sigma^2$. Here we fix $\sigma^2=1$ and let $\mu_{k}=2k$ with $1\leq k \leq K$. There also exists an observed covariate $X_{ij} \in \mR$ for each node pair $(i,j)$. The covariate $X_{ij}$ is independently generated from the standard normal distribution. Finally, we follow \cite{hoff2002latent} and assume $P(a_{ij}=1\mid Z_{i}, Z_{j}, X_{ij})=\big\{1+\exp(-c_{ij})\big\}^{-1}$, where $c_{ij}=\alpha_{ij}+\beta X_{ij}-N\big|Z_{i}-Z_{j}\big|/K$.
Here, $(\alpha_{ij},\beta)$ are predefined parameters.
We fix $\beta=1$ and consider $\alpha_{ij}=5$ if node $i$ and node $j$ belong to the same cluster (i.e., $c_i=c_j$) and $\alpha_{ij}=1$ otherwise.
This leads to the entire network structure $A$. For illustration, Figure \ref{f:lsm} present two toy networks ($N=100$) generated from LSM with $N/K=5$ and $N/K=25$, respectively. As shown, both toy networks have clear cluster structures.

To investigate the performance of subnetwork estimation, we consider three different network sizes $N=(10000,30000,50000)$, and fix the subsampling ratio to be $n/N=0.01$. For a fixed $N$, three different cluster sizes are also studied. They are, respectively, $N/K =$ 10, 20 and 50. The detailed simulation results with the error terms following the {\sc EXP} error distribution are summarized in Table \ref{t:LSM}, and we omit the simulation results with the {\sc NORM} error distribution to save space. By Table \ref{t:LSM}, we find the results under the LSM network structure are quantitatively similar to those under the SBM network structure. Specifically, the bias $\flat$ of the subnetwork estimator $\wh\rho_{\mS}$ is sufficiently small when the network size is relatively large.
For example, all biases $\flat$ are close to zero under network structure with $N/K=10$. The SE and $\SE$ have similar values, and both decrease as the network size $N$ (also the sample size $n$) increases, which indicates the statistical consistency of $\wh\rho_{\mS}$.
The empirical coverage
\begin{landscape}
\begin{table}[t]
\caption{The detailed simulation results under the LSM network structure, the {\sc EXP} error distribution, and different network sizes, cluster sizes, and network coefficients. The bias $\flat$, estimated standard error $\wh{\rm SE}$, true standard error SE, and the empirical coverage probability ECP are reported for both the SNOW and CS methods. The average CPU computational time is also reported in seconds.}
\label{t:LSM}
\centering
\scriptsize
\begin{tabular}{cc|ccccc|ccccc|ccccc|ccccc}
\hline
\hline
\multirow{2}{*}{$N$} &\multirow{2}{*}{Method} &\multicolumn{5}{c|}{$\rho=0$} &\multicolumn{5}{c|}{$\rho=0.2$} &\multicolumn{5}{c|}{$\rho=0.4$} & \multicolumn{5}{c}{$\rho=0.6$} \\
\cline{3-22}
 & &$\flat$ &$\wh{\rm SE}$ &SE &ECP &CPU &$\flat$ &$\wh{\rm SE}$ &SE &ECP &CPU &$\flat$ &$\wh{\rm SE}$ &SE &ECP &CPU &$\flat$ &$\wh{\rm SE}$ &SE &ECP &CPU\\
\hline
\multicolumn{22}{c}{$N/K=10$}	\\
\multirow{2}{*}{10000} 	&SNOW	&-0.008	&0.134	&0.125	&95.8\%	&0.01	&-0.012	&0.123	&0.116	&96.2\%	&0.02	&-0.014	&0.106	&0.101	&96.0\%	&0.01	&-0.011	&0.081	&0.089	&96.0\%	&0.02	\\
	&CS	&-0.016	&0.142	&0.142	&94.4\%	&0.02	&-0.017	&0.132	&0.132	&94.6\%	&0.02	&-0.014	&0.114	&0.115	&94.2\%	&0.02	&-0.006	&0.089	&0.101	&94.4\%	&0.03	\\
	&	&	&	&	&	&	&	&	&	&	&	&	&	&	&	&	&	&	&	&	&	\\
\multirow{2}{*}{30000} 	&SNOW	&-0.003	&0.076	&0.077	&93.8\%	&0.04	&-0.004	&0.069	&0.071	&93.4\%	&0.05	&-0.005	&0.059	&0.061	&94.0\%	&0.05	&-0.005	&0.029	&0.030	&94.4\%	&0.06	\\
	&CS	&-0.006	&0.082	&0.080	&95.4\%	&0.05	&-0.006	&0.075	&0.073	&95.6\%	&0.05	&-0.002	&0.065	&0.063	&95.2\%	&0.06	&0.002	&0.050	&0.048	&95.6\%	&0.07	\\
	&	&	&	&	&	&	&	&	&	&	&	&	&	&	&	&	&	&	&	&	&	\\
\multirow{2}{*}{50000} 	&SNOW	&-0.002	&0.058	&0.060	&94.2\%	&0.09	&-0.003	&0.053	&0.055	&94.0\%	&0.09	&-0.003	&0.046	&0.048	&93.4\%	&0.10	&-0.003	&0.035	&0.037	&92.4\%	&0.09	\\
	&CS	&-0.002	&0.064	&0.064	&95.8\%	&0.10	&-0.002	&0.059	&0.059	&95.4\%	&0.11	&0.001	&0.050	&0.052	&95.2\%	&0.12	&0.005	&0.039	&0.040	&93.2\%	&0.12	\\
\hline
\multicolumn{22}{c}{$N/K=20$}	\\
\multirow{2}{*}{10000} 	&SNOW	&-0.001	&0.137	&0.130	&95.2\%	&0.01	&-0.005	&0.126	&0.121	&95.2\%	&0.01	&-0.007	&0.108	&0.105	&94.4\%	&0.02	&-0.004	&0.083	&0.096	&94.0\%	&0.02	\\
	&CS	&-0.018	&0.145	&0.143	&95.4\%	&0.01	&-0.021	&0.135	&0.132	&95.2\%	&0.02	&-0.017	&0.117	&0.115	&96.4\%	&0.02	&-0.009	&0.092	&0.099	&95.8\%	&0.03	\\
	&	&	&	&	&	&	&	&	&	&	&	&	&	&	&	&	&	&	&	&	&	\\
\multirow{2}{*}{30000} 	&SNOW	&-0.003	&0.076	&0.073	&96.2\%	&0.03	&-0.003	&0.070	&0.066	&96.6\%	&0.04	&-0.003	&0.060	&0.057	&96.4\%	&0.04	&-0.003	&0.046	&0.045	&95.8\%	&0.05	\\
	&CS	&-0.003	&0.083	&0.084	&94.8\%	&0.04	&-0.002	&0.076	&0.077	&95.0\%	&0.06	&0.001	&0.066	&0.066	&94.6\%	&0.05	&0.005	&0.051	&0.051	&94.4\%	&0.06	\\
	&	&	&	&	&	&	&	&	&	&	&	&	&	&	&	&	&	&	&	&	&	\\
\multirow{2}{*}{50000} 	&SNOW	&0.002	&0.059	&0.057	&96.8\%	&0.08	&0.002	&0.054	&0.052	&96.8\%	&0.09	&0.001	&0.046	&0.044	&96.2\%	&0.09	&0.000	&0.035	&0.034	&94.8\%	&0.10	\\
	&CS	&0.001	&0.064	&0.060	&97.0\%	&0.10	&0.002	&0.059	&0.055	&97.4\%	&0.13	&0.005	&0.051	&0.048	&96.4\%	&0.11	&0.008	&0.039	&0.037	&95.2\%	&0.14	\\
\hline
\multicolumn{22}{c}{$N/K=50$}	\\
\multirow{2}{*}{10000} 	&SNOW	&0.003	&0.140	&0.134	&95.8\%	&0.01	&-0.002	&0.129	&0.124	&95.4\%	&0.02	&-0.005	&0.111	&0.109	&94.0\%	&0.02	&-0.005	&0.086	&0.092	&93.6\%	&0.03	\\
	&CS	&-0.003	&0.146	&0.141	&95.4\%	&0.01	&-0.005	&0.135	&0.131	&95.6\%	&0.01	&-0.003	&0.117	&0.113	&95.4\%	&0.02	&0.001	&0.092	&0.093	&95.0\%	&0.03	\\
	&	&	&	&	&	&	&	&	&	&	&	&	&	&	&	&	&	&	&	&	&	\\
\multirow{2}{*}{30000} 	&SNOW	&0.001	&0.076	&0.075	&94.6\%	&0.04	&-0.001	&0.070	&0.069	&93.6\%	&0.06	&-0.003	&0.060	&0.060	&94.2\%	&0.05	&-0.004	&0.046	&0.047	&93.8\%	&0.07	\\
	&CS	&-0.002	&0.083	&0.075	&97.4\%	&0.05	&-0.003	&0.076	&0.069	&97.2\%	&0.06	&-0.001	&0.066	&0.060	&96.8\%	&0.06	&0.002	&0.052	&0.047	&96.0\%	&0.07	\\
	&	&	&	&	&	&	&	&	&	&	&	&	&	&	&	&	&	&	&	&	&	\\
\multirow{2}{*}{50000} 	&SNOW	&-0.003	&0.059	&0.057	&95.4\%	&0.10	&-0.004	&0.054	&0.052	&96.0\%	&0.11	&-0.004	&0.046	&0.045	&95.6\%	&0.11	&-0.004	&0.035	&0.035	&95.2\%	&0.14	\\
	&CS	&0.001	&0.064	&0.063	&96.0\%	&0.09	&0.000	&0.059	&0.058	&96.0\%	&0.10	&0.003	&0.051	&0.050	&95.8\%	&0.12	&0.006	&0.040	&0.039	&93.6\%	&0.12	\\
\hline
\hline
\end{tabular}
\end{table}
\end{landscape}

\noindent
probabilities of all estimates are around the nominal level 95\%. Finally, the two network sampling methods SNOW and CS also perform similarly for the LSM network structures. Tiny differences can be detected when $\rho$ is relatively large or the cluster size $N/K$ is relatively large. However, similar with the SBM example, the comparable performance of CS is obtained under the assumption that the cluster structure is known in advance. Otherwise the performance of the CS method could be much worse.

\csubsection{The Sina Weibo Network}

\begin{figure}[h]
	\centering
\subfloat[Histogram of In-degree]{
		\includegraphics[width=0.49\textwidth]{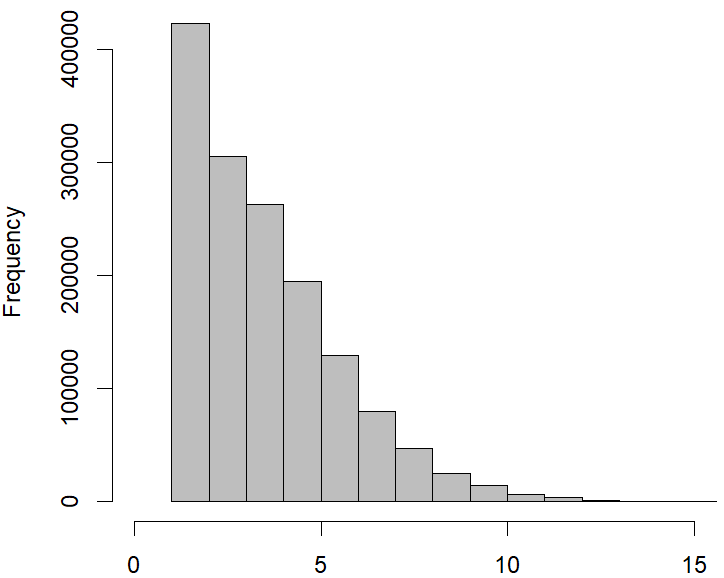}}\hfill
\subfloat[Network Visualisation]{
		\includegraphics[width=0.44\textwidth]{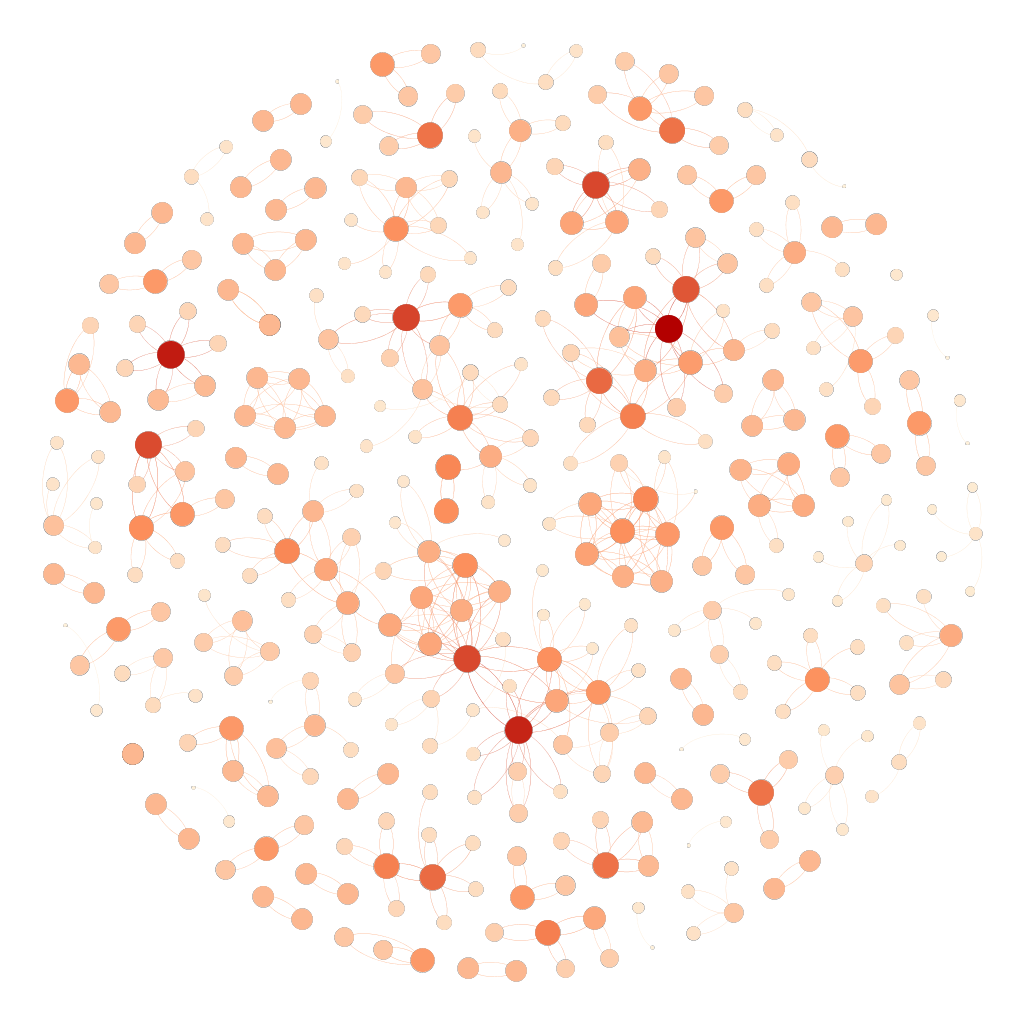}}\hfill
\caption{Left panel: the histogram of in-degrees for all users in the Weibo network. Right panel: network visualization using the first 500 users in the Weibo dataset. In the network structure, a dot denotes a node, and a line represents an edge. A deeper color and a larger dot indicate a larger in-degree.}
	\label{f:weibo}
\end{figure}

In this subsection we consider a real social network structure, which was collected from Sina Weibo (\textit{www.weibo.com}).
The detailed data collection process can be found in \cite{2017Estimating}.
The entire Weibo network has $N = 557,818$ nodes. Each node corresponds to one Weibo user. The follower-followee relationships among these users are used to construct the adjacency matrix $A$. This leads to a total of 1,496,399 edges and a network density of $4.809\times 10^{-6}$. Figure \ref{f:weibo}(a) presents the histogram of the in-degrees ($d_j=\sum_{i}a_{ij}$) for all users in this network. The distribution of in-degrees is very skewed, indicating that some users are more popular than others in the network.
To further illustrate the network structure, we take the first five hundred users as an example and present their network structure in Figure \ref{f:weibo}(b). As shown, the Weibo network is extremely sparse.
In addition, there are a number of small communities in the Weibo network.

By treating this network structure as if it were the entire network structure, a simulation study can be conducted.
Specifically, the response vector $\mY$ can be generated following the same procedure as described in Section 3.1. To compute the subnetwork estimator, we consider four different subsampling ratios as $n/N=(0.001, 0.005, 0.01, 0.015)$. For each subsampling ratio, the experiment is replicated $M=1,000$ times.
The simulation results with the error terms following the {\sc EXP} error distribution are summarized in Panel A in Table \ref{t:real}. The corresponding results with the {\sc NORM} error distribution are quantitatively similar and thus omitted. According to the simulation results in Panel A in Table \ref{t:real}, we find that $\wh\rho_{\mS}$ is consistent with an ignorable bias (close to zero) and a decreasing SE, as the sample size $n$ increases. The estimated standard error developed for $\wh\rho_{\mS}$ in Theorem 1 works quite well, as the difference between $\wh{\rm SE}$ and SE is very small.
In addition, the empirical coverage probabilities are always close to the nominal level of $95\%$. These results suggest that the subnetwork estimation method works very well for large-scale social networks such as Sina Weibo.

\begin{table}[h]
	\centering
	\caption{The detailed simulation results under the Sina Weibo network structure (Panel A) and the CC network structure (Panel B) with different subsampling ratios, network coefficient, and the {\sc EXP} error distribution.
	The bias $\flat$, estimated standard error $\wh{\rm SE}$, true standard error SE, and the empirical coverage probability ECP are reported for the SNOW method. The average CPU computational time is also reported in seconds.}
	\resizebox{\hsize}{!}{
	\begin{tabular}{cc|ccccc|ccccc}
		\hline  \hline
		\multirow{2}[4]{*}{$\rho$} & \multirow{2}[4]{*}{n/N} & \multicolumn{5}{c|}{Panel A: Weibo Network} & \multicolumn{5}{c}{Panel B: CC Network} \bigstrut\\
		\cline{3-12}          &       &$\flat$ &$\wh{\rm SE}$ & SE    & ECP   & CPU   &$\flat$ &$\wh{\rm SE}$ & SE    & ECP   & CPU \bigstrut\\
		\hline
		\multirow{4}[2]{*}{0}
&0.1\%	&0.001	&0.098	&0.097	&95.4\%	&0.14	&0.001	&0.107	&0.114	&94.3\%	&0.39	\\
&0.5\%	&-0.003	&0.043	&0.044	&94.3\%	&1.31	&-0.001	&0.038	&0.039	&95.7\%	&6.83	\\
&1.0\%	&0.000	&0.030	&0.031	&95.5\%	&5.62	&-0.001	&0.025	&0.026	&94.8\%	&28.16	\\
&1.5\%	&0.000	&0.024	&0.025	&94.7\%	&12.33	&0.000	&0.020	&0.020	&95.7\%	&64.73	\\
		\hline
		\multirow{4}[2]{*}{0.2}
&0.1\%	&0.000	&0.096	&0.095	&95.3\%	&0.15	&0.001	&0.106	&0.112	&94.1\%	&0.41	\\
&0.5\%	&-0.003	&0.042	&0.043	&94.4\%	&1.29	&-0.001	&0.038	&0.039	&95.7\%	&6.68	\\
&1.0\%	&0.000	&0.030	&0.030	&95.5\%	&5.58	&0.000	&0.025	&0.025	&95.4\%	&27.82	\\
&1.5\%	&0.001	&0.024	&0.025	&94.6\%	&12.29	&0.000	&0.020	&0.019	&95.6\%	&65.18	\\
		\hline
		\multirow{4}[2]{*}{0.4}
&0.1\%	&0.000	&0.091	&0.091	&95.4\%	&0.15	&0.005	&0.104	&0.108	&94.4\%	&0.43	\\
&0.5\%	&-0.002	&0.040	&0.041	&95.1\%	&1.39	&0.001	&0.037	&0.038	&95.4\%	&6.37	\\
&1.0\%	&0.001	&0.028	&0.029	&95.6\%	&5.75	&0.001	&0.024	&0.025	&95.3\%	&26.53	\\
&1.5\%	&0.002	&0.023	&0.023	&94.6\%	&12.58	&0.000	&0.019	&0.019	&95.6\%	&67.91	\\
		\hline
		\multirow{4}[2]{*}{0.6}
&0.1\%	&0.000	&0.085	&0.085	&95.2\%	&0.15	&0.013	&0.102	&0.104	&95.3\%	&0.41	\\
&0.5\%	&-0.001	&0.037	&0.038	&94.8\%	&1.38	&0.005	&0.036	&0.036	&95.3\%	&6.31	\\
&1.0\%	&0.001	&0.026	&0.027	&94.1\%	&5.83	&0.005	&0.024	&0.024	&94.3\%	&25.68	\\
&1.5\%	&0.003	&0.021	&0.022	&94.2\%	&12.90	&0.004	&0.018	&0.019	&94.8\%	&63.77	\\
		\hline
        \hline
	\end{tabular}%
}
	\label{t:real}%
\end{table}%

\csubsection{The CC Network}

\begin{figure}[h]
	\centering
\subfloat[Histogram of In-degree]{
		\includegraphics[width=0.49\textwidth]{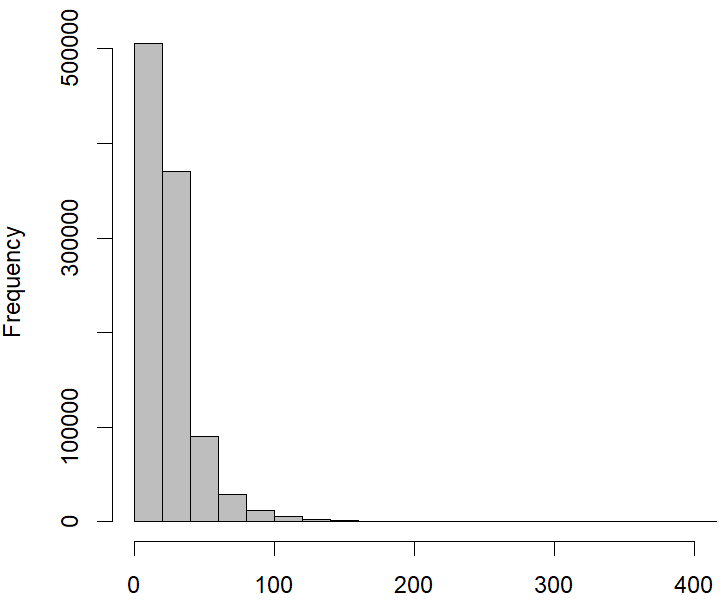}}\hfill
\subfloat[Network Visualisation]{
		\includegraphics[width=0.45\textwidth]{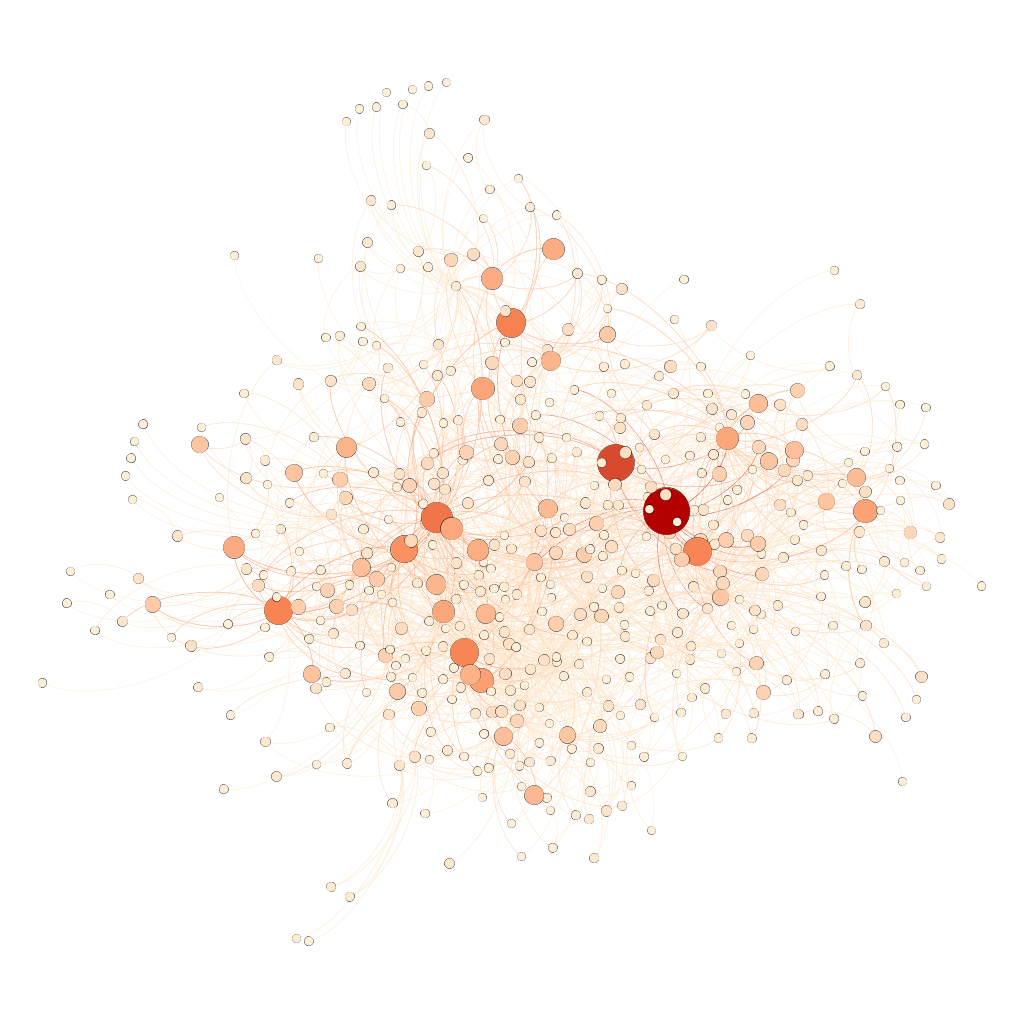}}\hfill
\caption{ Left panel: the histogram of in-degrees for all cells in the CC network. Right panel: network visualization using the sampled 500 cells by SNOW in the CC dataset. A dot denotes a node, and a line represents an edge. A deeper color and a larger dot indicate a larger in-degree.}
	\label{f:cc}
\end{figure}

In this subsection, we consider another real network structure, which is called the Megascale cell-cell similarity network. For convenience, we refer to it as the CC network. The CC network is publically available. The original dataset can be downloaded from \textit{http://snap.stanford.edu/biodata/index.html}.
The CC network dataset was designed for single-cell RNA sequencing of embryonic mouse brain cells. In the CC network, a node represents a cell in the mouse brain, and an edge represents the nearest neighbor similarities between two cells.
The existence of an edge indicates that the two cells have a similar gene expression as determined by a diffusion pseudotime analysis.
The entire network has a total of $N=1,018,524$ nodes and $\sum_{ij} a_{ij} = 24,735,503$ edges. The resulting network density is $2.38\times 10^{-5}$, which is extremely sparse. The histogram of in-degrees for all nodes in the CC network is given in Figure \ref{f:cc}(a).
As shown, the in-degrees vary within a wide range, and its histogram is very skewed, which indicates the existence of ``superpopular" cells in the network. To visualize the CC network structure, we select five hundred nodes using SNOW and present their network structure in Figure \ref{f:cc}(b).
As shown, the SNOW method can lead to a relatively dense network structure.

Based on the CC network structure, a simulation study similar to the one in the previous subsection is conducted. Specifically, the response vector $\mY$ is generated as described in Section 3.1. Similarly with the Weibo network, we consider four subsampling ratios as $n/N=(0.001, 0.005, 0.01, 0.015)$. Under each ratio, the experiment is replicated $M=500$ times.
The simulation results with the error terms following the {\sc EXP} error distribution are summarized in Panel B in Table \ref{t:real}. The results with the {\sc NORM} error distribution are quantitatively similar and thus omitted. As shown by Panel B in Table \ref{t:real}, the subnetwork estimates $\wh\rho_{\mS}$ also show satisfactory performance in the CC network. Specifically, the bias is close to zero, and the standard error SE drops towards zero as the sample size $n$ increases. The empirical coverage probabilities are well controlled around the nominal level $95\%$. Moreover, the difference between SE and $\wh{\rm SE}$ is very small, which suggests the estimated standard error derived from Theorem 1 can adequately approximate its true value.

\csubsection{Uncertainty Estimation Using Bootstrap}

In the above simulation studies, we compute the estimated standard error $\SE$ of the subnetwork estimator $\wh\rho_{\mS}$ according to Theorem 1 so that the uncertainty of the subnetwork estimator can be analytically quantified.
Another plausible solution in this regard is the bootstrap method, in which the network sampling should be replicated for multiple times.
Specifically, assume the number of bootstrap replications to be $B=20$.
Let $\wh\rho_{\mS}^{(m,b)}$ with $1\leq b \leq B$ be the subnetwork estimator in the $b$-th bootstrap replication of $m$-th simulation replication.
Then the standard error of the subnetwork estimator can be computed as $\widehat{\mbox{SE}}^{(m)}_{\rm bt}=\{B^{-1}\sum_b (\wh\rho_{\mS}^{(m,b)}-\bar\rho^{(m)}_{\rm bt})^2\}^{1/2}$, where $\bar\rho_{\rm bt}^{(m)} = B^{-1}\sum_b \wh\rho_{\mS}^{(m,b)}$.
Then we have the estimated bootstrap standard error as $\SE_{\rm bt}=M^{-1}\sum_m \SE_{\rm bt}^{(m)}$.
We can further construct a $95\%$ confidence interval for the true parameter $\rho$ as: $\text{CI}_{\rm bt}^{(m)}=(\wh\rho_{\mS}^{(m)}-z_{0.975}\wh{\mbox{SE}}_{\rm  bt}^{(m)}, \wh\rho_{\mS}^{(m)} + z_{0.975}\wh{\mbox{SE}}_{\rm bt}^{(m)})$. Then, the empirical coverage probability using the bootstrap method can be computed as $\text{ECP}_{\rm bt}=M^{-1}\sum_{m}I\big(\rho \in \text{CI}_{\rm bt}^{(m)}\big)$.

For illustration purpose, we take the SBM and LSM network structures as examples, and fix $\rho=0.2$, $n/N=0.01$, and $N/K=20$. We adopt the SNOW method to generate the subnetwork. The experiment is randomly replicated for a total of $M=500$ times. Table \ref{t:bootstrap} presents the simulation results of $\SE_{\rm bt}$, $\text{ECP}_{\rm bt}$, and the computational time $\text{CPU}_{\rm bt}$ by using the bootstrap method.
For comparison purpose, we also list the true standard error SE, the estimated standard error $\SE$ according to Theorem 1, and the corresponding ECP by using $\SE$. By Table \ref{t:bootstrap}, we find that both $\SE$ and $\widehat{\mbox{SE}}_{\rm bt}$ can approximate the true standard error (i.e., SE) fairly well. This makes the empirical coverage probabilities ECP and $\text{ECP}_{\rm bt}$ fairly close to the nominal level of 95\%. These results suggest that, the uncertainty of the subnetwork estimator can be well quantified by using the bootstrap method in these experimental settings. However, the computational time consumed by the bootstrap method increases linearly with the subsampling times $B$. Consequently, using the bootstrap method to quantify the uncertainty is less computationally efficient than the direct estimation method using Theorem 1. This is the price the bootstrap method has to pay to automatic statistical inference.

\begin{table}[h]
	\caption{ The detailed simulation results for the subnetwork estimator using the bootstrap method with the {\sc EXP} error distribution.
	The true standard error (SE), the estimated standard error ($\widehat{\text{SE}}_{\rm bt}$) and empirical coverage probability ($\text{ECP}_{\rm bt}$) by using the bootstrap method are reported. The estimated standard error ($\widehat{\text{SE}}$) and empirical coverage probability ($\text{ECP}$) using Theorem 1 are listed for comparison. The average CPU computational time is also reported in seconds.}
	\label{t:bootstrap}
 \centering
 \renewcommand\arraystretch{1.5}
 \small
 \begin{tabular}{rcc|ccc|ccc}
  \hline\hline
  \multirow{2}{*}{Network} &\multirow{2}{*}{$N$} & \multirow{2}{*}{SE} &\multicolumn{3}{c|}{Subnetwork Method} &\multicolumn{3}{c}{Bootstrap Method} \\
  \cline{4-9}
  &       &       & $\wh{\rm SE}$ & ECP   & CPU   & ${\widehat{\rm SE}}_{\rm bt}$ & ${\rm ECP}_{\rm bt}$ & ${\rm CPU}_{\rm bt}$ \\
  \hline
  & 10000 & \multicolumn{1}{c|}{0.127} & 0.131 & 95.80\% & 0.02  & 0.124 & 95.60\% & 0.12 \\
  \multicolumn{1}{l}{SBM} & 30000 & \multicolumn{1}{c|}{0.076} & 0.075 & 95.40\% & 0.04  & 0.070  & 93.60\% & 0.58 \\
  & 50000 & \multicolumn{1}{c|}{0.059} & 0.058 & 94.20\% & 0.08 & 0.055 & 95.00\% & 1.37 \\
  \hline
  & 10000 & 0.121 & 0.126 & 95.20\% & 0.01  & 0.120  & 94.20\% & 0.13 \\
  \multicolumn{1}{l}{LSM} & 30000 & 0.066 & 0.070  & 96.60\% & 0.04  & 0.066 & 93.80\% & 0.61 \\
  & 50000 & 0.052 & 0.054 & 96.80\% & 0.09  & 0.051 & 92.80\% & 1.48 \\
  \hline\hline
 \end{tabular}%
\end{table}%

\csection{CONCLUDING REMARKS}

Modern networks are often very large in size. To study the network dependence between different nodes, the spatial autoregressive (SAR) model has been popularly applied. Despite its popularity, major bottlenecks exist in the implementation of the SAR model on large-scale networks. This is because it is often impossible for independent studies to publicly collect all the network information due to policy limitations or high collection costs. In addition, even if the entire network is accessible, estimating a SAR model using the quasi-maximum likelihood estimator (QMLE) could be computationally infeasible. This is because the computation of QMLE, using Newton-Raphson algorithms, could be very expensive for large-scale networks because it requires $O(N^3)$ computational complexity \citep{2017Estimating,huang2019least}.

To address these challenges, we propose a subnetwork estimation method. By using the snowball sampling method, a subnetwork with $n\ll N$ nodes can be constructed. By using this method, collecting information for the subnetwork could be cost saving. By treating the sampled subnetwork as if it were the entire network, the QMLE can be subsequently computed, and its computational cost is also largely reduced. However, sampling a subnetwork would inevitably break the relationships between the nodes inside and outside of the subnetwork. Consequently, whether the subnetwork estimator can be considered a good approximation to the whole network estimator is an important question. To this end, the theoretical properties of the subnetwork estimator are investigated. We theoretically show that the subnetwork QMLE could be consistent and asymptotically normal, as long as the number of relationships occurring between nodes, inside and outside of the subnetwork, is small enough. Extensive numerical studies demonstrate the outstanding performance of the proposed method.

To conclude this work, we discuss some interesting directions for future research. First, the subnetwork estimation method is developed for SAR models without covariates. It is of great interest to extend the proposed method for SAR models with covariate information taken into consideration.
Second, we make a bootstrap extension for the subnetwork estimator to quantify its uncertainty, and demonstrate good finite performances. The theoretical properties of the bootstrap method is worth of consideration in the future. Third, the classical SAR model only assumes one constant autoregressive coefficient parameter. However in practice, the network effects could be varied for different nodes. Therefore how to extend the subnetwork estimation method to SAR models with more flexible network dependence structure is an interesting direction for future work. Last, we have numerically verified the technical conditions (C1) to (C3) for all the simulation experiments in Appendix B. We find that all the conditions are verified for the SBM and LSM network structures. However, for the real Weibo and CC networks, we unfortunately find the condition (C3) is violated. Even though, the subnetwork estimator still performs very well for both Weibo and CC networks. This suggests that the current condition (C3) is a sufficient but not necessary condition for the theoretical properties of the subnetwork estimator. Then investigation of more necessary conditions should be an interesting topic in the future.

\renewcommand \refname{\centerline{REFERENCES}}
\bibliographystyle{asa}
\bibliography{reference}

\newpage
\bc
{\bf\large APPENDIX}
\ec
\setcounter{equation}{0}
\renewcommand\theequation{A.\arabic{equation}}
\setcounter{table}{0}
\setcounter{figure}{0}
\renewcommand{\thetable}{A.\arabic{table}}
\renewcommand{\thefigure}{A.\arabic{figure}}

\scsection{Appendix A: Technical Proofs}
\scsubsection{Appendix A.1: Some Useful Lemmas}

To facilitate the proofs of this work, we need the following three lemmas.
Lemma \ref{L1: matrix inequality} directly follows \cite{seber2008matrix}. Therefore, we omit the proof of Lemma \ref{L1: matrix inequality}, and only present the detailed proofs of Lemma \ref{L2: E var} and Lemma \ref{L3: L1 asy property} below.

\bel
\label{L1: matrix inequality}
For any $N \times N$ matrix $A =\big( a_{ij} \big)$, we have:
(a) $\tr \big\{ \diag^2(A) \big\} \le \tr \big(A^\top A \big)$, and
(b) $\tr^2(A) \le N \tr \big( A^2 \big)$, with the equality holds if and only if all the eigenvalues of $A$ are equal. Particularly, if $A$ is symmetric, then we have $\tr^2(A) \le \rank(A) \tr\big(A^2\big)$.
(c) For any $N \times N$ matrix $B$, we have $\tr^2 \big(AB\big) \le \tr \big(A^\top A\big) \tr \big(B^\top B\big)$. In particular, we have $\tr\big(A^2\big) \le \tr\big(A^\top A\big)$.
Furthermore, we have (d) $\tr\big(A^\top A\big) \le rank(A)\lambda_{\max}(A^\top A)$.
\eel

\bel
\label{L2: E var}
Let $\mE = (\varepsilon_1, \dots, \varepsilon_N)^\top \in \mR^N$ be a random vector with $E(\mE) = 0$ and $\var(\mE) = \sigma^2 I_{N}$.
We further assume that $E \big(\varepsilon_{i_{1}}^{g_{1}} \varepsilon_{i_{2}}^{g_{2}}\varepsilon_{i_{3}}^{g_{3}} \varepsilon_{i_{4}}^{g_{4}} \big)
= E \big(\varepsilon_{i_{1}}^{g_{1}}\big) E \big(\varepsilon_{i_{2}}^{g_{2}}\big)E \big(\varepsilon_{i_{3}}^{g_{3}}\big)  E \big(\varepsilon_{i_{4}}^{g_{4}}\big)$ for indices $i_{1}, i_{2}, i_{3}, i_{4} \in\{1,2, \ldots, N\}$ and for any integers $g_{v} \ge 0$ with $\sum_{v=1}^{4} g_{v} \le 4$.
Define $\mu_4 = E(\varepsilon_i^4)$.
Then, for any $N \times N$ matrix $M= \big( M_{i j} \big) $, we have: (a) $E \big( \mE^{\top} M \mE \big) = \sigma^2 \tr(M)$ and
(b) $\var \big( \mE^{\top} M \mE \big) = (\mu_{4}-3 \sigma^4) \tr \big\{ \diag^2(M) \big\} + \sigma^4 \{ \tr \big( M^{2} \big) + \tr \big( M^{\top}M \big) \}$.
Furthermore, write $\mE = \big( \mE_1^{\top}, \mE_2^{\top} \big)^{\top} $, where $\mE_1 \in \mR^{n}$ and $\mE_2 \in \mR^{N-n}$. Then, for any $n \times (N-n)$ matrix $B$, we have:
(c) $E \big( \mE_1^{\top} B \mE_2 \big) = 0$ and
(d) $\var \big( \mE_1^{\top} B \mE_2 \big) = \sigma^{4} \tr\big(B^{\top}B\big)$.
\eel

{\sc Proof of Lemma \ref{L2: E var}.} We first work on $\mE^{\top} M \mE$. For (a), it is easy to verify that $E \big(\mE^{\top} M \mE \big) = \sigma^2 \tr(M)$. For (b), we need calculate $\var \big(\mE^{\top} M \mE \big)$.
Because $\var \big( \mE^{\top} M \mE \big) = E (\mE^{\top} M \mE )^{2}-E^{2} ( \mE^{\top} M \mE )$, we first calculate $E (\mE^{\top} M \mE )^{2}$ and obtain
\beqrs
&& E \Big( \mE^{\top} M \mE \Big)^{2}
= E\Big(\sum_{i=1}^{N} \sum_{j=1}^{N} \sum_{k=1}^{N} \sum_{l=1}^{N} M_{i j} M_{k l} \varepsilon_{i} \varepsilon_{j} \varepsilon_{k} \varepsilon_{l}\Big)
\\&=& E \Big\{\sum_{i=1}^{N} M_{ii}^{2} \varepsilon_{i}^{4} + \Big( \sum_{i=1}^{N} \sum_{j = 1, j\neq i}^{N} M_{ii} M_{jj} + M_{ij} M_{ji} + M_{ij}^{2} \Big) \varepsilon_i^2 \varepsilon_j^2\Big\}
\\&=& \Big(\mu_{4}-3 \sigma^4\Big) \tr \Big\{ \diag^2\big(M\big) \Big\} + \sigma^4 \Big\{ \tr^{2} \big( M \big) + \tr \big( M^{2} \big) + \tr \big( M^{\top}M \big)  \Big\}.
\eeqrs
Therefore, we have $\var\big( \mE^{\top} M \mE \big) = E \big( \mE^{\top} M \mE \big)^{2} - E^{2} \big( \mE^{\top} M \mE \big) = \big(\mu_{4}-3 \sigma^4\big) \tr \big\{ \diag^2(M) \big\} + \sigma^4 \{ \tr \big( M^{2} \big) + \tr \big( M^{\top}M \big) \}$.
Next, we investigate the properties of $\mE_1^{\top} B \mE_2$.
For (c), we need calculate $E \big( \mE_1^{\top} B \mE_2 \big)$.
Because $E \big( \mE_1 \mE_2^{\top} \big) = 0$, it is easy to verify that $E \big( \mE_1^{\top} B \mE_2 \big) = 0$.
For (d), we need compute $\var \big(\mE_1^{\top} B \mE_2 \big)$.
We have $\var \big( \mE_1^{\top} B \mE_2 \big) = E \big(  \mE_1^{\top} B \mE_2 \big)^2 = \tr\big\{ B^\top E\big( \mE_1 \mE_1^{\top} \big) B E\big( \mE_2 \mE_2^{\top} \big) \big\} = \sigma^4 \tr\big( B^\top B \big)$.
This completes the proof.

\bel
\label{L3: L1 asy property}
Define $\mE_1 = \big(\varepsilon_1,\dots, \varepsilon_n \big) \in \mR^n$ with $E \big(\mE_1\big) = 0$ and $\var \big(\mE_1\big) = \sigma^2I_{n}$.
Define $ \mathbb{Q}\big(\mE\big) = \mE_1^\top \mathbb{M} \mE_1$, where $\mathbb{M} = M_{\mS}-\tr \big(M_{\mS}\big)I_n / n$ and $M_{\mS} = W_{11} \big(I_{11}-\rho W_{11} \big)^{-1}$. Then we have $\mathbb{Q}(\mE) / \sqrt{n} \rightarrow_d N \big(0,\sigma^2_{1Q}\big)$ as $n \to \infty$, where $\sigma^2_{1Q} = \lim_{n \to \infty} \var\big\{ \mathbb{Q}(\mE)\big\}/n$.
\eel

{\sc Proof of Lemma \ref{L3: L1 asy property}.}
We apply the martingale difference theorem to verify the asymptotic normality of $\mathbb{Q}\big(\mE\big)$ in the following two step.
In the first step, we need to construct a martingale difference array $\big\{ \mathbb{Q}_i\big(\mE\big) \big\}_{i=1}^{n}$ such that $\mathbb{Q}\big(\mE\big) = \sum_{i=1}^{n}\mathbb{Q}_i\big(\mE\big)$. Subsequently, we employ the martingale difference theorem to obtain $\mathbb{Q}(\mE) \rightarrow_d N \big(0,\sigma^2_{1Q}\big)$ as $n$ goes to infinity. In the second step, we need to verify the conditions required by the martingale difference theorem.

{\sc Step 1.} Define $\mathcal{F}_i$ as the $\sigma$-field generated by $\big\{\varepsilon_{j}: 1 \le j \le i\big\}$. We then construct a martingale difference array $\big\{ \mathbb{Q}_i\big(\mE\big) \big\}_{i=1}^{n}$ to make $\mathbb{Q}\big(\mE\big) = \sum_{i=1}^{n}\mathbb{Q}_i\big(\mE\big)$. Specifically, we define $\mathbb{Q}_i\big(\mE\big) =\mathbb{Q}_{i1}\big(\mE\big) + \mathbb{Q}_{i2}\big(\mE\big) + \mathbb{Q}_{i3}\big(\mE\big)$, where $\mathbb{Q}_{i1}\big(\mE\big) = \mathbb{M}_{ii} \big(\varepsilon_{i}^{2}-\sigma^{2}\big)$,
$\mathbb{Q}_{i2}\big(\mE\big) = \sum_{j=1}^{i-1} \mathbb{M}_{i j} \varepsilon_{i} \varepsilon_{j}$,  $\mathbb{Q}_{i3}\big(\mE\big) =  \sum_{j=1}^{i-1} \mathbb{M}_{j i} \varepsilon_{i} \varepsilon_{j}$, and $\mathbb{M}_{ij}$ is the $(i,j)$-th element in $\mathbb{M}$.
Hence, we could verify $\mathbb{Q}\big(\mE\big)=\sum_{i=1}^{N} \mathbb{Q}_i\big(\mE\big)$ and $E\big\{\mathbb{Q}_i\big(\mE\big) \mid \mathcal{F}_{i-1}\big\}=0$.
To apply the martingale difference theorem, it suffices to show the following two conditions:
\beqr
\label{eq: mds condition 1}
&& \sum_{i=1}^{n} E\Big\{\mathbb{Q}^{4}_i\big(\mE\big)\Big\} \big/n^2 \rightarrow 0,
\\ && \label{eq: mds condition 2}
\sum_{i=1}^{n} E\Big\{ \mathbb{Q}^2_i\big(\mE\big) \mid \mathcal{F}_{i-1} \Big\} \big/n \rightarrow_{p} \sigma_{1Q}^2.
\eeqr

We first verify \eqref{eq: mds condition 1}.
By the Cauchy-Schwarz inequality, it suffices to verify that $\sum_{i=1}^{n} E\big\{\mathbb{Q}^{4}_{ik}\big(\mE\big)\big\} /n^2 \rightarrow 0$ for $k = 1,2,3$. The proof of the case $k = 3$ is similar with that of $k = 2$ and thus omitted. For $k=1$, we have $E\big\{\mathbb{Q}^4_{i1}\big(\mE\big)\big\} = \mathbb{M}_{ii}^4 E\big(\varepsilon_{i}^{2}-\sigma^{2}\big)^4 \le c_{\varepsilon} \mathbb{M}_{ii}^4$, where $c_{\varepsilon}$ is a positive constant. Therefore, we obtain that $\sum_{i=1}^{n} E\big\{\mathbb{Q}^4_{i1}\big(\mE\big)\big\} /n^2 \le c_{\varepsilon} \sum_{i=1}^{n} \mathbb{M}_{ii}^4 /n^2 = O(1/n) \rightarrow 0$ as $n \to \infty$.
For $k=2$, we have
\beqrs
n^{-2}\sum_{i=1}^n E \Big\{\mathbb{Q}^4_{i2}\big(\mE\big)\Big\}
&=& n^{-2}\sum_{i=1}^n \sum_{j_{1}, j_{2}, j_{3}, j_{4}<i} \mathbb{M}_{ij_1} \mathbb{M}_{ij_2} \mathbb{M}_{ij_3} \mathbb{M}_{ij_4} E ( \varepsilon_{i}^{4} )  E ( \varepsilon_{j_1}\varepsilon_{j_2}\varepsilon_{j_3}\varepsilon_{j_4})\\
&\le& n^{-2}\mu_4 \sigma^{4} \sum_{i=1}^n \sum_{j_{1} j_{2}<i, j_{1} \neq j_{2}} \mathbb{M}_{i j_1}^2 \mathbb{M}_{i j_2}^2
+ n^{-2} \mu_4^{2}\sum_{i=1}^n \sum_{j<i} \mathbb{M}_{i j}^{4}
\\&\le& c_Q n^{-2} \tr \Big\{\diag^2 \Big( \big|\mathbb{M}\big|\big|\mathbb{M}\big|^\top \Big) \Big\}
\le c_{Q} n^{-2} \tr\Big( \big| \mathbb{M}\big| \big|\mathbb{M}\big|^{\top} \Big)^2,
\eeqrs
where $c_{Q}=\max \big\{\mu_4\sigma^4, \mu_4^2 \big\}$. Then we can prove \eqref{eq: mds condition 1} holds if $n^{-2} \tr\big( |\mathbb{M}| |\mathbb{M}|^{\top} \big)^2 \to 0$ as $n$ goes to infinity. 

We next verify \eqref{eq: mds condition 2} holds.
Because $\big\{ \mathbb{Q}_i(\mE) \big\}_{i=1}^n$ is a martingale difference sequence, it is easy to prove $E\big[ \sum_{i=1}^{n} E\big\{\mathbb{Q}_{i}^{2}(\mE) \mid \mathcal{F}_{i-1}\big\}/n \big]
= \sum_{i=1}^{n} E\big\{ \mathbb{Q}_{i}^{2}(\mE) \big\} /n
= E\big\{ \mathbb{Q}^{2}(\mE)\big\}/n \rightarrow \sigma_{1 Q}^{2}$ as $n$ goes to infinity. We next show that as $n \rightarrow \infty$, we have $\var\big[\sum_{i=1}^{n} E\{\mathbb{Q}_{i}^{2}(\mE) \mid \mathcal{F}_{i-1} \}\big] \big/n^2 \rightarrow 0$. By the Cauchy-Schwarz inequality, we only need to verify $n^{-2}\var\big[ \sum_{i=1}^{n} E\big\{ \mathbb{Q}_{ik}^2(\mE) \mid \mathcal{F}_{i-1} \big\}\big] \rightarrow 0$ as $n \rightarrow \infty$ for $k = 1, 2, 3$. The proof of the case $k=3$ is similar with that of $k=2$ and thus omitted.
For $k=1$, we could easily prove that $\var\big[ E\big\{ \mathbb{Q}_{i1}^2(\mE) \mid \mathcal{F}_{i-1} \big\}\big]=0$. We then prove the case for $k=2$.
By Lemma \ref{L2: E var}(b), we could compute
$\var\big[ \sum_{i=1}^{n} E\big\{ \mathbb{Q}_{i2}^2(\mE) \mid \mathcal{F}_{i-1} \big\}\big]/n^2
= \sigma^{4} \var \big( \mE_1^{\top} \sum_{i=1}^{n} \mathbb{I}_{i-1} \mathbb{M}_{i,\cdot} \mathbb{M}_{i,\cdot}^{\top} \mathbb{I}_{i-1} \mE_1 \big)/n^2
= \sigma^4\big(\mu_4-3\sigma^4\big) \tr \big\{ \diag^2 \big( \sum_{i=1}^{n} \mathbb{I}_{i-1} \mathbb{M}_{i,\cdot} \mathbb{M}_{i,\cdot}^{\top} \mathbb{I}_{i-1} \big)\big\}/n^2 \\+ 2\sigma^8 \tr\big(\sum_{i=1}^{n} \mathbb{I}_{i-1} \mathbb{M}_{i,\cdot} \mathbb{M}_{i,\cdot}^{\top} \mathbb{I}_{i-1}\big)^{2}/n^2 $,
where $\mathbb{M}_{i,\cdot} \in \mR^n$ is the $i$-th row of $\mathbb{M}$, $\mathbb{I}_{i-1}=\sum_{j=1}^{i-1} e_{j} e_{j}^{\top}$ and $e_{j} \in \mR^{n}$ is a zero vector with only the $j$-th element being 1. Note that we have $\tr \big\{ \diag^2 \big( \sum_{i=1}^{n} \mathbb{I}_{i-1} \mathbb{M}_{i,\cdot} \mathbb{M}_{i,\cdot}^{\top} \mathbb{I}_{i-1} \big)\big\} \le \tr\big(\sum_{i=1}^{n} \mathbb{I}_{i-1} \mathbb{M}_{i,\cdot} \mathbb{M}_{i,\cdot}^{\top} \mathbb{I}_{i-1}\big)^{2}$. Therefore by Lemma \ref{L1: matrix inequality}, it suffices to prove $\tr\big(\sum_{i=1}^{n} \mathbb{I}_{i-1} \mathbb{M}_{i,\cdot} \mathbb{M}_{i,\cdot}^{\top} \mathbb{I}_{i-1}\big)^{2} /n^2 \rightarrow 0$ as $n \to \infty$. Define $A = \big(a_{ij}\big) \in \mathbb{R}^{n_1 \times n_2}$ and $B=\big(b_{ij}\big) \in \mathbb{R}^{n_{1} \times n_{2}}$ are two arbitrary matrices. Then we define $A \preccurlyeq B$ if $a_{i j} \le b_{i j}$ for any $1 \le i  \le n_1$ and $1 \le j  \le n_2$. Due to $\big|\mathbb{I}_{i-1} \mathbb{M}_{i,\cdot} \mathbb{M}_{i,\cdot}^{\top} \mathbb{I}_{i-1} \big| \preccurlyeq \big| \mathbb{M}_{i,\cdot} \big|\big| \mathbb{M}_{i,\cdot} \big|^{\top}$, we could show that
$\tr\big(\sum_{i=1}^{n} \mathbb{I}_{i-1} \mathbb{M}_{i,\cdot} \mathbb{M}_{i,\cdot}^{\top} \mathbb{I}_{i-1}\big)^{2}
\le \sum_{i_{1}, i_{2}}\big(\big| \mathbb{M}_{i_1, \cdot} \big|^{\top}\big|\mathbb{M}_{i_2, \cdot} \big|\big)\big(\big|\mathbb{M}_{i_2, \cdot} \big|^{\top}\big|\mathbb{M}_{i_1, \cdot} \big|\big)
\le \tr\big( \big|\mathbb{M}\big| \big|\mathbb{M}\big|^\top \big)^2$. Then we can prove \eqref{eq: mds condition 2} holds if $\tr\big( \big|\mathbb{M}\big| \big|\mathbb{M}\big|^\top \big|\mathbb{M}\big| \big|\mathbb{M}\big|^\top \big)/n^2 \to 0$ as $n$ goes to infinity.
As a result, we would verify the condition $n^{-2} \tr\big( | \mathbb{M}| |\mathbb{M}|^{\top} \big)^2 \to 0$ in {\sc Step 2}.

{\sc Step 2.} To make \eqref{eq: mds condition 1} and \eqref{eq: mds condition 2} hold, it suffices to verify, as $n$ goes to infinity,
\beqr
\label{eq: trace condition2}
\tr\Big( |\mathbb{M}| |\mathbb{M}|^\top |\mathbb{M}| |\mathbb{M}|^\top \Big)/n^2 \to 0.
\eeqr
Recall we have $\mathbb{M} = M_{\mS}-\tr \big(M_{\mS}\big)I_n / n$. By the Cauchy-Schwarz inequality, it only suffices to verify the following two conditions: $\tr( | M_{\mS}| |M_{\mS}|^{\top} |M_{\mS}| | M_{\mS}|^{\top} )/n^2 \to 0$ and $\tr\big( | \tr(M_{\mS})I_{n}/n|| \tr(M_{\mS})I_{n}/n|^{\top} | \tr(M_{\mS})I_{n}/n | | \tr(M_{\mS})I_{n}/n|^{\top}\big )/n^2 \to 0$ when $n$ goes to infinity. For the second term, it is obvious that $\tr\big( | \tr(M_{\mS})I_{n}/n|
| \tr( M_{\mS}) I_{n} /n|^{\top} | \tr(M_{\mS})I_{n}/n|\\ |\tr(M_{\mS})I_{n}/n|^{\top} \big)/n^2 = n^{-5}\tr^4\big(M_{\mS}\big)\to 0$ as $n \to \infty$. Then we only need to demonstrate $\tr( | M_{\mS}| |M_{\mS}|^{\top} |M_{\mS}| | M_{\mS}|^{\top} )/n^2 \to 0$ as $n \to \infty$.

Define $\mathcal{W}_{0}=\sum_{m=0}^{K} W^{m}_{11}+\mathbf{1}_{n} \pi_1^{\top}$, where $\pi_1 \in \mR^{n}$ is the sub-vector of $\pi$ corresponding to the sampled $n$ nodes in the sub-network. We need to prove there exists a positive constant $c_a$ such that $\big|M_{\mS}\big| \preccurlyeq c_a W_{11} \mathcal{W}_{0}$. Recall $M_{\mS} = W_{11} \big(I_{n}-\rho W_{11} \big)^{-1}$. Therefore, it suffices to show $\big|(I_{n} - \rho W_{11} )^{-1}\big| \preccurlyeq c_a \mathcal{W}_{0}$ since any $(i,j)$-th element in $W_{11}$ is no smaller than zero.
By the condition (C1), we know there exists a positive constant $c_w$ such that for a sufficient large $K$, we have $W^{k} \preccurlyeq c_w \mathbf{1}_N \pi^{\top}$ for $k \ge K$. Thus for the subnetwork, we have $W_{11}^{k} \preccurlyeq c_w \mathbf{1}_n \pi_1^{\top}$ for $k \ge K$. Therefore, we have
\beqrs
&& \Big|(I_{n} - \rho W_{11})^{-1}\Big| = \sum_{m=0}^{K} \rho^{m} W_{11}^{m}+\sum_{m>K} \rho^{m} W_{11}^{m}
\\&\preccurlyeq& \sum_{m=0}^{K} \rho^{m} W^{m}_{11} + c_{w} \mathbf{1}_{n} \pi_1^{\top}\big(\sum_{m>K} \rho^{m}\big)
\preccurlyeq \sum_{m=0}^{K} W^{m}_{11} + c_{\rho} c_{w} \mathbf{1}_{n} \pi_1^{\top} \preccurlyeq c_a \mathcal{W}_{0}.
\eeqrs
Then $\big|M_{\mS}\big| \preccurlyeq c_a W_{11} \mathcal{W}_{0}$
could be subsequently verified.
By Lemma \ref{L1: matrix inequality}(d), we know
\beqrs
\tr\Big( |M_{\mS}| |M_{\mS}|^\top |M_{\mS}||M_{\mS}|^\top \Big) \big/n^2
&\le& c_a^4\tr\Big( \mathcal{W}_{0}^\top W_{11}^{\top} W_{11} \mathcal{W}_{0} \mathcal{W}_{0}^\top W_{11}^{\top} W_{11} \mathcal{W}_{0} \Big) \big/n^2
\\ &\le& c_a^4\lambda^2_{\max} \Big(\mathcal{W}_{0}^\top W_{11}^{\top} W_{11} \mathcal{W}_{0} \Big)
 \big/n.
\eeqrs
We then need to verify $\lambda_{\max}^2 \big(\mathcal{W}_{0}^{\top} W_{11}^{\top} W_{11} \mathcal{W}_{0} \big) /n \to 0$.
By the Cauchy-Schwarz inequality, we could calculate that
\beqrs
\lambda_{\max }\Big(\mathcal{W}_{0}^{\top} W_{11}^{\top} W_{11} \mathcal{W}_{0} \Big)
&\le& c_{wq}  \sum_{m=0}^{K} \lambda^{m+1}_{\max} \Big( W_{11}^{\top} W_{11} \Big)
+ n c_{wq} \lambda_{\max}\Big( W_{11}^{\top} W_{11} \Big) \lambda_{\max}\Big(\pi_1 \pi^{\top}_1 \Big),
\eeqrs
where $c_{wq}$ is a finite constant.
Then the order of $\lambda_{\max }\big(\mathcal{W}_{0}^{\top} W_{11}^{\top} W_{11} \mathcal{W}_{0} \big)$ is determined by the orders of $\sum_{m=0}^{K} \lambda_{\max}^{m+1} \big( W_{11}^{\top} W_{11}\big)$ and $n \pi_1^{\top}\pi_1$. By the condition (C2), we know there exists a positive constant $c_{\max}$ such that $ \lambda_{\max}\big(W^\top W\big) \le c_{\max}$. Then for the first term, we have $\sum_{m=0}^{K} \lambda_{\max}^{m+1} \big( W_{11}^{\top} W_{11}\big) \le \sum_{m=0}^{K} \lambda^{m+1}_{\max} \big( W^{\top} W\big) \le (K+1)\max_{0 \le m\le K}\big(c_{\max}^{m},1\big)=O(1)$. For the second term, by the condition (C1), we have $n \pi_1^{\top}\pi_1 \le n \pi^{\top}\pi \le n/N^{1-\tau} \le n/n^{1-\tau} = n^{\tau}$.
As a result, we know $\lambda_{\max }\big( \mathcal{W}_{0}^{\top} W_{11}^{\top} W_{11} \mathcal{W}_{0} \big) = O \big( n^{\tau} \big)$.
Subsequently, when $ 0 \le \tau < 1/2$, we could verify $\lambda_{\max}^2 \big( \mathcal{W}_{0}^{\top} W_{11}^{\top} W_{11} \mathcal{W}_{0} \big) /n = O \big( n^{2\tau - 1} \big)$, which converges to $0$ as $n$ tends to infinity.
Then the required condition \eqref{eq: trace condition2} is obtained. This accomplishes the proof of Lemma \ref{L3: L1 asy property}.

\scsubsection{Appendix A.2: The Derivation Details about $\dot{\mathcal{L}}(\rho)$ and $\ddot{\mathcal{L}}(\rho)$}

We provide here the computational details about $\dot{\mathcal{L}}(\rho)$ and $\ddot{\mathcal{L}}(\rho)$. By \eqref{eq:whole-likelihood}, we can write $\mathcal{L}(\rho)$ as $\mathcal{L}(\rho)=\mathcal{L}_1(\rho)+\mathcal{L}_2(\rho)+\mathcal{L}_3(\rho)$, where
$\mathcal{L}_1(\rho) = -N \big\{\ln (2\pi) + 1 \big\}/2$, $\mathcal{L}_2(\rho) = -N \ln \big\{ \widetilde{\sigma}^2(\rho) \big\} /2$ and $\mathcal{L}_3(\rho) = \ln \big|I_{N}-\rho W \big|$.
Then it suffices to compute $\dot{\mathcal{L}}_k(\rho) = \partial{\mathcal{L}_k(\rho)} / \partial{\rho}$ and $\ddot{\mathcal{L}}_k(\rho) = \partial^2{\mathcal{L}_k(\rho)} / \partial{\rho^2}$ for $1\le k \le 3$.

{\sc Compute $\dot{\mathcal{L}} (\rho)$.}
It is obvious that $\dot{\mathcal{L}}_1(\rho) =0$. We then only need to focus on $\dot{\mathcal{L}}_2(\rho)$ and $\dot{\mathcal{L}}_3(\rho)$.
We start with $\dot{\mathcal{L}}_2(\rho)$. It could be easily verified that $\dot{\mathcal{L}}_2(\rho) = - N \dot{\widetilde{\sigma}}^{2}(\rho) / \big\{ 2 \widetilde{\sigma}^{2} (\rho) \big\} $, where $\dot{\widetilde{\sigma}^{2}} (\rho)$ denotes the first-order derivative of $\widetilde{\sigma}^{2}(\rho)$ with respect to $\rho$.
Recall $\widetilde{\sigma}^2(\rho) = \mY^\top \big( I_{N}-\rho W \big)^\top \big(I_{N} - \rho W \big) \mY/N$. By simple calculations, we have $\dot{\widetilde{\sigma}^{2}}(\rho) = - 2 \mY^{\top} W^{\top} \big(I_{N}-\rho W \big) \mY / N$.
Consequently, we have $\dot{\mathcal{L}}_2(\rho) = - N \dot{\widetilde{\sigma}}^{2}(\rho) \big/ \{ 2 \widetilde{\sigma}^{2} (\rho) \}
= \{\mY^{\top} W^{\top} (I_{N}-\rho W ) \mY \} \big/ \widetilde{\sigma}^{2}(\rho).$

We next study $\mathcal{L}_3(\rho) = \ln \big|I_{N}-\rho W \big|$. Define $f(x)$ as a matrix function of scalar $x$ and $\dot f(x)$ as the first-order derivative of $f(x)$ with respect to $x$. By \cite{minka2000old}, we have $\partial{\ln \big|f(x)\big|} / \partial{x} = \tr\big\{ f^{-1}(x)\dot f(x) \big\}$. Then the first-order derivative of $\mathcal{L}_3(\rho)$ with respect to $\rho$ could be derived as $\dot{\mathcal{L}}_3(\rho) = - \tr \big\{ \big(I_{N}-\rho W \big)^{-1} W \big\}$.
Combining $\dot{\mathcal{L}}_1(\rho)$, $\dot{\mathcal{L}}_2(\rho)$ and $\dot{\mathcal{L}}_3(\rho)$, we have
$$
\dot{\mathcal{L}} (\rho) = \Big\{\mY^{\top} W^{\top} \Big(I_{N}-\rho W\Big) \mY \Big\} \big/ \widetilde{\sigma}^{2}(\rho)
-\tr\Big\{ \Big(I_{N}-\rho W \Big)^{-1} W \Big\} .
$$

{\sc Compute $\ddot{\mathcal{L}} (\rho)$.} Based on $\dot{\mathcal{L}} (\rho)$, we could further derive the second-order derivative $\ddot{\mathcal{L}} (\rho)$ with respect to $\rho$. We also have $\ddot{\mathcal{L}}_1(\rho)=0$ and only need to calculate $\ddot{\mathcal{L}}_2 (\rho)$ and $\ddot{\mathcal{L}}_3 (\rho)$.
By applying simple derivative techniques for scalars, we can derive $\ddot{\mathcal{L}}_2 (\rho)$ as $\ddot{\mathcal{L}}_2 (\rho) = 2 \{\mY^{\top} W^{\top} (I_{N}-\rho W ) \mY\}^2 \big/ \{ N\widetilde{\sigma}^{4} (\rho) \}
- \mY^{\top} W^{\top} W \mY \big/ \widetilde{\sigma}^{2} (\rho)$.

Next we work on $\ddot{\mathcal{L}}_3 (\rho)$. By \cite{petersen2008matrix}, we have $\partial{\tr \big\{ f(x)\big\} }/\partial{x} = \tr \big\{ \dot f(x)\big\}$, where $f(x)$ is a matrix function of scalar $x$.
Therefore, we have $\ddot{\mathcal{L}}_3(\rho) = -\tr \big[ \partial{ \big\{ (I_{N} - \rho W )^{-1}W \big\}} / \partial{\rho} \big]$.
Then we focus on the first-order derivative of $\big(I_{N} - \rho W\big)^{-1}$. By \cite{selby1973standard}, for any matrix function $f(x)$ of scalar $x$, we have $\partial{f(x)^{-1}}/ \partial{x} = -f(x)^{-1} \dot{f}(x) f(x)^{-1}$. Then we can derive $\partial{ \big\{ (I_{N} - \rho W)^{-1} \big\} } / \partial{\rho} = (I_{N}-\rho W )^{-1} W (I_{N}-\rho W )^{-1} $. As a result, we could obtain $\ddot{\mathcal{L}}_3 (\rho) = -\tr \big\{ (I_{N}-\rho W )^{-1} W \big\}^{2}$.
Combining the results of $\ddot{\mathcal{L}}_1(\rho)$, $\ddot{\mathcal{L}}_2(\rho)$ and $\ddot{\mathcal{L}}_3(\rho)$, we have
\beqrs
\ddot{\mathcal{L}} (\rho) &=& 2 \Big\{\mY^{\top} W^{\top} \Big(I_{N}-\rho W \Big) \mY \Big\}^2 \big/ \Big\{ N\widetilde{\sigma}^{4} (\rho) \Big\}
- \mY^{\top} W^{\top} W \mY \big/ \widetilde{\sigma}^{2} (\rho)
\\&& -\tr \Big\{ \Big(I_{N}-\rho W \Big)^{-1} W \Big\}^{2}.
\eeqrs
It completes the computation for $\dot{\mathcal{L}} (\rho)$ and $\ddot{\mathcal{L}} (\rho)$.

\scsubsection{Appendix A.3: Proof of Theorem \ref{Theorem 1}}

Based on the working model \eqref{eq: sub model}, we can derive the profiled log-likelihood function on the subnetwork as \eqref{eq: sub loss}. Then we obtain the subnetwork QMLE for $\rho$ as $\wh{\rho}_{\mS} = \argmax_\rho \mathcal{L}_{\mS}(\rho)$. 
However, as we mentioned in Section 2, the true model for $\mY_1$ is derived by $\mY_1 = \Lambda_1 \mE_1 + \Lambda_1 \Lambda_2 \mE_2$,
where $\Lambda_1 = \big( I_{11}-\rho W_{11} \big)^{-1} \big(I_{11} - \mathbb{D}\big)^{-1}$
, $\Lambda_2 = \rho W_{12} \big(I_{22}-\rho W_{22}\big)^{-1}$ and $\mathbb{D} = \rho^2 W_{12} \big(I_{22}-\rho W_{22}\big)^{-1} W_{21} (I_{11}-\rho W_{11})^{-1}$.
Then we need to use the true model for $\mY_1$ to explore the asymptotic behaviors of $\wh{\rho}_{\mS}$. Specifically, to show that $\wh\rho_{\mS}$ is $\sqrt{n}$-consistent, it suffices to verify there exists some constant $C>0$ such that
\beq
\label{eq: circle inequality}
\sup _{t \in \mR, |t|=C} \mathcal{L}_{\mS} \Big( \rho + t \big/ \sqrt{n} \Big)< \mathcal{L}_{\mS} \big( \rho \big)
\eeq
with probability tending to 1 as $n$ goes to infinity \citep{fan2001variable}. By Taylor's expansion, we have
\beq
\label{eq: Taylor Expansion}
\mathcal{L}_{\mS} \Big( \rho + t/ \sqrt{n} \Big) - \mathcal{L}_{\mS} \big( \rho \big)  = \Big\{ t \dot{\mathcal{L}}_{\mS} ( \rho ) / \sqrt{n} +  t^2 \ddot{\mathcal{L}}_{\mS} ( \rho ) / \big(2n\big) \Big\} \big\{ 1+o_p(1) \big\}.
\eeq
We can demonstrate that $\dot{\mathcal{L}}_{\mS}(\rho)/ \sqrt{n} = O_p(1)$ in Appendix A.5, and $\ddot{\mathcal{L}}_{\mS}(\rho)/n$ converges to a negative scalar in probability in Appendix A.6. Then we could verify \eqref{eq: circle inequality} holds for a sufficiently large $C$. Due to the convexity of $\mathcal{L}_{\mS}(\rho)$, we obtain that $\sup _{\|t\|>C} \mathcal{L}_{\mS} \big(\rho+n^{-1/2} t \big) < \mathcal{L}_{\mS}(\rho)$.
As $\mathcal{L}_{\mS}(\rho)$ is maximized at $\wh\rho_{\mS}$, we know $\wh\rho_{\mS}$ lies in the ball $\big\{ \rho + n^{-1/2} t:\|t\| \le C \big\}$. In other words, we have $\big| \wh{\rho}_{\mS} \big|= O_{p} \big( n^{-1/2} \big)$.

Given $\wh{\rho}_{\mS}$ is $\sqrt{n}$-consistent, it enables us to apply the Taylor's expansion to obtain the following asymptotic approximation,
\beqr
\label{eq: rho Taylor expansion}
\sqrt{n} \Big( \wh{\rho}_{\mS} - \rho \Big) = \Big\{ \ddot{\mathcal{L}}_{\mS} \big( \rho^{*} \big) \big/ n \Big\}^{-1} \Big\{ \dot{\mathcal{L}}_{\mS} \big(\rho\big) \big/ \sqrt{n} \Big\},
\eeqr
where $\rho^{*}$ lies between $\rho$ and $\wh{\rho}_{\mS}$. In Appendix A.5, we have found $\dot{\mathcal{L}}_{\mS}(\rho)/ \sqrt{n} = O_p(1)$ and further demonstrated $\dot{\mathcal{L}}_{\mS} \big(\rho\big) \big/ \sqrt{n} \rightarrow_{d} N\big(0,\sigma_{1\mS}^2\big)$. In Appendix A.6, we have demonstrated $\ddot{\mathcal{L}}_{\mS} (\rho)/n \rightarrow_{p} -\sigma_{2\mS}^2$ when $n$ goes to infinity. Based on the above results, we can derive $\sqrt{n} \big( \wh{\rho}_{\mS} - \rho \big) \rightarrow_{d} N\big(0, \sigma_{2\mS}^{-4} \sigma_{1\mS}^2 \big)$. This completes the proof of Theorem \ref{Theorem 1}.

\scsubsection{Appendix A.4: The Theoretical Properties of $\widetilde{\sigma}_{\mS}^2(\rho)$}

Because $\widetilde{\sigma}_{\mS}^2(\rho)$ is involved in $\dot{\mathcal{L}}_{\mS}(\rho)/ \sqrt{n}$ and $\ddot{\mathcal{L}}_{\mS}(\rho)/n$, we need to investigate the theoretical properties of $\widetilde{\sigma}_{\mS}^2(\rho)$ first. Recall $\widetilde{\sigma}_{\mS}^{2}(\rho) = \mY_1^{\top} \big( I_{11}-\rho W_{11} \big) ^{\top} (I_{11}-\rho W_{11}) \mY_1 /n$. To obtain the true behaviors of $\widetilde{\sigma}_{\mS}^2(\rho)$, we substitute the true model $\mY_1 = \Lambda_1 \mE_1 + \Lambda_1 \Lambda_2 \mE_2$ into $\widetilde{\sigma}_{\mS}^{2}(\rho)$. This leads to $\widetilde{\sigma}_{\mS}^{2}(\rho)= \big(\mE_1 + \Lambda_2 \mE_2 \big)^\top \Lambda_1^\top \big( I_{11}-\rho W_{11} \big)^{\top} \big( I_{11}-\rho W_{11} \big) \Lambda_1 \big(\mE_1 + \Lambda_2 \mE_2 \big) / n$. Recall $\mathbb{D} = \rho^2 W_{12} \big(I_{22}-\rho W_{22}\big)^{-1} W_{21} \big(I_{11}-\rho W_{11}\big)^{-1}$ and $\delta = \mathbb{D} + \mathbb{D}^\top - \mathbb{D}\mathbb{D}^\top$. Then we define $\mathbb{B}_{\sigma} = \Lambda_1^\top \big( I_{11}-\rho W_{11} \big) ^{\top} (I_{11}-\rho W_{11}) \Lambda_1 - I_{11} =(I_{11}-\delta)^{-1} - I_{11}$. Based on $\mathbb{B}_{\sigma}$, we can write $\widetilde{\sigma}_{\mS}^2(\rho) = \widetilde{\sigma}_{\mS1}^2(\rho) + 2\widetilde{\sigma}_{\mS2}^2(\rho) + \widetilde{\sigma}_{\mS3}^2(\rho) + \widetilde{\sigma}_{\mS4}^2(\rho)+
2\widetilde{\sigma}_{\mS5}^2(\rho) + \widetilde{\sigma}_{\mS6}^2(\rho)$, where $\widetilde{\sigma}_{\mS1}^2(\rho) = \mE_1^{\top} \mE_1/n$, $\widetilde{\sigma}_{\mS2}^2(\rho) = \mE_1^{\top} \Lambda_2 \mE_2/n$, $\widetilde{\sigma}_{\mS3}^2(\rho) = \mE_2^{\top} \Lambda_2^\top \Lambda_2 \mE_2/n$,
$\widetilde{\sigma}_{\mS4}^2(\rho) = \mE_1^{\top} \mathbb{B}_{\sigma} \mE_1 /n$,
$\widetilde{\sigma}_{\mS5}^2(\rho) = \mE_1^{\top} \mathbb{B}_{\sigma} \Lambda_2 \mE_2/n$ and
$\widetilde{\sigma}_{\mS6}^2(\rho) = \mE_2^{\top} \Lambda_2^\top \mathbb{B}_{\sigma} \Lambda_2 \mE_2 /n$.
We then investigate the orders of $\widetilde{\sigma}_{\mS i}^2(\rho)$ for $1 \le i \le 6$.

We first compute the order of $\widetilde{\sigma}^{2}_{\mS1}(\rho)$.
Recall $\widetilde{\sigma}_{\mS1}^2(\rho) = \mE_1^{\top} \mE_1/n$. We have $ E \big\{ \widetilde{\sigma}^{2}_{\mS1}(\rho) \big\} = \sigma^{2}$ and
$\var \big\{ \widetilde{\sigma}^{2}_{\mS1} (\rho) \big\}
= \big(\mu_4-\sigma^4 \big) /n$ by Lemma \ref{L2: E var}(a) and (b) in Appendix A.1. Hence, we know $\widetilde{\sigma}^{2}_{\mS1} (\rho) \rightarrow_p \sigma^2$ when $n$ tends to infinity.
We then calculate the order of $\widetilde{\sigma}^{2}_{\mS2} (\rho)$.
By Lemma \ref{L2: E var}(c) and (d) in Appendix A.1, we have $E \big\{ \widetilde{\sigma}^{2}_{\mS2} (\rho) \big\} = 0$ and $\var \big\{ \widetilde{\sigma}^{2}_{\mS2} (\rho) \big\} = \sigma^4 \tr\big( \Lambda_2^\top \Lambda_2\big)/n^2$.
We then focus on the order of $\var \big\{ \widetilde{\sigma}^{2}_{\mS2} (\rho) \big\}$. By Lemma \ref{L1: matrix inequality}(d) in Appendix A.1, we have $\tr\big( \Lambda_2^\top \Lambda_2 \big) \le \tr\big(W_{12}^{\top} W_{12}\big) \lambda_{\max}\big\{ \big(I_{22} - \rho W_{22}^{\top} \big)^{-1} \big(I_{22} - \rho W_{22}\big)^{-1} \big\}$, which has the order of $O\big(n^\kappa\big)$ with $0\le \kappa < 1/2$ by condition (C3.2). As a consequence, we have $\var \big\{ \widetilde{\sigma}^{2}_{\mS2} (\rho) \big\} = o\big( 1/n^{2-\kappa} \big) = o\big( 1/n \big)$. Together with $E \big\{ \widetilde{\sigma}^{2}_{\mS2} (\rho) \big\}=0$, we have $\widetilde{\sigma}^{2}_{\mS2} (\rho) = o_p\big( 1/\sqrt{n} \big)$.

We next focus on the order of $\widetilde{\sigma}^{2}_{\mS4} (\rho)$. We can first derive
$E \big( \widetilde{\sigma}^{2}_{\mS4} \big)
= \sigma^2 \tr\big( \mathbb{B}_{\sigma} \big)/n$ and $\var \big( \widetilde{\sigma}^{2}_{\mS4} \big) = \big[ \big( \mu_{4}-3 \sigma^4 \big) \tr \big\{ \diag^2\big( \mathbb{B}_{\sigma} \big)\big\} + 2\sigma^4 \tr \big( \mathbb{B}_{\sigma}^{2} \big) \big]/n^2$.
Then we evaluate the orders of $E \big( \widetilde{\sigma}^{2}_{\mS4} \big)$ and $\var \big( \widetilde{\sigma}^{2}_{\mS4} \big)$, separately.
We first calculate the order of $E \big( \widetilde{\sigma}^{2}_{\mS4} \big)$.
Note that $\tr\big( \mathbb{B}_{\sigma} \big)
= \sum_{k=1}^{\infty} \tr\big(\delta^k\big)
\le \tr\big(\delta^2\big) \sum_{k=0}^{\infty} \rho^k(\delta) + \tr\big(\delta\big)
= \tr\big(\delta^2\big)/\big(1-\rho(\delta)\big)
+ \tr\big(\delta\big)$.
It suffices to verify $\rho(\delta)<1$, $\tr\big(\delta^2\big) =o(\sqrt{n})$ and $\tr\big(\delta\big) =o(\sqrt{n})$.
Then we could derive $E \big( \widetilde{\sigma}^{2}_{\mS4} \big) = o\big( 1/\sqrt{n} \big)$.
We start with $\rho(\delta)<1$.
Note that $\rho(\delta) = 1-\rho\{(I_{11}-D)(I_{11}-D^\top)\} \le 1-\{1-\rho(D)\}^2$.
Then we focus on prove $\rho(D)<1$.
By condition (C3.1), we have $\rho^2(D)\le \rho^2 \lambda_{\max}(W_{12}^\top W_{12}) \lambda_{\max}(W_{21}^\top W_{21}) \lambda_{\max} \{(I_{11}-\rho W_{11}^\top)^{-1}(I_{11}-\rho W_{11})^{-1}\} \lambda_{\max} \{(I_{22}-\rho W_{22}^\top)^{-1}(I_{22}-\rho W_{22})^{-1}\} \le \rho^2 c_{\min}^{-2} \lambda_{\max}(W_{12}^\top W_{12}) \lambda_{\max}(W_{21}^\top W_{21}) <1$.
Thus, we obtain $\rho(\delta)<1$.
Next, we prove $\tr\big(\delta^2\big) =o(\sqrt{n})$ by condition (C3).
Note that $\tr(\delta^2)=2\tr(D^\top D)+2\tr(D^2)-2\tr(DD^\top D)-2\tr(DDD^\top)+\tr(DD^\top D D^\top)$.
Then we compute these terms respectively.
First, we have $\tr(D^\top D)=\lambda_{\max}\{(I_{11}-\rho W_{11}^\top)^{-1}(I_{11}-\rho W_{11})^{-1}\} \lambda_{\max}\{(I_{22}-\rho W_{22}^\top)^{-1}(I_{22}-\rho W_{22})^{-1}\} \tr(W_{12}^\top W_{12}) \lambda_{\max}(W_{12}^\top W_{12})=o(\sqrt{n})$ by condition (C3).
Then, $\tr(D^2)\le \tr(D^\top D)=o(\sqrt{n})$.
Meanwhile, we compute $\tr(DD^\top D) \le \tr^{1/2}(D^\top D)\tr^{1/2}(D^\top DD^\top D) \le \tr(D^\top D) \lambda_{\max}(D^\top D) = o(\sqrt{n})$ by Lemma 1(c).
Similarly, we have $\tr(DD^\top D)= o(\sqrt{n})$ and $\tr(DD^\top D D^\top)= o(\sqrt{n})$.
Combining the above results, we prove $\tr(\delta^2)=o(\sqrt{n})$.
Lastly, we work on $\tr(\delta) = 2\tr(D)-tr(DD^\top)= o(\sqrt{n})$.
By Lemma 1(c), we have $|\tr(D)| \le \rho^2 \tr^{1/2}\{ (I_{22}-\rho W_{22}^\top)^{-1} W_{12}^\top W_{12} (I_{22}-\rho W_{22})^{-1} \} \tr^{1/2}\{ (I_{11}-\rho W_{11}^\top)^{-1} W_{21}^\top W_{21} (I_{11}-\rho W_{11})^{-1} \} = O(1)
\\ \tr^{1/2}\{ W_{12}^\top W_{12}\} \tr^{1/2}\{ W_{21}^\top W_{21}\} = o(\sqrt{n})$.
Since we have $tr(DD^\top)= o(\sqrt{n})$, we prove $\tr(\delta)= o(\sqrt{n})$.
we have $\tr\big( \mathbb{B}_{\sigma} \big) = o(\sqrt{n})$.
Then we could derive $E \big( \widetilde{\sigma}^{2}_{\mS4} \big) = o\big( 1/\sqrt{n} \big)$.
We next work on the order of $\var \big( \widetilde{\sigma}^{2}_{\mS4} \big)$.
Note that $\tr\big( \mathbb{B}_{\sigma}^{2}\big) = \sum_{i=1}^{n} \lambda_i^2(\delta)/\{1-\lambda_i(\delta)\}^2 \le \tr(\delta^2)/c_{\delta}^2$, where $c_{\delta}$ is a positive constant.
Also by condition (C3), we have $\tr\big( \mathbb{B}_{\sigma}^{2}\big) = O\big( n^\kappa \big)$ with $0\le \kappa < 1/2$. Then we have $\var \big( \widetilde{\sigma}^{2}_{\mS4} \big) =O\big(1/n^{2-\kappa}\big) = o\big(1/n\big)$. Together with $E \big( \widetilde{\sigma}^{2}_{\mS4} \big) =o\big( 1/\sqrt{n}\big)$, we have $\widetilde{\sigma}^{2}_{\mS4} = o_p\big( 1/\sqrt{n} \big)$.

Using similar techniques, we can derive $\widetilde{\sigma}^2_{\mS3}(\rho)= o_p\big( 1/\sqrt{n} \big)$, $\widetilde{\sigma}^2_{\mS5}(\rho)= o_p\big( 1/\sqrt{n} \big)$, and $\widetilde{\sigma}^2_{\mS6}(\rho)= o_p\big( 1/\sqrt{n} \big)$. The derivation details are omitted to save space. Based on the above results for $\widetilde{\sigma}^2_{\mS1}(\rho)$ to $\widetilde{\sigma}^2_{\mS6}(\rho)$, we have $\widetilde{\sigma}^{2}_{\mS}(\rho) = \widetilde{\sigma}^{2}_{1\mS}(\rho) \big\{1+o_p(1)\big\}\rightarrow_p \sigma^{2}$ as $n$ goes to infinity.

\scsubsection{Appendix A.5: The Theoretical Properties of $\dot{\mathcal{L}}_{\mS}(\rho) \big/ \sqrt{n}$}

We first derive the formula of $\dot{\mathcal{L}}_{\mS}(\rho) \big/ \sqrt{n}$. Similarly with \eqref{eq: first_d}, we can write $\dot{\mathcal{L}}_{\mS} (\rho) =
\{\mY_1^{\top} W_{11}^{\top} (I_{11}-\rho W_{11}) \mY_{1} \} / \widetilde{\sigma}_{\mS}^{2}(\rho) -\tr \{ W_{11} (I_{11}-\rho W_{11} )^{-1} \}$. Define $M_{\mS} = W_{11} \big(I_{11}-\rho W_{11} \big)^{-1}$ and $q_{\mS}(\mE) = \mY_1^{\top} W_{11}^{\top} \big(I_{11}-\rho W_{11} \big) \mY_1$. Then we can write $\dot{\mathcal{L}}_{\mS}(\rho) / \sqrt{n} = Q_{\mS}(\mE)/\widetilde{\sigma}^{2}_{\mS}(\rho)$, where
$ Q_{\mS}(\mE) = q_{\mS}(\mE)/\sqrt{n} -\tr \big(M_{\mS} \big)\widetilde{\sigma}^{2}_{\mS}(\rho) /\sqrt{n}$. Next, we conduct the following two steps to study the theoretical properties of $\dot{\mathcal{L}}_{\mS}(\rho) \big/ \sqrt{n}$.

{\sc Step 1.}
We start with $q_{\mS} (\mE)$.
To obtain the true theoretical behaviors of $q_{\mS} (\mE)$, we first substitute the true model $\mY_1 = \Lambda_1 \mE_1 + \Lambda_1 \Lambda_2 \mE_2$ into $q_{\mS} (\mE)$. Define $\mathbb{B}_{q} = \Lambda_1^\top \big(I_{11} -\rho W_{11}^{\top}\big) W_{11}\Lambda_1 - M_{\mS}$. We then have $\mathbb{B}_{q} = \big( I_{11} - \mathbb{D}^\top \big)^{-1} M_{\mS} \big( I_{11} - \mathbb{D} \big)^{-1} - M_{\mS} = \big( I_{11} - \mathbb{D}^\top \big)^{-1} \big( M_{\mS}\mathbb{D} + \mathbb{D}^\top M_{\mS} - \mathbb{D}^\top M_{\mS}\mathbb{D}\big) \big( I_{11} - \mathbb{D} \big)^{-1}$.
Then we can write $q_{\mS} (\mE) = q_{\mS1}(\mE) + 2 q_{\mS2}(\mE) + q_{\mS3}(\mE) + q_{\mS4}(\mE)+ 2 q_{\mS5}(\mE) + q_{\mS6}(\mE)$, where $q_{\mS1}(\mE) = \mE_1^{\top} M_{\mS} \mE_1 $,
$q_{\mS2}(\mE) = \mE_1^{\top} M_{\mS} \Lambda_2 \mE_2$,
$q_{\mS3}(\mE) = \mE_2^{\top} \Lambda_2^\top M_{\mS} \Lambda_2 \mE_2$,
$q_{\mS4}(\mE) = \mE_1^\top \mathbb{B}_{q} \mE_1$,
$q_{\mS5}(\mE) = \mE_1^{\top} \mathbb{B}_{q} \Lambda_2\mE_2$ and
$q_{\mS6}(\mE) = \mE_2^{\top} \Lambda_2^\top \mathbb{B}_{q} \Lambda_2 \mE_2$. We then evaluate the orders of each term.

We first focus on $q_{\mS1}(\mE)$. It is easy to derive $E \big\{ q_{\mS1}(\mE) \big\} =\sigma^2 \tr \big(M_{\mS} \big) = O(n)$ and
$\var\big\{ q_{\mS1}(\mE) \big\}
= \big( \mu_{4}-3 \sigma^4 \big) \tr \big\{ \diag^2\big( M_{\mS} \big)\big\}
+ \sigma^4 \big\{ \tr \big( M_{\mS}^{2} \big) + \tr \big( M_{\mS}^{\top}M_{\mS} \big) \big\}=O(n)$.
Thus we have $q_{\mS1}(\mE) = O_p(n)$.
Next, we evaluate the order of $q_{\mS2}(\mE)$. By Lemma \ref{L2: E var}(c) and (d) in Appendix A.1, we could obtain $E \big\{ q_{\mS2}(\mE) \big\} = 0$ and $\var \big\{ q_{\mS2}(\mE) \big\} = \sigma^4 \tr \big( \Lambda_2^\top M_{\mS}^\top M_{\mS} \Lambda_2\big)$. Next, we have $\tr \big( \Lambda_2^\top M_{\mS}^\top M_{\mS} \Lambda_2\big)
\le \tr\big( \Lambda_2^\top\Lambda_2 \big) \lambda_{\max}\big( M_{\mS}^\top M_{\mS} \big) = O\big(n^\kappa \big)$ with $0 \le \kappa < 1/2$. Based on the above results, we have $q_{\mS2}(\mE) = o_p\big(\sqrt{n}\big)$.

We then calculate the order of $q_{\mS4}(\mE)$. By Lemma \ref{L2: E var}(a) and (b) in Appendix A.1, we have $E \big\{ q_{\mS4}(\mE) \big\}
= \sigma^2 \tr\big( \mathbb{B}_{q} \big)$ and $\var \big\{ q_{\mS4}(\mE) \big\} = \big( \mu_{4}-3 \sigma^4 \big) \tr \big\{ \diag^2\big( \mathbb{B}_{q} \big)\big\} + \sigma^4 \tr \big( \mathbb{B}_{q}^{2} \big) + \sigma^4\tr \big( \mathbb{B}_{q}^{\top}\mathbb{B}_{q} \big)$.
Then we work on the orders of $E \big\{ q_{\mS4}(\mE)\big\}$ and $\var \big\{ q_{\mS4}(\mE)\big\}$ separately.
We first focus on the order of $E \big\{ q_{\mS4}(\mE)\big\}$.
Recall $\mathbb{B}_{q} = \big( I_{11} - \mathbb{D}^\top \big)^{-1}
\big( M_{\mS}\mathbb{D} + \mathbb{D}^\top M_{\mS} - \mathbb{D}^\top M_{\mS}\mathbb{D}\big) \big( I_{11} - \mathbb{D} \big)^{-1}$.
We have $E \big\{ q_{\mS4}(\mE) \big\} = \sigma^2 \tr\big\{ \big( I_{11} - \mathbb{D}^\top \big)^{-1} M_{\mS}\mathbb{D} \big( I_{11} - \mathbb{D} \big)^{-1} \big\}
+ \sigma^2 \tr\big\{ \big( I_{11} - \mathbb{D}^\top \big)^{-1} \mathbb{D}^\top M_{\mS} \big( I_{11} - \mathbb{D} \big)^{-1} \big\}
- \sigma^2 \tr\big\{ \big( I_{11} - \mathbb{D}^\top \big)^{-1} \mathbb{D}^\top M_{\mS}\mathbb{D} \big( I_{11} - \mathbb{D} \big)^{-1} \big\}$. For the sake of similarity, we only compute the order of the first term in $E \big\{ q_{\mS4}(\mE) \big\}$. By Lemma \ref{L1: matrix inequality} in Appendix A.1, we have $\big| \tr\big\{ \big( I_{11} - \mathbb{D}^\top \big)^{-1} M_{\mS}\mathbb{D} \big( I_{11} - \mathbb{D} \big)^{-1} \big\} \big|
\le \rho^2 \tr^{1/2}\big( W_{12}^\top W_{12} \big) \tr^{1/2}\big( W_{21}^\top W_{21} \big) \rho \big\{\big( I_{11} -\delta\big)^{-1} \big\} \lambda_{\max}^{1/2} \big( W_{11}^\top W_{11} \big)
\lambda_{\max} \big\{\big( I_{11} -\rho W_{11}^\top\big)^{-1} \\ \big( I_{11} -\rho W_{11}\big)^{-1} \big\} \lambda_{\max}^{1/2} \big\{\big( I_{22} -\rho W_{22}^\top\big)^{-1} \big( I_{22} -\rho W_{11}\big)^{-1} \big\}  $, which has the order of $O(n^\kappa)$ with $0 \le \kappa < 1/2$ by condition (C3). Therefore, we have $E \big\{ q_{\mS4}(\mE) \big\} =o\big(\sqrt{n}\big)$. We then work on $\var \big\{ q_{\mS4}(\mE) \big\}$. We first focus on $\tr \big( \mathbb{B}_{q}^{\top}\mathbb{B}_{q} \big)$. By Lemma \ref{L1: matrix inequality}(a) and (c) in Appendix A.1, we have $\tr \big( \mathbb{B}_{q}^{\top}\mathbb{B}_{q} \big) \le \lambda_{\max}^2 \big\{\big( I_{11}-\delta\big)^{-1}\big\} \tr\big\{  \big(M_{\mS}\mathbb{D} + \mathbb{D}^\top M_{\mS} - \mathbb{D}^\top M_{\mS} \mathbb{D}\big)^{\top} \big(M_{\mS}\mathbb{D} + \mathbb{D}^\top M_{\mS} - \mathbb{D}^\top M_{\mS} \mathbb{D} \big) \big\}$. Then by the Cauchy-Schwarz inequality, we only need to compute the orders of $\tr\big( \mathbb{D}^{\top} M_{\mS}^{\top} M_{\mS}\mathbb{D} \big)$, $\tr\big( M_{\mS}^{\top} \mathbb{D} M_{\mS}\mathbb{D} \big)$ and $\tr\big( \mathbb{D}^\top M_{\mS}^{\top} \mathbb{D} \mathbb{D}^\top M_{\mS} \mathbb{D} \big)$, which are all $O\big( n^\kappa \big)$ with $0 \le \kappa < 1/2$.
For the order of $\tr\big( \mathbb{D}^{\top} M_{\mS}^{\top} M_{\mS}\mathbb{D} \big)$, we have $\tr\big( \mathbb{D}^{\top} M_{\mS}^{\top} M_{\mS}\mathbb{D} \big) \le \tr\big( \mathbb{D}^{\top}\mathbb{D} \big) \lambda_{\max}\big( M_{\mS}^\top M_{\mS} \big) = O\big( n^\kappa \big)$, where $0 \le \kappa < 1/2$. Similarly, we can prove $\tr\big( M_{\mS}^{\top} \mathbb{D} M_{\mS}\mathbb{D} \big)$ has the order $O\big( n^\kappa \big)$. Finally, for the order of $\tr\big( \mathbb{D}^\top M_{\mS}^{\top} \mathbb{D} \mathbb{D}^\top M_{\mS} \mathbb{D} \big)$, we have $\tr\big( \mathbb{D}^\top M_{\mS}^{\top} \mathbb{D} \mathbb{D}^\top M_{\mS} \mathbb{D} \big) \le \tr\big( \mathbb{D}^{\top}\mathbb{D} \big) \lambda_{\max}\big( \mathbb{D}^{\top}\mathbb{D} \big) \lambda_{\max}\big( M_{\mS}^\top M_{\mS} \big)= O\big( n^\kappa \big)$. As a result, we have $\var \big\{ q_{\mS4}(\mE) \big\}= O\big( n^\kappa \big) = o(n)$. Together with $E \big\{ q_{\mS4}(\mE) \big\} =o\big(\sqrt{n}\big)$, we then have $q_{\mS4}(\mE) = o_p\big(\sqrt{n}\big)$.

Using similar techniques, we can compute  $q_{\mS3}(\mE) = o_p\big(\sqrt{n}\big)$, $q_{\mS5}(\mE) = o_p\big(\sqrt{n}\big)$ and $q_{\mS6}(\mE) = o_p\big(\sqrt{n}\big)$, the detailed deviations of which are omitted. Finally, based on the above results, we have $q_{\mS} (\mE)= o_p\big(\sqrt{n}\big)$.

{\sc Step 2.} We focus on $Q_{\mS}(\mE)$ in this step. Recall $ Q_{\mS}(\mE) = \big\{q_{\mS}(\mE) -\tr \big(M_{\mS} \big)\widetilde{\sigma}^{2}_{\mS}(\rho)\big\} /\sqrt{n}$. In addition, we have split $q_{\mS} (\mE) = q_{\mS1}(\mE) + 2 q_{\mS2}(\mE) + q_{\mS3}(\mE) + q_{\mS4}(\mE)+ 2 q_{\mS5}(\mE) + q_{\mS6}(\mE)$ in {\sc Step 1}. Then we have $Q_{\mS}(\mE)
= \big\{ q_{\mS1}(\mE) -\widetilde{\sigma}_{\mS}^2(\rho) \tr \big(M_{\mS} \big) \big\} /\sqrt{n}
+ 2q_{\mS2}(\mE)/\sqrt{n}
+ q_{\mS3}(\mE)/\sqrt{n}
+ q_{\mS4}(\mE)/\sqrt{n}
+ 2q_{\mS5}(\mE)/\sqrt{n}
+ q_{\mS6}(\mE)/\sqrt{n}$. We then study the order of each term in $Q_{\mS}(\mE)$. By Lemma \ref{L3: L1 asy property} in Appendix A.1, we could prove $\big\{ q_{\mS1}(\mE) -\widetilde{\sigma}_{\mS}^2(\rho)\tr \big(M_{\mS} \big) \big\} /\sqrt{n} \\=\mE_1^\top \big\{M_{\mS}-\tr(M_{\mS})I_{11}/n\big\}\mE_1/\sqrt{n} \rightarrow_d N\big(0, \sigma^4\sigma_{1\mS}^2 \big)$. Furthermore, we have verified $q_{\mS k}(\mE)/\sqrt{n}= o_p(1)$ for $k=2,...,6$ in {\sc Step 1}. Then we have $Q_{\mS}\big(\mE\big) = n^{-1/2} \big\{ q_{\mS1}(\mE) -\widetilde{\sigma}_{\mS}^2(\rho) \tr \big(M_{\mS} \big) \big\}  \big\{1+o_p(1)\big\}$ and $Q_{\mS}\big(\mE\big) \rightarrow_{d} N\big(0, \sigma^4\sigma_{1\mS}^2\big)$.

We have studied the theoretical properties of $\widetilde{\sigma}_{\mS}^{2}(\rho)$ in Appendix A.4. Based on the results in Appendix A.4, we can derive $\widetilde{\sigma}_{\mS}^{-2}(\rho)= \sigma^{-2} -\sigma^{-4} \big( \widetilde{\sigma}_{\mS}^{2}(\rho) - \sigma^{2} \big) + O_p(1/n) = \sigma^{-2} \big\{1+o_p(1)\big\}$ by Taylor's expansion.
Then by Slutsky's theorem, we could obtain $\dot{\mathcal{L}}_{\mS}(\rho) \big/ \sqrt{n} = (\sqrt{n}\sigma^2)^{-1} \mE_1^\top \big\{M_{\mS} -\tr(M_{\mS}) I_{11}/n \big\} \mE_1   \{ 1+o_p(1) \}$, which implies $\dot{\mathcal{L}}_{\mS}(\rho) / \sqrt{n} \rightarrow_{d} N\big(0, \sigma_{1\mS}^2\big)$.

\scsubsection{Appendix A.6: The Theoretical Properties of $\ddot{\mathcal{L}}_{\mS}(\rho) \big/ n$}

We try to prove $\ddot{\mathcal{L}}_{\mS} (\rho)/n \rightarrow_{p} - \sigma_{2\mS}^2$ as $n$ tends to infinity in this step. To this end,
we first derive the formula of $\ddot{\mathcal{L}}_{\mS}(\rho) \big/ n$. Similar with \eqref{eq: second_d}, we have $\ddot{\mathcal{L}}_{\mS} (\rho)= - \mY_1^{\top} W_{11}^{\top} W_{11} \mY_1 / \widetilde{\sigma}_{\mS}^{2} (\rho) - \tr \{ W_{11} (I_{11}-\rho W_{11} )^{-1} \}^{2} + 2 \{\mY_1^{\top} W_{11}^{\top} (I_{11}-\rho W_{11} ) \mY_1\}^2 / \{ n\widetilde{\sigma}_{\mS}^{4} (\rho) \}$. Define $p_{\mS}(\mE) = \mY_{1}^{\top} W_{11}^{\top} W_{11} \mY_{1}$. We can decompose $\ddot{\mathcal{L}}_{\mS} (\rho)/n = -P_{1\mS}(\mE)+ P_{2\mS}(\mE) -P_{3\mS}(\mE) + P_{4\mS}(\mE)$, where
$P_{1\mS} (\mE) = p_{\mS}(\mE) / \big( n\sigma^{2} \big) + \tr\big( M_{\mS} \big)^{2}/n$,
$P_{2\mS} (\mE) = 2 q_{\mS}^2(\mE) / \big(n^2 \sigma^{4}\big)$,
$P_{3\mS}(\mE) = p_{\mS}(\mE) \big\{ \widetilde{\sigma}_{\mS}^{-2}(\rho) - \sigma^{-2} \big\} /n$ and
$P_{4\mS}(\mE) = 2 q_{\mS}^2(\mE)
\big\{ \widetilde{\sigma}_{\mS}^{-4}(\rho) - \sigma^{-4} \big\} / n^2$.
By the Slutsky's theorem and the mapping theorem, it is obvious that $P_{3\mS}(\mE) = o_p\big\{P_{1\mS}(\mE) \big\}$ and $P_{4\mS}(\mE) = o_p\big\{P_{2\mS}(\mE) \big\}$. Then it suffices to show that $-P_{1\mS}(\mE)+P_{2\mS}(\mE)\rightarrow_p -\sigma_{2\mS}^2$ as  $n$ goes to infinity in the following two steps.

{\sc Step 1.} We focus on $P_{1\mS}(\mE)$ in this step. To this end, we first study the order of $p_{\mS}(\mE)$. Define $\mathbb{B}_{p} = \Lambda_1^\top W_{11}^{\top} W_{11} \Lambda_1- M_{\mS}^\top M_{\mS} = \big(I_{11}-\mathbb{D}^\top\big)^{-1} \big(M_{\mS}^\top M_{\mS} \mathbb{D} + \mathbb{D}^\top M_{\mS}^\top M_{\mS} - \mathbb{D}^\top M_{\mS}^\top M_{\mS} \mathbb{D}\big) \big(I_{11}-\mathbb{D}\big)^{-1}$. By substituting the true model of $\mY_1$ into $p_{\mS}(\mE)$, we could write $p_{\mS} (\mE) = p_{\mS1}(\mE) + 2p_{\mS2}(\mE) + p_{\mS3}(\mE) + p_{\mS4}(\mE) + 2p_{\mS5}(\mE) + p_{\mS6}(\mE)$, where $p_{\mS1}(\mE) = \mE_1^{\top} M_{\mS}^\top M_{\mS} \mE_1$,
$p_{\mS2}(\mE) = \mE_1^{\top} M_{\mS}^\top M_{\mS} \Lambda_2 \mE_2$,
$p_{\mS3}(\mE) = \mE_2^{\top} \Lambda_2^\top M_{\mS}^\top M_{\mS} \Lambda_2 \mE_2$,
$p_{\mS4}(\mE) = \mE_1 \mathbb{B}_{p} \mE_1 $,
$p_{\mS5}(\mE) = \mE_1^{\top} \mathbb{B}_{p} \Lambda_2 \mE_2$ and
$p_{\mS6}(\mE) = \mE_2^{\top} \Lambda_2^\top \mathbb{B}_{p} \Lambda_2 \mE_2$.
In addition, we could write that $P_{1\mS} (\mE) = \big\{ p_{\mS1}(\mE)+ \sigma^2\tr\big(M_{\mS}^{2} \big) \big\} / \big( n\sigma^{2} \big)
+ 2p_{\mS2}(\mE) / \big( n\sigma^{2} \big) + p_{\mS3}(\mE) / \big( n\sigma^{2} \big) + p_{\mS4}(\mE) / \big( n\sigma^{2} \big)+ 2p_{\mS5}(\mE) / \big( n\sigma^{2} \big) + p_{\mS6}(\mE) / \big( n\sigma^{2} \big)$. We then study the order of each term separately.

We first focus on $p_{\mS1}(\mE) / \big( n\sigma^{2} \big)$.
By Lemma \ref{L2: E var}(a) and (b) in Appendix A.1, we have $E\big\{p_{\mS1}(\mE) / \big( n\sigma^{2} \big)\big\} = \tr\big(M_{\mS}^\top M_{\mS} \big) /n$ and $\var \big\{p_{\mS1}(\mE) / \big( n\sigma^{2} \big)\big\} = 2 \tr\big(M_{\mS}^\top M_{\mS} \big)^{2} /n^2 + \big( \mu_{4}/\sigma^4 -3 \big) \tr\big\{ \diag^2\big(M_{\mS}^{\top} M_{\mS} \big) \big\} /n^2$.
Because we know $\lambda_{\max}\big(M_{\mS}^\top M_{\mS}\big) \le \lambda_{\max}\big(W_{11}^\top W_{11}\big)
\\ \lambda_{\max}\big\{ \big(I_{11} -\rho W_{11}^{\top}\big)^{-1}\big(I_{11} -\rho W_{11}\big)^{-1} \big\} = O\big(1\big)$ by condition (C2), we then have
$\tr \big(M_{\mS}^{\top}M_{\mS} \big)^2
\\ \le n\lambda_{\max}^2 \big(M_{\mS}^{\top}M_{\mS} \big) \le nc^4_{\max} = O\big(n\big)$ and
$\tr \big\{ \diag^2 \big(M_{\mS}^\top M_{\mS}\big) \big\} \le \tr \big(M_{\mS}^{\top}M_{\mS} \big)^2 = O\big(n\big)$ by Lemma \ref{L1: matrix inequality}(a) and (d) in Appendix A.1.
Therefore, we can prove $p_{\mS1}(\mE)/ \big( n\sigma^{2} \big)$ has the order $O_p(1)$.

We next investigate the order of $p_{\mS2}(\mE)$. By applying Lemma \ref{L2: E var}(c) and (d) in Appendix A.1, we have
$E \big\{ p_{\mS2}(\mE) \big\} = 0$ and
$\var \big\{ p_{\mS2}(\mE) \big\}
=\sigma^4 \tr\big\{ \Lambda_2^\top \big(M_{\mS}^\top M_{\mS}\big)^2 \Lambda_2 \big\}$.
It is noteworthy that $\tr\big\{ \Lambda_2^\top \big(M_{\mS}^\top M_{\mS}\big)^2 \Lambda_2 \big\}
\le \lambda_{\max}^2\big(W_{11}^\top W_{11}\big) \lambda_{\max}\big(W_{12}^\top W_{12}\big) \lambda_{\max}^2\big\{\big(I_{11}-\rho W_{11}^\top \big)^{-1}\big(I_{11}-\rho W_{11} \big)^{-1}\big\} \lambda_{\max}\big\{ \big(I_{22}-\rho W_{22}^{\top}\big)^{-1}
\big(I_{22}-\rho W_{22}\big)^{-1} \big\} = O\big(n^\kappa\big)$ with $0 \le \kappa < 1/2$ by condition (C3.2). Thus we have $\var \big\{ p_{\mS2}(\mE) \big\}= o(n)$. Together with $E \big\{ p_{\mS2}(\mE) \big\} = 0$, we have $p_{\mS2}(\mE) = o_p(\sqrt{n})$.

We then work on the order of $p_{\mS4}(\mE)$.
By Lemma \ref{L2: E var}(a) and (b) in Appendix A.1, we could verify that
$E \big\{ p_{\mS4}(\mE) \big\} = \sigma^2 \tr\big( \mathbb{B}_{p} \big)$ and $\var \big\{ p_{\mS4}(\mE) \big\} = \big( \mu_{4}-3 \sigma^4 \big) \tr \big\{ \diag^2\big( \mathbb{B}_{p} \big)\big\} + 2\sigma^4 \tr \big( \mathbb{B}_{p}^{2} \big)$. Then we study the orders of $E \big\{ p_{\mS4}(\mE) \big\}$ and $\var \big\{ p_{\mS4}(\mE) \big\}$ separately.
For the order of $E \big\{ p_{\mS4}(\mE) \big\}$, we have $\tr\big( \mathbb{B}_{p} \big) =
\tr\big\{\big(I_{11}-\mathbb{D}^\top\big)^{-1} M_{\mS}^\top M_{\mS} \mathbb{D} \big(I_{11}-\mathbb{D}\big)^{-1}\big\}
+ \tr\big\{\big(I_{11}-\mathbb{D}^\top\big)^{-1} \mathbb{D}^\top M_{\mS}^\top M_{\mS} \big(I_{11}-\mathbb{D}\big)^{-1}\big\}
- \tr\big\{\big(I_{11}-\mathbb{D}^\top\big)^{-1} \mathbb{D}^\top M_{\mS}^\top M_{\mS} \mathbb{D} \big(I_{11}-\mathbb{D}\big)^{-1}\big\}
$. We first focus on $\tr\big\{ \big( I_{11} - \mathbb{D}^\top \big) M_{\mS}^\top M_{\mS}\mathbb{D} \big( I_{11} - \mathbb{D} \big) \big\}$.
By Lemma \ref{L1: matrix inequality} in Appendix A.1, we have $\big| \tr\big\{ \big( I_{11} - \mathbb{D}^\top \big) M_{\mS}^\top M_{\mS}\mathbb{D} \big( I_{11} - \mathbb{D} \big) \big\} \big|
\le \rho^2 \lambda_{\max} \big( W_{11}^\top W_{11} \big)
\lambda_{\max} \big\{\big( I_{11} -\rho W_{11}^\top\big)^{-1} \big( I_{11} -\rho W_{11}\big)^{-1} \big\}
\rho \big\{\big( I_{11} -\delta\big)^{-1} \big\} \tr^{1/2}\big( \mathbb{D}^\top \mathbb{D} \big)$, which has the order $O\big(n^\kappa\big)$ with $0 \le \kappa < 1/2$ by condition (C3.1) and (C3.2).
Using the similar techniques, we could compute $\tr\big\{\big(I_{11}-\mathbb{D}^\top\big)^{-1} \mathbb{D}^\top M_{\mS}^\top M_{\mS} \big(I_{11}-\mathbb{D}\big)^{-1}\big\}$ and
$\tr\big\{\big(I_{11}-\mathbb{D}^\top\big)^{-1} \mathbb{D}^\top M_{\mS}^\top M_{\mS} \mathbb{D} \big(I_{11}-\mathbb{D}\big)^{-1}\big\}$ are both of order $O\big(n^\kappa\big)$. Based on the above results, we have $E \big\{ p_{\mS4}(\mE) \big\}=o\big(\sqrt{n}\big)$. We then calculate the order of $\var \big\{ p_{\mS4}(\mE) \big\}$, which only requires to compute the order of $\tr\big( \mathbb{B}_{p}^{2} \big)$.
Note that $\tr\big( \mathbb{B}_{p}^{2} \big) = \tr \big\{\big(I_{11}-\mathbb{D}^\top\big)^{-1} \big( M_{\mS}^\top M_{\mS} \mathbb{D} + \mathbb{D}^\top M_{\mS}^\top M_{\mS} - \mathbb{D}^\top M_{\mS}^\top M_{\mS} \mathbb{D} \big) \big(I_{11}-\mathbb{D}\big)^{-1}\big\}^2\le \tr \big( M_{\mS}^\top M_{\mS} \mathbb{D} + \mathbb{D}^\top M_{\mS}^\top M_{\mS} - \mathbb{D}^\top M_{\mS}^\top M_{\mS} \mathbb{D} \big)^2 \lambda_{\max}^2 \big\{\big( I_{11}-\delta \big)^{-1} \big\}$. Since $\lambda_{\max}^2 \big\{\big( I_{11}-\delta \big)^{-1} \big\}$ could be bounded by a positive constant, then it suffices to explore the orders of $\tr \big( M_{\mS}^\top M_{\mS} \mathbb{D} \big)^{2}$, $\tr\big( \mathbb{D}^\top M_{\mS}^\top M_{\mS} \big)^{2}$ and $\tr\big( \mathbb{D}^\top M_{\mS}^\top M_{\mS} \mathbb{D} \big)^{2}$ separately by the Cauchy-Schwarz inequality.
By the condition (C2) and (C3), we can compute $\tr\big( M_{\mS}^\top M_{\mS} \mathbb{D} \big)^{2} \le \tr\big\{ \mathbb{D}^\top \big(M_{\mS}^\top M_{\mS}\big)^{2} \mathbb{D} \big\} = O\big(n^\kappa\big)$ and $\tr\big( \mathbb{D}^\top M_{\mS}^\top M_{\mS} \mathbb{D} \big)^{2} \le \lambda_{\max}^{2}\big(M_{\mS}^\top M_{\mS}\big) \lambda_{\max}\big( \mathbb{D}^\top \mathbb{D} \big) \tr\big( \mathbb{D}^\top \mathbb{D} \big)
\\= O\big(n^\kappa\big)$, where $0 \le \kappa <1/2$.
Hence we can obtain $\tr\big( \mathbb{B}_{p}^{2} \big) = O\big(n^\kappa\big)$ and $\var \big\{ p_{\mS4}(\mE) \big\} = o\big(n\big)$. As a result, we have $p_{\mS4}(\mE) = o_p\big(\sqrt{n}\big)$.

By similar techniques, we could obtain $p_{\mS3}(\mE)=o_p\big(\sqrt{n}\big)$, $p_{\mS5}(\mE)=o_p\big(\sqrt{n}\big)$ and $p_{\mS6}(\mE)=o_p\big(\sqrt{n}\big)$. The computational details of these three terms are omitted to save space.
Combining the above results for $p_{\mS1}(\mE)$ to $p_{\mS6}(\mE)$, we could finally obtain $P_{1\mS} (\mE) = \big\{p_{\mS1}(\mE)/ \big( n\sigma^{2} \big)+ \tr\big(M_{\mS}^{2} \big)/n \big\} \big\{ 1+o_p(1)\big\}$.

{\sc Step 2.} We focus on $P_{2\mS}(\mE) $ in this step. Recall $q_{\mS}(\mE) = \sqrt{n} Q_{\mS}(\mE) + \tr(M_{\mS})\widetilde{\sigma}_{\mS}^{2}(\rho)$. We then have $P_{2\mS}(\mE) = 2 q_{\mS}^2(\mE) / \big(n^2 \sigma^{4}\big) = 2 Q_{\mS}^2(\mE) / \big( n\sigma^{4} \big)
+ 2\tr^2\big(M_{\mS}\big) \widetilde{\sigma}_{\mS}^{4}(\rho) / \big(n^2\sigma^{4}\big)
+ 4 Q_{\mS}(\mE) \tr(M_{\mS}) \widetilde{\sigma}_{\mS}^{2}(\rho)/\big(n^{3/2}\sigma^4\big)$. Then each term in $P_{2\mS}(\mE)$ is to be studied separately.
Note that we have proved $Q_{\mS}(\mE) = O_p(1)$ in {\sc Step 2} in Appendix A.5. We then have $2 Q_{\mS}^2(\mE) / \big( n\sigma^{4} \big) = o_p(1)$ and $4 Q_{\mS}(\mE) \tr(M_{\mS}) \widetilde{\sigma}_{\mS}^{2}(\rho)/\big(n^{3/2}\sigma^4\big) = o_p(1)$. Together with $\widetilde{\sigma}^2_{\mS}(\rho) \rightarrow_{p} \sigma^2$, we have $P_{2\mS}(\mE) = 2\tr^2(M_{\mS}) \big\{1+ o_p(1)\big\}/n^2$ consequently.

Based on the results in the above two steps, we can derive $\ddot{\mathcal{L}}_{\mS} (\rho)/n = \{ 2\tr^2 (M_{\mS})/n - \mE_1^\top M_{\mS}^{\top} M_{\mS} \mE_1 / \sigma^2 - \tr(M^2_{\mS} ) \}\{1+ o_p(1)\} /n$. It suggests that $\ddot{\mathcal{L}}_{\mS}(\rho)/n$ converges to $-\sigma_{2\mS}^2$ in probability as $n \to \infty$, which is a negative scalar by Lemma \ref{L1: matrix inequality}.

\setcounter{equation}{0}
\renewcommand\theequation{B.\arabic{equation}}
\setcounter{table}{0}
\setcounter{figure}{0}
\renewcommand{\thetable}{B.\arabic{table}}
\renewcommand{\thefigure}{B.\arabic{figure}}

\scsection{Appendix B: Technical Condition Verification}

\begin{figure}[h]
	\centering
	\subfloat[The SBM Network]{
		\includegraphics[width=0.48\textwidth]{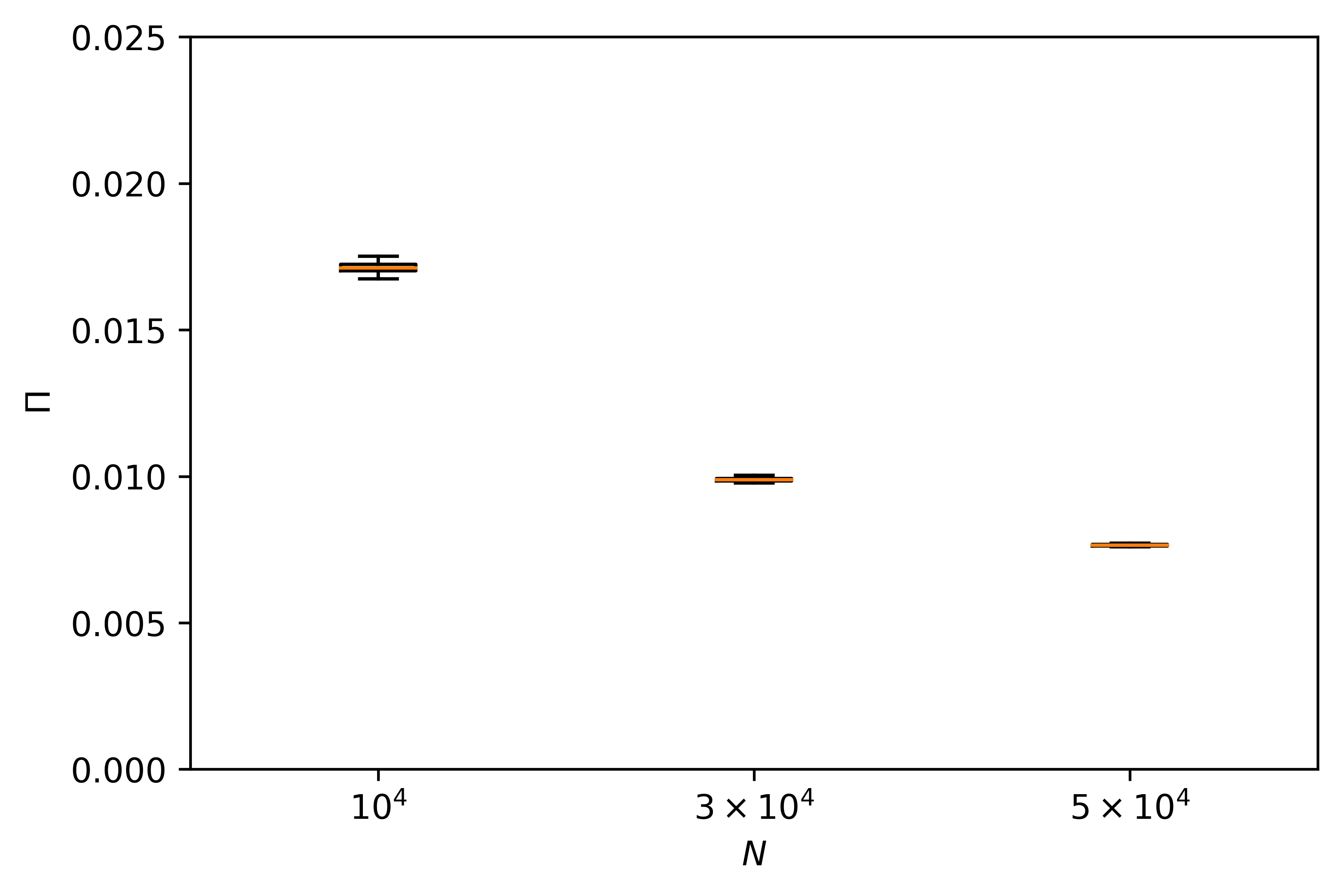}}\hfill
	\subfloat[The LSM Network]{
		\includegraphics[width=0.48\textwidth]{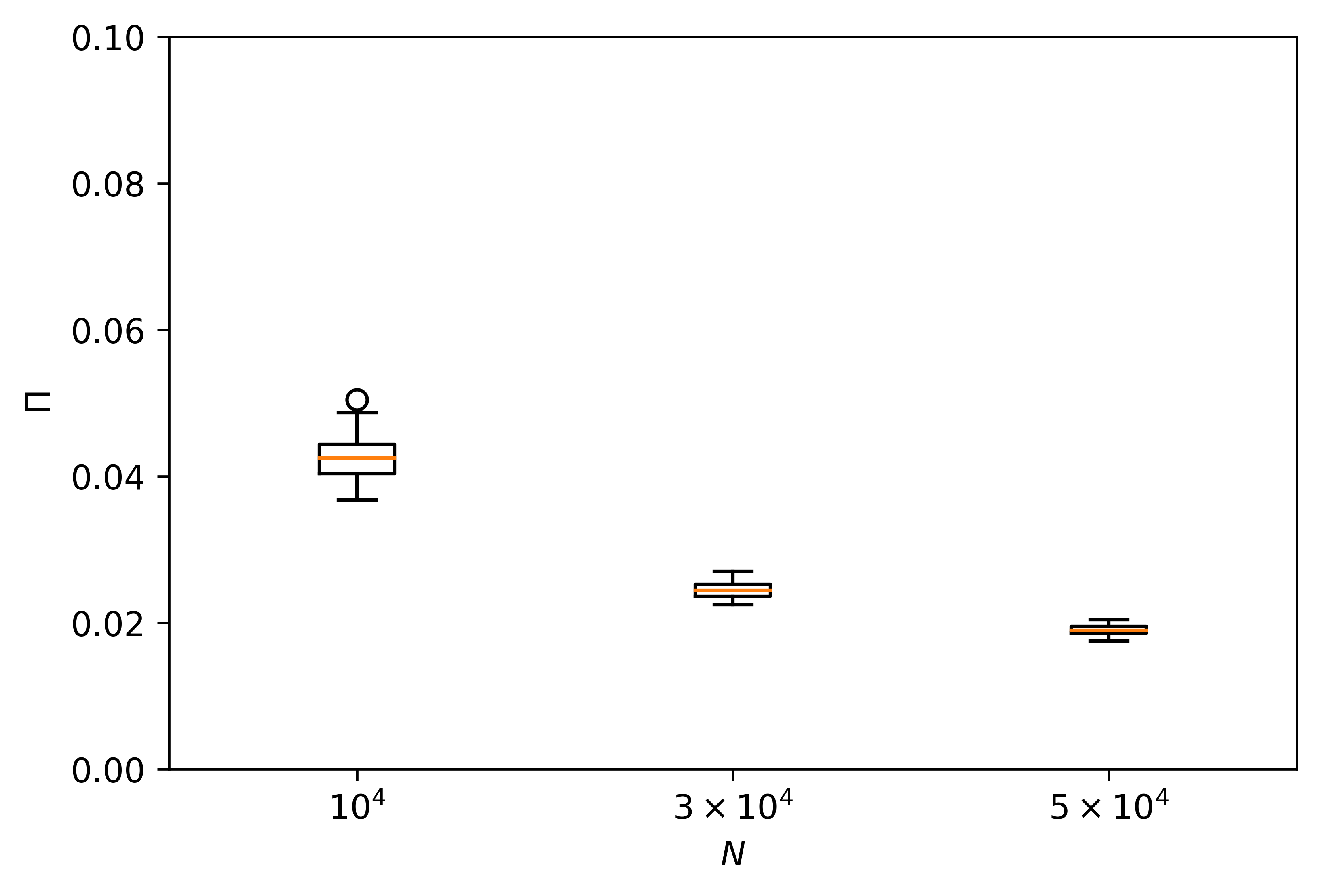}}\hfill
	\caption{ The boxplots of $\Pi$ values for the SBM network (left panel) and the LSM network (right panel). }
	\label{f:C1}
\end{figure}

We consider how to verify the technical conditions given in (C1)--(C3).
We start with the condition (C1). By (C1), we wish to have $\sqrt{N}\pi^{\top}\pi \rightarrow 0$ as $N \rightarrow \infty$, where $\pi$ is the stationary distribution of the $W$ matrix.
For illustration purpose, we fix $N/K=20$ for both SBM and LSM. To verify this condition, we compute $\pi$ for every simulated network structure in Sections 3.2 and 3.3.
Next, compute $\Pi=\sqrt{N}\pi^{\top}\pi$. We randomly repeated this experiment for a total of $M=100$ replications. This leads to a total of $M=100$ $\Pi$ values, which are then boxplotted in Figure \ref{f:C1}. By Figure \ref{f:C1}, we find that the $\Pi$ values steadily decrease toward 0 as $N$ diverges to infinity. This verifies the condition (C1) for the two synthetic network examples studied in simulation.
For the two real networks (i.e., the Weibo network and CC network), we conduct the same computation given the whole observed networks. The resulted $\Pi$ values for the Weibo network and CC network are $8.76\times10^{-8}$ and $1.03\times10^{-10}$, respectively. It seems that they are already extremely small.

\begin{figure}[h]
	\centering
	\subfloat[The SBM Network]{
		\includegraphics[width=0.48\textwidth]{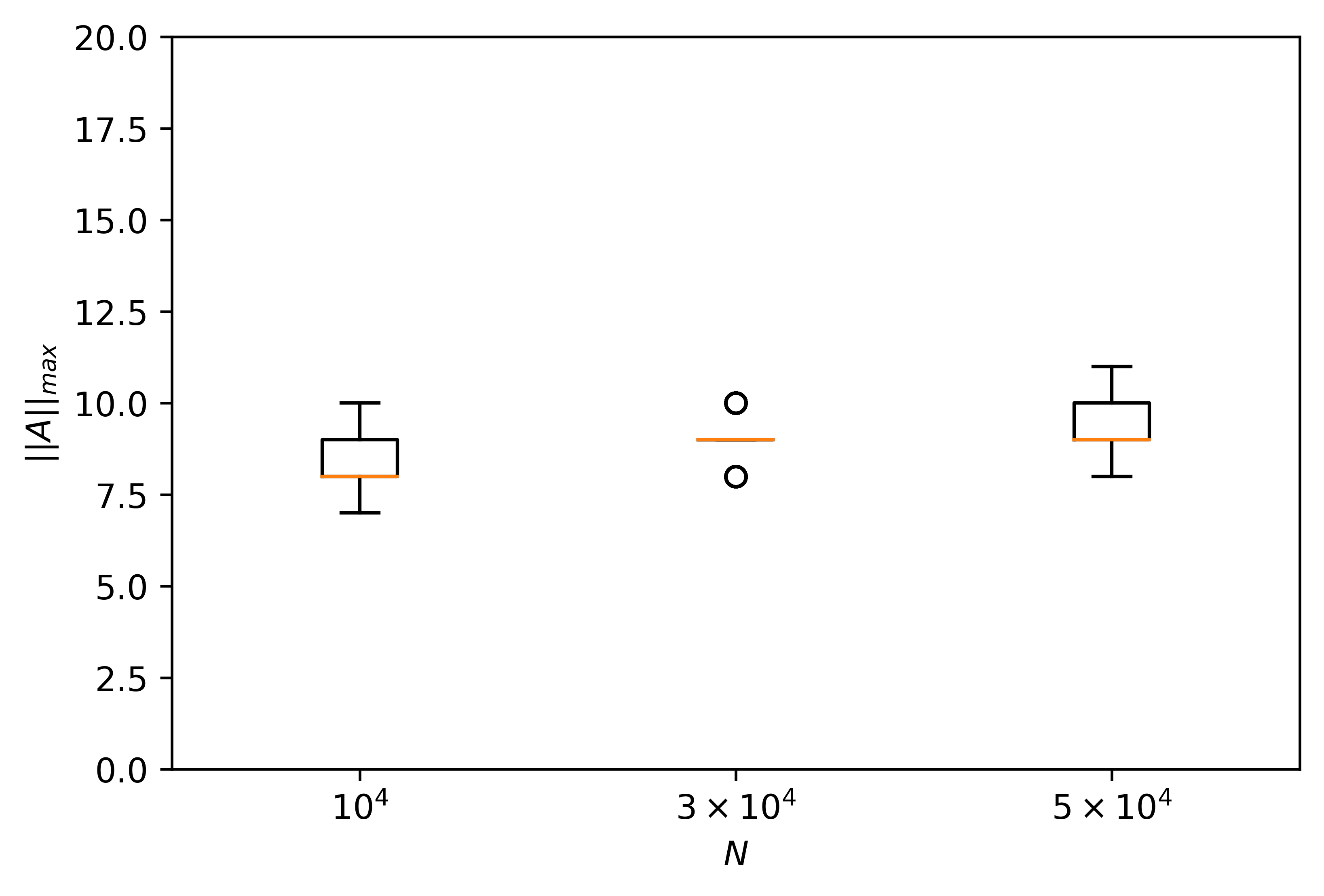}}\hfill
	\subfloat[The LSM Network]{
		\includegraphics[width=0.48\textwidth]{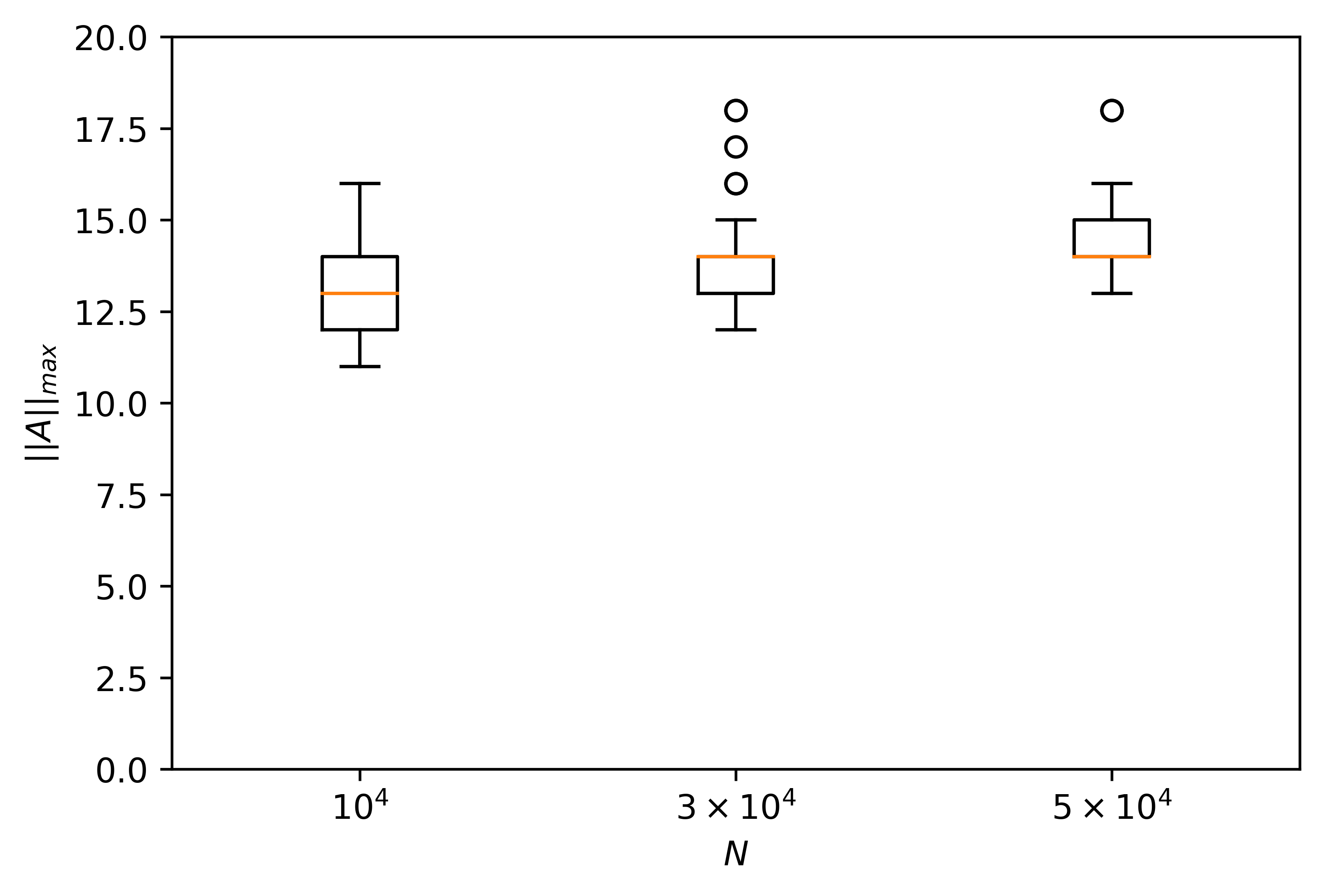}}\hfill
	\caption{ The boxplots of $\|A\|_{\max}$ values for the SBM network (left panel) and the LSM network (right panel). }
	\label{f:C2}
\end{figure}

We next study the condition (C2). By condition (C2), we wish to have $\|A\|_{\max}$ uniformly bounded away from infinity as $N$ diverges toward infinity, where $W$ is the weighting matrix of the whole network.
To verify this condition, we compute $\|A\|_{\max}$ for every simulated network structure $N/K=20$ for both SBM and LSM.
The experiment is randomly repeated for $M=100$ replications.
This leads to a total of $M=100$ $\|A\|_{\max}$ values, which are then boxplotted in Figure \ref{f:C2}.
By Figure \ref{f:C2}, we find the $\|A\|_{\max}$ values are all bounded away from infinity as $N$ diverges to infinity.
This verifies the condition (C2) for the the SBM and LSM network examples in simulation studies.
For the Weibo network and CC network, we conduct the same computation. The resulted $\|A\|_{\max}$ for the Weibo network and CC network are 9814 and 602, respectively.
As shown, the $\|A\|_{\max}$ value in the Weibo network is much larger than that in the CC network. This is because in the Weibo network, there exist superstars which have a huge number of followers.
However, the two values seem still bounded away from infinity in the real networks.

\begin{figure}[ht]
	\centering
	\subfloat[The SBM Network]{
		\includegraphics[width=0.48\textwidth]{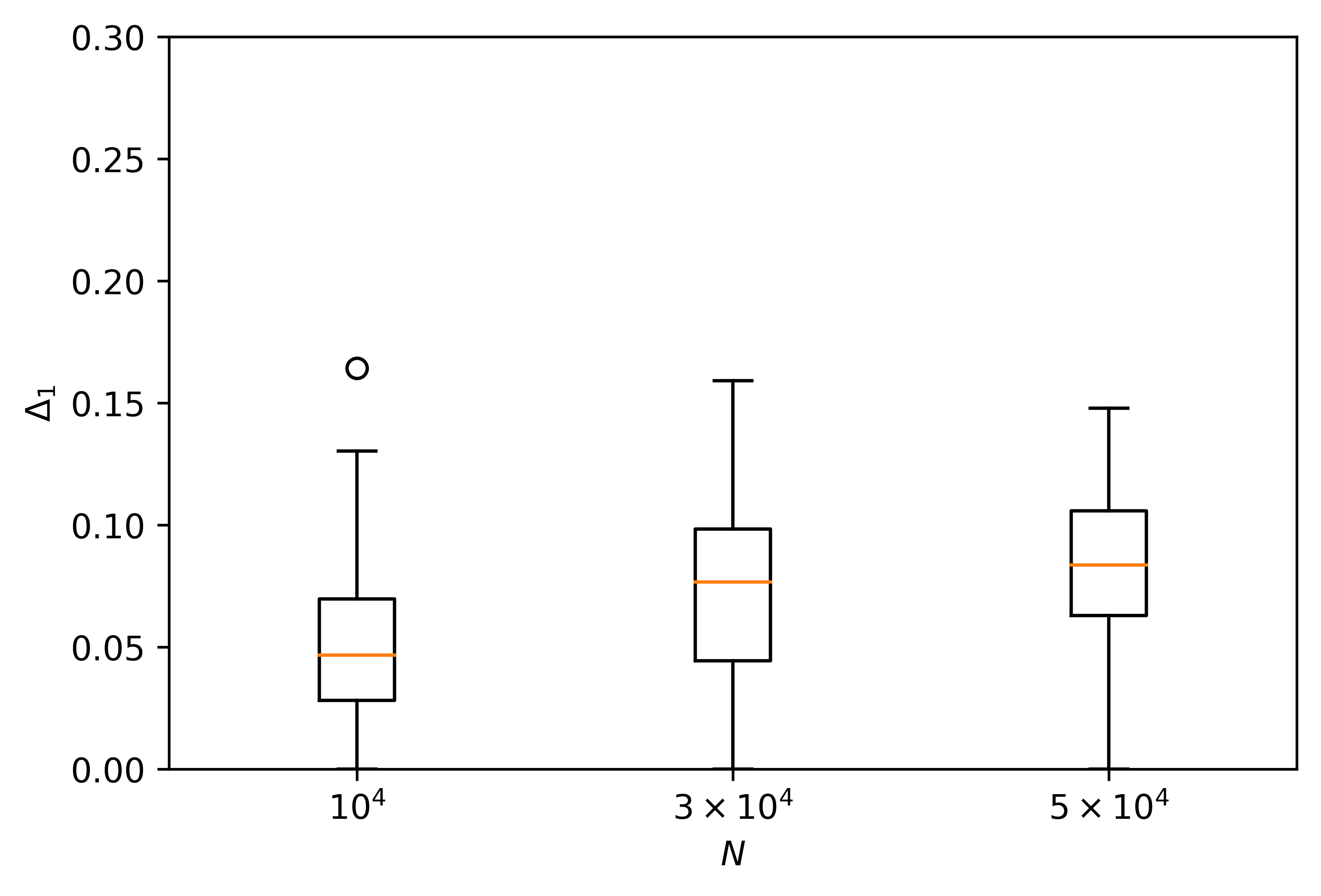}}\hfill
	\subfloat[The LSM Network]{
		\includegraphics[width=0.48\textwidth]{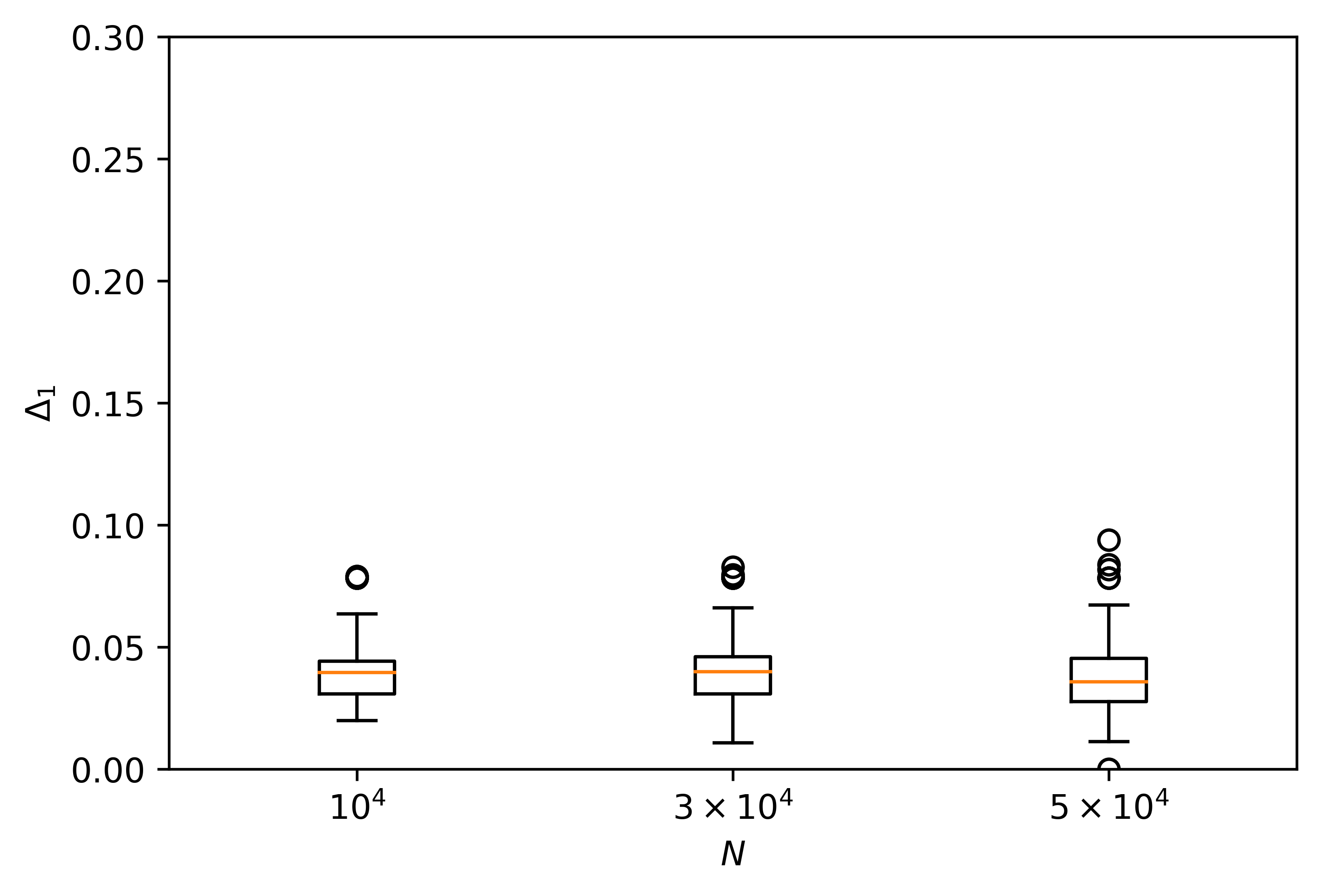}}\hfill
	\caption{The boxplots of $\Delta_1$ values for the SBM network (left panel) and the LSM network (right panel). }
	\label{f:C31}
\end{figure}

Last, we study the condition (C3), which has two subconditions. We start with condition (C3.1). Define $\Delta_1=\rho^2c_{\min}^{-2} \lambda_{\max}\big(W_{12}^\top W_{12}\big) \lambda_{\max}\big(W_{21}^\top W_{21}\big)$, where $c_{\min}$ is the minimal value of $\lambda_{\min} \big\{ \big(I_{ii}-\rho W_{ii}^\top\big) \big(I_{ii}-\rho W_{ii}\big)  \big\}$ for $i=1$ and 2. Then by condition (C3.1), we wish to have $\Delta_1 < 1$.
To verify this condition, we consider SBM and LSM as network examples and fix $N/K=20$. The SNOW method is applied for network sampling. Meanwhile, we still set $n/N=0.01$.
For each generated network, we compute $\Delta_1$. The experiment is randomly repeated for a total of $M=100$ replications.
The resulting $\Delta_1$ values under these two network structures are then box-plotted in Figure \ref{f:C31}. It is obvious that, $\Delta_1$ is uniformly bounded by 1 as $n$ diverges to infinity. We also conduct the same computation for the Weibo and CC networks. The $\Delta_1$ values for the Weibo network and CC network are 2358.613 and 6.158, respectively.
This suggests the condition (C3.1) is well satisfied for the SBM and LSM networks.
However, it seems not the case for the Weibo network and the CC network, as the value of $\Delta_1$ is much larger than 1.
However, although this condition is not verified for the two real networks, we surprisingly find that the subnetwork estimator $\wh{\rho}_{\mS}$ still performs very well empirically as shown in Table \ref{t:real}. This finding suggests that, the conditions given in (C1)--(C3) are sufficient conditions but not necessary.

\begin{figure}[h]
	\centering
	\subfloat[The SBM Network]{
		\includegraphics[width=0.48\textwidth]{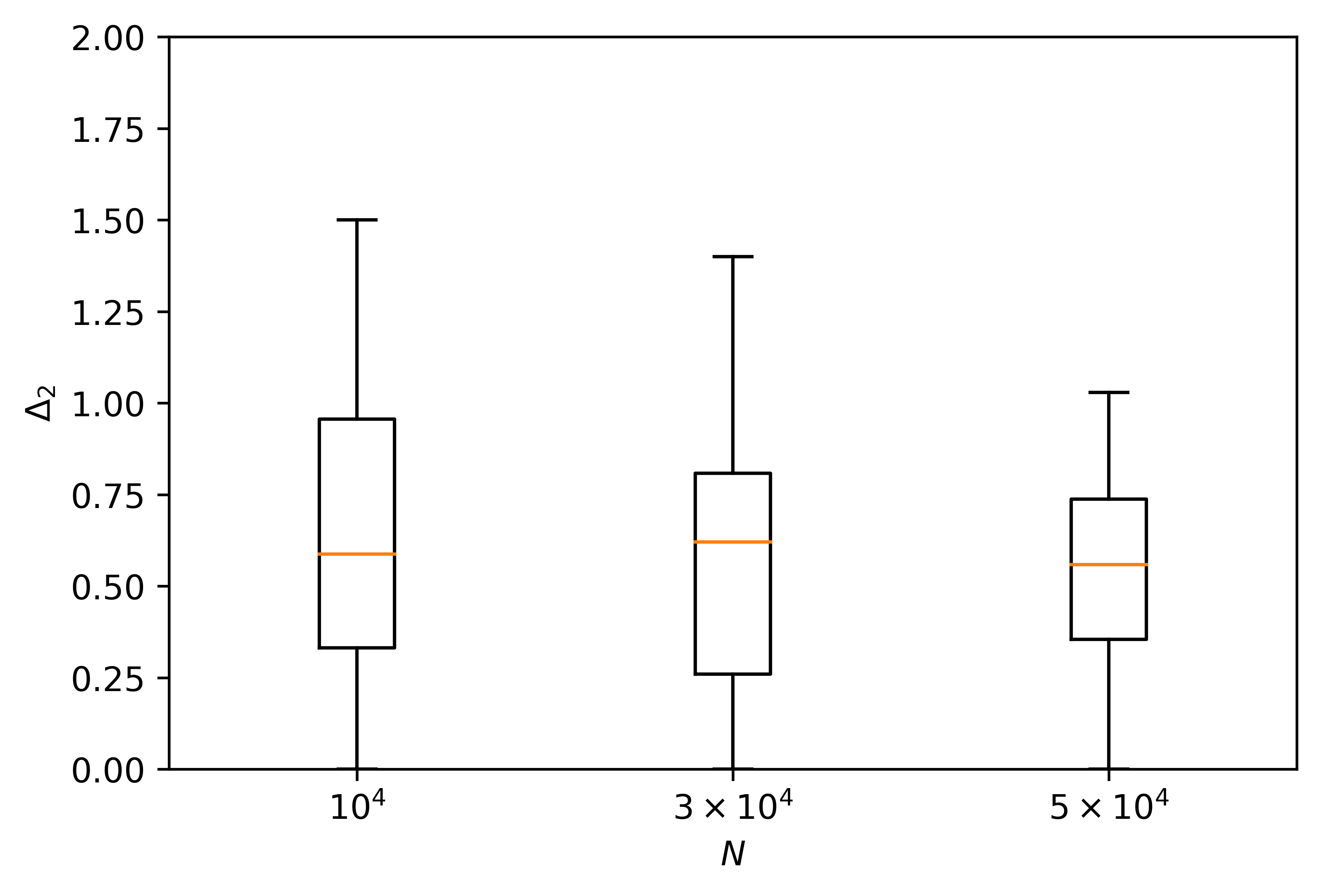}}\hfill
	\subfloat[The LSM Network]{
		\includegraphics[width=0.48\textwidth]{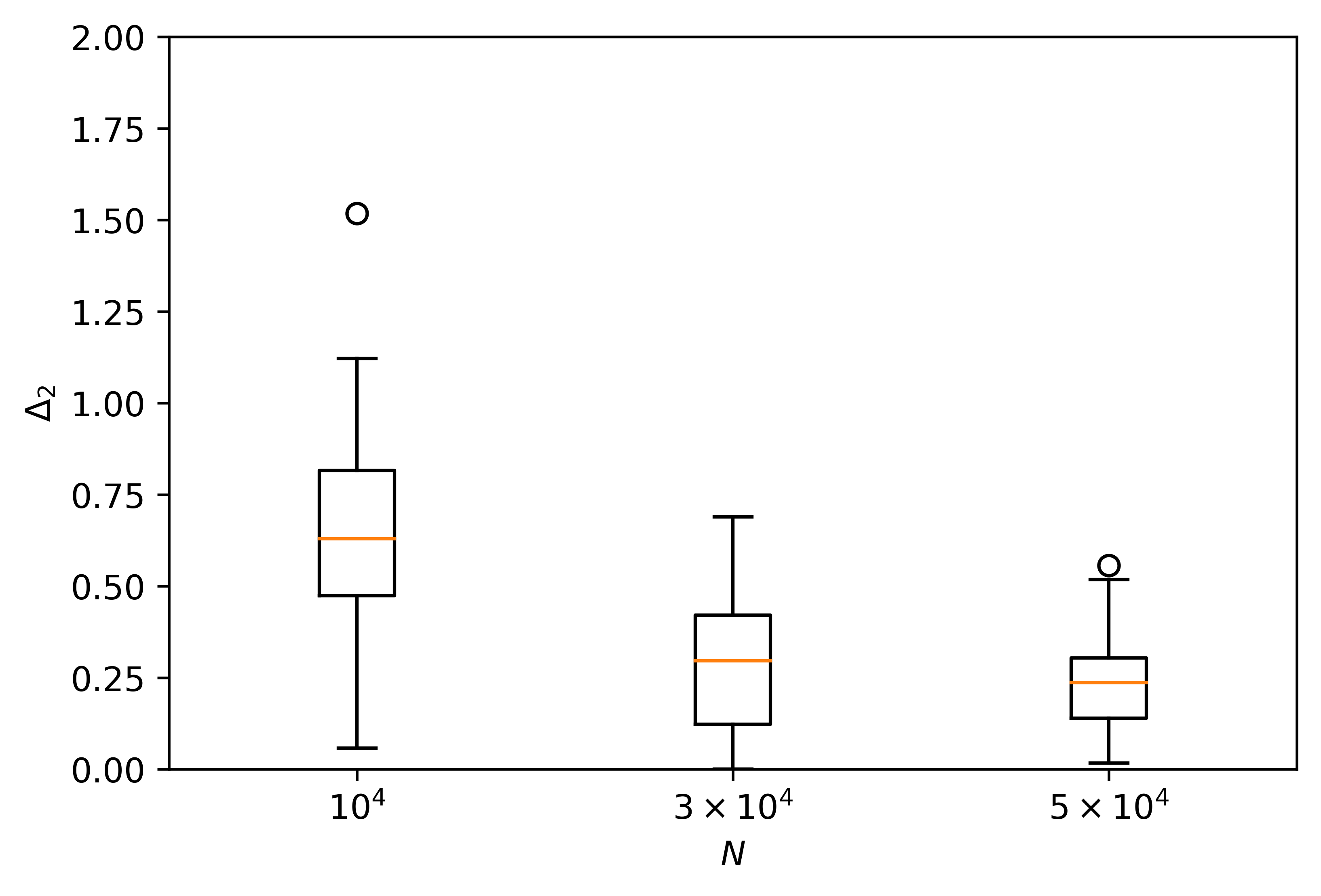}}\hfill
	\caption{ The boxplots of $\Delta_2$ for the SBM network (left panel), the LSM network (right panel).}
	\label{f:C32}
\end{figure}

Finally, we study the condition (C3.2).
By condition (C3.2), we wish to have $\Delta_2=\big\{ \tr\big(W_{12}^{\top}W_{12} \big)+\tr\big(W_{21}^{\top}W_{21} \big) \big\}/\sqrt{n} \rightarrow 0$ as $n \rightarrow \infty$.
Similarly, we compute $\Delta_2$ for the generated SBM and LSM networks. The experiment is randomly repeated for a total of $M=100$ replications.
The resulting $\Delta_2$ values under these two network structures are then box-plotted in Figure \ref{f:C32}. It is obvious that, $\Delta_2 \to 0$ as $n$ diverges to infinity.
We also conduct the same computation for the Weibo and CC networks.
The resulted $\Delta_2$ values for the Weibo network and CC network are 138.674 and 38.124, respectively.
It seems that the $\Delta_2$ values are not small at all. This suggests that, the condition (C3.2) is not satisfied for the two real networks. However, as demonstrated in Table \ref{t:real}, we find that the resulting subnetwork estimator still performs well.
This suggests that the condition (C3.2) might be unnecessarily straightforward, and thus should be further relaxed in the future study.

\setcounter{equation}{0}
\renewcommand\theequation{C.\arabic{equation}}
\setcounter{table}{0}
\setcounter{figure}{0}
\renewcommand{\thetable}{C.\arabic{table}}
\renewcommand{\thefigure}{C.\arabic{figure}}

\scsection{Appendix C: Other Network Sampling Methods}

For a more comprehensive evaluation, we consider here six other network sampling methods.
They are, respectively, the breadth-first-search (BFS) and depth-first-search (DFS) methods \citep{2001Introduction}, the forest fire (FF) method \citep{leskovec2006sampling}, the SNOW-$k$ method \citep{goodman1961snowball}, the random walk with restart (RWR) and the random walk with random jump (RWJ) methods \citep{al2016methods}.
It is remarkable that, the BFS method seems to be the same as our SNOW method, which is discussed in Section 2.3. Therefore, we omit this method here. This leads to a total of five benchmark subsampling methods for further evaluation. Their implementation details are given below.

The DFS method \citep{2001Introduction} starts with a seed node $i_1$. Denote $\mS^*$ to be the set contained all selected nodes and start with $\mS^*=\{i_1\}$. Then we randomly select one node connected with the seed node (denoted by $i_2$) and update $\mS^*$ to be $\mS^*=\{i_1,i_2\}$. Regard node $i_2$ as the new seed node and randomly select one node connected with $i_2$ and not contained in $\mS^*$. Repeat this procedure until $|\mS^*|=n$ or all nodes connected with the current seed node (denoted by $i_t$) are already contained in $\mS^*$. If the latter situation happens, then conduct backtracking by regarding $i_{t-1}$ as the seed node again and select another node connected with $i_{t-1}$. The FF method and SNOW-$k$ method generally follow the similar selection procedure with SNOW. The only difference lies in how to select nodes that are connected with the seed nodes in each iteration. In this regard, the SNOW method selects \textit{all} nodes connected with \textit{each} seed node. The FF method \citep{leskovec2006sampling} applies a probabilistic procedure. For each seed node $i$, it randomly select $r_i$ nodes connected with this seed node, where $r_i$ follows a geometric distribution with probability $p_{\rm FF}$. The SNOW--$k$ method \citep{goodman1961snowball} randomly selects a fixed number of nodes (denoted by $k$) connected with each seed node. The RWR and RWJ methods \citep{al2016methods} generally use the similar selection procedure with DFS, but apply a two-step strategy. Specifically, for the current seed node (denoted by $i$), they first select a latent indicator $Z_i$ with probability $p_{\rm RW}$. If $Z_i=1$, then we randomly select one node connected with the seed node $i$. If $Z_i=0$, the RWR method would not select any node, but restart from the first seed node; while the RWJ method would randomly select one node from all the non-selected nodes.

\begin{table}[h]
\centering
\caption{The detailed simulation results under the SBM network structure (Panel A) and the LSM network structure (Panel B) with different $N$s, network sampling methods, and the {\sc EXP} error distribution. The bias $\flat$, estimated standard error $\wh{\rm SE}$, true standard error SE, and the empirical coverage probability ECP are reported. The average CPU computational time is also reported in seconds.}
\resizebox{\hsize}{!}{
\begin{tabular}{c|c|ccccc|ccccc}
	\hline\hline
	\multirow{2}[4]{*}{Methods} & \multirow{2}[4]{*}{$N$} & \multicolumn{5}{c|}{Panel A: SBM Network}              & \multicolumn{5}{c}{Panel B: LSM Network} \bigstrut\\
	\cline{3-12}          &       & $\flat$     & $\wh{\rm SE}$ & SE    & ECP   & CPU   & $\flat$     & $\wh{\rm SE}$ & SE    & ECP   & CPU \\
\hline
\multirow{3}{*}{DFS}	&10000	&-0.006	&0.134	&0.139	&94.6\%	&0.02	&-0.002	&0.126	&0.128	&93.0\%	&0.03	\\
	&30000	&0.002	&0.079	&0.076	&95.6\%	&0.19	&0.002	&0.075	&0.069	&96.0\%	&0.11	\\
	&50000	&-0.003	&0.061	&0.060	&95.4\%	&0.32	&0.001	&0.058	&0.058	&95.4\%	&0.30	\\
\hline												
\multirow{3}{*}{FF}	&10000	&-0.004	&0.129	&0.129	&94.2\%	&0.02	&-0.014	&0.127	&0.126	&94.2\%	&0.04	\\
	&30000	&-0.009	&0.075	&0.072	&96.0\%	&0.09	&-0.002	&0.072	&0.071	&94.6\%	&0.09	\\
	&50000	&-0.004	&0.058	&0.059	&93.8\%	&0.23	&-0.006	&0.056	&0.056	&94.8\%	&0.18	\\
\hline												
\multirow{3}{*}{SNOW-$k$}	&10000	&-0.010	&0.127	&0.122	&95.4\%	&0.02	&-0.017	&0.123	&0.121	&94.0\%	&0.02	\\
	&30000	&-0.001	&0.073	&0.075	&93.4\%	&0.09	&-0.001	&0.070	&0.072	&93.2\%	&0.09	\\
	&50000	&-0.004	&0.057	&0.058	&93.8\%	&0.22	&-0.004	&0.055	&0.050	&96.2\%	&0.18	\\
\hline												
\multirow{3}{*}{RWR}	&10000	&-0.008	&0.171	&0.170	&94.4\%	&0.03	&0.002	&0.154	&0.153	&93.6\%	&0.03	\\
	&30000	&-0.007	&0.098	&0.095	&95.8\%	&0.13	&0.001	&0.089	&0.082	&95.6\%	&0.17	\\
	&50000	&0.002	&0.076	&0.078	&93.2\%	&0.32	&0.003	&0.068	&0.066	&95.6\%	&0.42	\\
\hline											
\multirow{3}{*}{RWJ}	&10000	&0.005	&0.187	&0.194	&93.0\%	&0.04	&-0.015	&0.167	&0.157	&96.0\%	&0.03	\\
	&30000	&0.003	&0.108	&0.106	&94.6\%	&0.18	&-0.004	&0.096	&0.099	&94.2\%	&0.19	\\
	&50000	&-0.005	&0.084	&0.083	&95.8\%	&0.44	&0.004	&0.074	&0.076	&93.6\%	&0.47	\\
\hline \hline
\end{tabular}%
}
\label{t: sm5}%
\end{table}%

Once the network sampling methods are given, the simulation examples can be considered in the same way as in Section 3. For illustration purpose, we take the SBM and LSM network structures as examples, and fix $\rho=0.2$ and $N/K=20$. We set $p_{\rm FF}=0.25$ for the FF method, $k=5$ for the SNOW--$k$ method, and $p_{\rm RW}=0.75$ for the RWR and RWJ methods. The experiment is randomly replicated for a total of $M=500$ times. The detailed results are summarized in Table \ref{t: sm5}. We find that, the results are quantitatively similar with those in Tables \ref{t:SBM} and \ref{t:LSM}. Simply sparking, the proposed subnetwork estimation method can produce encouraging results by using these network sampling methods. We also investigate the performance of these network sampling methods in the bootstrap method discussed in Section 3.6. Table \ref{t:bootstrap2} presents the detailed results. As shown, the finite sample performance of the bootstrap method remains to be satisfactory by using these network sampling methods.

\begin{table}[h]
	\caption{ The detailed simulation results for the subnetwork estimator using the bootstrap method with different network sampling methods and the {\sc EXP} error distribution. The true standard error (SE), the estimated standard error ($\widehat{\text{SE}}_{\rm bt}$) and empirical coverage probability ($\text{ECP}_{\rm bt}$) by using the bootstrap method are reported. The average CPU computational time is also reported in seconds.}
	\label{t:bootstrap2}
 \centering
 \footnotesize
 \begin{tabular}{cc|cccc|cccc}
  \hline\hline
  \multirow{2}{*}{Methods} &\multirow{2}{*}{$N$} &\multicolumn{4}{c|}{Panel A: SBM Network} &\multicolumn{4}{c}{Panel B: LSM Network} \\
  \cline{3-6}
  \cline{7-10}
  &       &SE      & ${\widehat{\rm SE}}_{\rm bt}$ & ${\rm ECP}_{\rm bt}$ & ${\rm CPU}_{\rm bt}$   & SE & ${\widehat{\rm SE}}_{\rm bt}$ & ${\rm ECP}_{\rm bt}$ & ${\rm CPU}_{\rm bt}$ \\
  \hline
\multirow{3}{*}{DFS}	&10000	&0.139	&0.122	&94.4\%	&0.68	&0.128	&0.114	&93.0\%	&0.46	\\
	&30000	&0.076	&0.073	&93.2\%	&3.01	&0.069	&0.071	&94.2\%	&2.40	\\
	&50000	&0.060	&0.058	&93.4\%	&7.07	&0.058	&0.056	&93.0\%	&5.65	\\
\hline										
\multirow{3}{*}{FF}	&10000	&0.129	&0.120	&94.8\%	&0.24	&0.126	&0.119	&96.2\%	&0.29	\\
	&30000	&0.072	&0.071	&93.2\%	&1.19	&0.071	&0.068	&94.2\%	&1.38	\\
	&50000	&0.059	&0.055	&96.4\%	&2.93	&0.056	&0.054	&94.0\%	&3.09	\\
\hline										
\multirow{3}{*}{SNOW--$k$}	&10000	&0.122	&0.119	&94.4\%	&0.25	&0.121	&0.115	&94.2\%	&0.28	\\
	&30000	&0.075	&0.070	&93.6\%	&1.11	&0.072	&0.067	&93.6\%	&1.25	\\
	&50000	&0.058	&0.054	&95.4\%	&3.70	&0.050	&0.052	&94.2\%	&3.93	\\
\hline										
\multirow{3}{*}{RWR}	&10000	&0.170	&0.162	&95.2\%	&0.39	&0.153	&0.141	&93.0\%	&0.50	\\
	&30000	&0.095	&0.093	&96.2\%	&2.29	&0.082	&0.083	&94.2\%	&3.21	\\
	&50000	&0.078	&0.073	&93.0\%	&5.80	&0.066	&0.065	&94.8\%	&7.82	\\
\hline										
\multirow{3}{*}{RWJ}	&10000	&0.194	&0.172	&94.2\%	&0.54	&0.157	&0.155	&95.0\%	&0.38	\\
	&30000	&0.106	&0.101	&96.0\%	&2.95	&0.099	&0.091	&95.6\%	&2.25	\\
	&50000	&0.083	&0.078	&94.4\%	&7.61	&0.076	&0.070	&94.4\%	&5.47	\\
  \hline\hline
 \end{tabular}%
\end{table}%

\end{CJK}
\end{document}